\documentclass[10pt,journal]{IEEEtran}
\usepackage{cite,calc}
\usepackage[cmex10]{amsmath} % ensure not too small fonts (so Type 1 still)
\usepackage{amsfonts,amssymb,graphics,subfigure,epsfig,graphics}
\usepackage[mathcal]{euscript} % redefine \mathcal to be \EuScript;
\newtheorem{thm}{Theorem}
\newtheorem{lem}{Lemma}

\newtheorem{cor}{Corollary}
\newtheorem{fact}{Fact}
\newtheorem{claim}{Claim}

\newtheorem{defn}{Definition}

\newtheorem{exa}{Example}
\newtheorem{remark}{Remark}

\DeclareMathOperator{\poly}{poly}

\newcommand{\ds}{\displaystyle}

\newcommand{\defeq}{\triangleq}

\newcommand{\asyncexp}{\alpha}
\newcommand{\asyncexpt}{\asyncexp^\mathrm{T}}
\newcommand{\asyncexptl}{\asyncexp^\mathrm{T}_-}
\newcommand{\asyncexptu}{\asyncexp^\mathrm{T}_+}

\newcommand{\asyncexpub}{\asyncexp_+}
\newcommand{\asyncexplb}{\asyncexp_-}

\newcommand{\asynclev}{A}

\newcommand\cX{\mathcal{X}}
\newcommand\cY{\mathcal{Y}}
\newcommand{\cA}{\mathcal{A}}
\newcommand{\cB}{\mathcal{B}}
\newcommand{\cC}{\mathcal{C}}
\newcommand{\cD}{\mathcal{D}}
\newcommand{\cE}{\mathcal{E}}
\newcommand{\cF}{\mathcal{F}}

\newcommand{\cN}{\mathcal{N}}
\newcommand{\cP}{\mathcal{P}}
\newcommand{\cS}{\mathcal{S}}

\newcommand{\cT}{\mathcal{T}}

\newcommand{\comp}{\mathrm{c}}

\newcommand{\pr}{{\mathbb{P}}}
\newcommand{\ex}{{\mathbb{E}}}
\newcommand{\E}{\cE}
\newcommand{\openone}{\leavevmode\hbox{\small1\normalsize\kern-.33em1}}

\newcommand{\thres}{\asyncexp_\circ}

\newcommand{\eps}{\epsilon}
\newcommand{\veps}{\eps}
\newcommand{\DB}{D_\mathrm{B}}
\newcommand{\delay}{\Delta}  % or use \Delta or T

\begin{document}

% \title{Basic Limits on Asynchronous Communication \\ over Memoryless Channels}

%\title{On the Capacity of Asynchronous Channels\\ and the
%  Suboptimality of Training}
%\title{Asynchronous Communication:\\ 
%Capacity Bounds and Training Performance}
%\title{On Asynchronous Capacity and the Suboptimality of Training}
\title{Asynchronous Communication:\\ 
Capacity Bounds and Suboptimality of Training}

% The Asynchronous Capacity of Channels:  Bounds and Training Performance
%On the Asynchronous Capacity of a Channel\\ and the  Suboptimality of Training
% On the Capacity Region of Asynchronous Communication
% Training-Based Synchronization for Communication is Inefficient
%

\author{Aslan Tchamkerten, Venkat Chandar, and Gregory
W.~Wornell~\IEEEmembership{Fellow,~IEEE}

\thanks{This work was supported in part by an Excellence Chair Grant
from the French National Research Agency (ACE project), and by the
National Science Foundation under Grant No.~CCF-1017772.   This work
was presented in part at the IEEE International Symposium on Information Theory,
Toronto, Canada, July 2008~\cite{TCW2}, and at the IEEE Information Theory Workshop,
Taormina, Italy, October 2009~\cite{CTW2}.}  \thanks{A.~Tchamkerten is with the
Department of Communications and Electronics, Telecom
ParisTech, 75634 Paris Cedex 13, France.
(Email: aslan.tchamkerten@telecom-paristech.fr).}  \thanks{V.~Chandar
is with MIT Lincoln Laboratory, Lexington, MA 02420 (Email:
vchandar@mit.edu).}  \thanks{G.~W.~Wornell is with the Department of
Electrical Engineering and Computer Science, Massachusetts Institute
of Technology, Cambridge, MA 02139 (Email: gww@mit.edu).}}

\maketitle

\begin{abstract} 
Several aspects of the problem of asynchronous point-to-point
communication without feedback are developed when the source is highly
intermittent.  In the system model of interest, the codeword is
transmitted at a random time within a prescribed window whose length
corresponds to the level of asynchronism between the transmitter and
the receiver. The decoder operates sequentially and communication rate
is defined as the ratio between the message size and the elapsed time
between when transmission commences and when the decoder makes a
decision.

For such systems, general upper and lower bounds on capacity as a
function of the level of asynchronism are established, and are shown
to coincide in some nontrivial cases.  From these bounds, several
properties of this asynchronous capacity are derived.  In addition,
the performance of training-based schemes is investigated. It is shown
that such schemes, which implement synchronization and information
transmission on separate degrees of freedom in the encoding, cannot
achieve the asynchronous capacity in general, and that the penalty is
particularly significant in the high-rate regime.
\end{abstract}


\section{Introduction}
\label{intro}

\IEEEPARstart{I}{nformation-theoretic} analysis of communication systems
frequently ignores synchronization issues. In many applications where large
amounts of data are to be transmitted,  such simplifications may
be justified. Simply prepending a suitable synchronization preamble to the
initial data incurs negligible overhead yet ensures that the transmitter and the
receiver are synchronized.  In turn,  various coding techniques
(e.g., graph based codes, polar codes) may guarantee delay optimal communication for
data transmission in the
sense that they can achieve the capacity of the synchronous channel.

In quantifying the impact due to a lack of
synchronization  between a transmitter and a receiver, it
is important to note that asynchronism is a relative
notion that depends on the size of the data to be
transmitted. For instance, in the above ``low
asynchronism'' setting it is implicitly assumed that the
data is large with respect to the timing uncertainty. 

In a growing number of applications, such as many involving sensor networks,
data is transmitted in a bursty manner. An example would be a sensor in a monitoring system.  By
contrast with the previous setting, here timing uncertainty is
large with respect to the data to be transmitted.

To communicate in such ``high asynchronism'' regimes, one can use the
traditional preamble based communication scheme for each block.  Alternatively,
one can pursue a fundamentally different strategy in which synchronization is
integrated into the encoding of the data, rather than separated from it.

To evaluate the relative merits of such diverse strategies, and more generally
to explore fundamental performance limits, we recently introduced a general
information-theoretic model for asynchronous communication in \cite{TCW}. This
model extends Shannon's original communication model~\cite{Sha2} to include
asynchronism. In this model, the message is encoded into a codeword of fixed
length, and this codeword starts being sent across a discrete memoryless
channel at a time instant that is randomly and uniformly distributed over some predefined
transmission window.  The size of this window is known to transmitter and
receiver, and the level of asynchronism in the system is governed by the size
of the window with respect to the codeword length. Outside the information
transmission period, whose duration equals the codeword length, the transmitter
remains idle and the receiver observes noise, i.e., random output
symbols. The receiver uses a sequential decoder whose scope is twofold: decide
when to decode and what message to declare.

The performance measure is the communication rate which is defined as the ratio between the
message size and the average delay between when transmission starts and when the
message is decoded.  Capacity is the supremum of achievable rates, i.e., 
rates for which vanishing error probability can be guaranteed in the limit of
long codeword length.

The scaling between the transmission window and the codeword length that
meaningfully quantifies the level of asynchronism in the system turns out to be
exponential, i.e., $A=e^{\alpha n}$ where $A$ denotes the size of the
transmission window, where $n$ denotes the codeword length, and where $\alpha$
denotes the asynchronism exponent. Indeed, as discussed in \cite{TCW}, if $A$
scales subexponentially in $n$, then asynchronism doesn't impact communication: the
asynchronous capacity is equal to the capacity of the synchronous channel. By
contrast, if the window size scales superexponentially, then the asynchrony is
generally catastrophic.  Hence, exponential asynchronism is the interesting
regime and we aim to compute capacity as a function of the asynchronism exponent.

For further motivation and background on the model, including a
summary of related models (e.g., the insertion, deletion, and
substitution channel model, and the detection and isolation model) we
refer to \cite[Section II]{TCW}. Accordingly, we omit such material
from the present paper.

The first main result in \cite{TCW} is the characterization of the
synchronization threshold, which is defined as the largest
asynchronism exponent for which it is still possible to guarantee
reliable communication---this result is recalled in
Theorem~\ref{ulimit} of Section~\ref{results}.

The second main result in \cite{TCW} (see \cite[Theorem 1]{TCW}) is a
lower bound to capacity. A main consequence of this bound is that for
any rate below the capacity of the synchronous channel it is possible
to accommodate a non-trivial asynchronism level, i.e., a positive
asynchronism exponent.

While this work focuses on rate, an alternative performance metric is
the minimum energy (or, more generally, the minimum cost) needed to
transmit one bit of information asynchronously. For this metric,
\cite{CTT,CTTj} establishes the capacity per unit cost for the above
bursty communication setup.

We now provide a brief summary of the results contained in this paper:
\begin{itemize} \item {\it{General capacity lower bound, Theorems~\ref{ach}
and~\ref{tgaussian}.}} Theorem~\ref{ach} provides a lower
bound to capacity which is obtained by considering a coding scheme
that performs synchronization and information transmission jointly. The derived
bound results
in a much simpler and often much better lower bound than the one obtained in
\cite[Theorem 1]{TCW}. Theorem~\ref{ach}, which holds for arbitrary discrete
memoryless channels, also holds for a natural Gaussian setting, which yields
Theorem~\ref{tgaussian}.

\item {\it{General capacity upper bound, Theorem~\ref{convs}.}} This bound and
the above lower bound, although not tight in general, provide interesting and
surprising insights into the asynchronous capacity. For instance, 
Corollary~\ref{limittocap} says that, in general, it is possible to
reliably achieve a communication rate equal to the capacity of the synchronous
channel while operating at a strictly positive asynchronism exponent. In other
words, it is possible to accommodate both a high rate and an exponential
asynchronism. 

Another insight is provided by Corollary~\ref{disc}, which relates to
the very low rate communication regime. This result says that, in
general, one needs to (sometimes significantly) back off from the
synchronization threshold in order to be able to accommodate a
positive rate. As a consequence, capacity as a function of the
asynchronism exponent does not, in general, strictly increase as the
latter decreases.
\item {\it{Capacity for channels with infinite synchronization threshold,
Theorem~\ref{alphinfin}.}} For the class of channels for which there
exists a particular channel input which can't be confused with noise, a
closed-form expression for capacity is established.
\item {\it Suboptimality of training based schemes, Theorem~\ref{thm2}, Corollaries~\ref{cotocap}
and \ref{subtr}.} These results show that communication strategies that separate
synchronization from information transmission do not achieve the asynchronous
capacity in general. 
\item{\it{Good synchronous codes, Theorem~\ref{prop:expurgation}.}} This result
may be independent interest and
relates to synchronous communication. It says that any codebook that achieves a nontrivial error
probability contains a large subcodebook, whose rate is almost the same as the
rate of the original codebook, and whose error probability decays exponentially with the
blocklength with a suitable decoder. This result, which is a byproduct of our
analysis, is a stronger version of \cite[Corollary~1.9,
  p.~107]{CK} and its proof amounts to a tightening of some of the
arguments in the proof of the latter. 
\end{itemize}

It is worth noting that most of our proof techniques differ in some significant
respects from more traditional capacity analysis for synchronous
communication---for example, we make little use of Fano's inequality for
converse arguments. The reason for this is that there are decoding error events
specific to asynchronous communication. One such event is when the decoder, unaware
of the information transmission time, declares a message before
transmission even starts.

An outline of the paper is as follows.  Section~\ref{notation}
summarizes some notational conventions and standard results we make
use of throughout the paper.  Section~\ref{moper} describes the communication
model of interest.  Section~\ref{results} contains our main results, and
Section~\ref{pfofresults} is devoted to the proofs. Section~\ref{concludingrem} contains some concluding remarks.

\section{Notation and Preliminaries}
\label{notation}

In general, we reserve upper case letters for random variables (e.g.,
$X$) and lower case letters to denote their corresponding sample
values (e.g., $x$), though as is customary, we make a variety of
exceptions.  Any potential confusion is generally avoided by context.
In addition, we use $x_i^j$ to denote the sequence
$x_i,x_{i+1},\dots,x_j$, for $i\le j$.  Moreover, when $i=1$ we use
the usual simpler notation $x^n$ as an alternative to $x_1^n$.
Additionally, $\defeq $ denotes ``equality by definition.''

Events (e.g., $\cE$) and sets (e.g., $\cS$) are denoted using
caligraphic fonts, and if $\cE$ represents an event, $\cE^\comp$
denotes its complement.  As additional notation, $\pr[\cdot]$ and
$\ex[\cdot]$ denote the probability and expectation of their
arguments, respectively, $\|\cdot\|$ denotes the $L_1$ norm of its
argument, $|\cdot|$ denotes absolute value if its argument is numeric,
or cardinality if its argument is a set, $\lfloor\cdot\rfloor$ denotes
the integer part of its argument, $a\wedge b\defeq\min\{a,b\}$, and
$x^+\defeq\max\{0,x\}$.  Furthermore, we use $\subset$ to denote nonstrict
set inclusion, and use the Kronecker notation $\openone(\cA)$ for the
function that takes value one if the event $\cA$ is true and
zero otherwise.

We also make use of some familiar order notation for asymptotics (see,
e.g., \cite[Chapter~3]{CLRS}).  We use $o(\cdot)$ and $\omega(\cdot)$
to denote (positive or negative) quantities that grow strictly slower
and strictly faster, respectively, than their arguments; e.g., $o(1)$
denotes a vanishing term and $n/\ln n= \omega(\sqrt{n})$.  We also use
$O(\cdot)$ and $\Omega(\cdot)$, defined analogously to $o(\cdot)$ and
$\omega(\cdot)$, respectively, but without the strictness constraint.
Finally, we use $\poly(\cdot)$ to denote a function that does not grow
or decay faster than polynomially in its argument.

We use $\pr(\cdot)$ to denote the probability of its argument, and use $\cP^\cX$, $\cP^\cY$, and $\cP^{\cX,\cY}$ to denote the set of
distributions over the finite alphabets $\cX$, $\cY$, and $\cX\times
\cY$ respectively, and use $\cP^{\cY|\cX}$ to denote the set of
conditional distributions of the form $V(y|x)$ for $(x,y)\in\cX\times
\cY$.

For a memoryless channel characterized by channel law
$Q\in\cP^{\cY|\cX}$, the probability of the output sequence
$y^n\in\cY^n$ given an input sequence $x^n\in\cX^n$
is
\begin{equation*}
Q(y^n|x^n) \defeq  \prod_{i=1}^n Q(y_i|x_i).
\end{equation*}
Throughout the paper,  $Q$ always refers to the underlying
channel and $C$ denotes its synchronous capacity.

Additionally, we use $J_\cX$ and $J_\cY$ to denote the left and
right marginals, respectively, of the joint distribution
$J\in\cP^{\cX,\cY}$, i.e.,
\begin{equation*}
J_{\cX}(x)\defeq\sum_{y\in \cY} J(x,y)\quad\text{and}\quad
J_{\cY}(y)\defeq\sum_{x\in \cX} J(x,y).
\end{equation*}

We define all information measures relative to the natural logarithm.
Thus, the entropy associated
with $P\in \cP^\cX$ is\footnote{In the
  definition of all such information measures, we use the usual
  convention $0\ln (0/0)=0$.}
\begin{equation*}
H(P) \defeq -\sum_{x\in \cX}P(x)\ln P(x),
\end{equation*}
and the conditional entropy associated with $Q\in \cP^{\cY|\cX}$ and
$P\in \cP^\cX$ is
\begin{equation*}
H(Q|P) \defeq -\sum_{x\in \cX}P(x)\sum_{y\in
\cY}Q(y|x)\ln Q(y|x).
\end{equation*}
Similarly, the mutual information induced by $J(\cdot,\cdot)\in
\cP^{\cX,\cY}$ is
\begin{equation*}
I(J) \defeq
\sum_{(x,y)\in\cX\times\cY} J(x,y) \ln \frac{J(x,y)}{J_\cX(x)J_\cY(y)},
\end{equation*}
so
\begin{equation*}
I(PQ) \defeq
\sum_{x\in\cX} P(x) \sum_{y\in\cY} Q(y|x) \ln
\frac{Q(y|x)}{(PQ)_\cY(y)}
\end{equation*}
for $P\in\cP^\cX$ and $W\in\cP^{\cY|\cX}$.
Furthermore, the information divergence (Kullback-Leibler distance) between
$P_1\in\cP^\cX$ and $P_2\in \cP^\cX$ is
\begin{equation*}
D(P_1\|P_2) \defeq  \sum_{x\in \cX}P_1(x)\ln \frac{P_1(x)}{P_2(x)},
\end{equation*}
and conditional information divergence is denoted using
\begin{align*}
D(W_1\|W_2|P)
&\defeq \sum_{x\in \cX} P(x) \sum_{y\in \cY}
W_1(y|x)\ln \frac{W_1(y|x)}{W_2(y|x)}\\
&\defeq D(PW_1\|PW_2),
\end{align*}
where $P\in \cP^\cX$ and $W_1,W_2\in \cP^{\cY|\cX}$. As a
specialized notation, we use
\begin{equation*}
\DB(\eps_1\|\eps_2) \defeq \eps_1 \ln
\left(\frac{\eps_1}{\eps_2}\right)+(1-\eps_1)\ln
\left(\frac{1-\eps_1}{1-\eps_2}\right)
\end{equation*}
to denote the divergence between Bernoulli distributions with
parameters $\eps_1,\eps_2\in[0,1]$.
\iffalse In our analysis, we make use of the usual notion of strong typicality
\cite{CT,CK}.  In particular, a sequence $y^n$ is strongly typical
with respect to the distribution $P\in\cP^\cY$ for some (implicit)
parameter $\mu\in(0,1)$ if $|\hat{P}_{y^n}(y)-P(y)|<\mu$ for all
$y\in\cY$.
\fi

We make frequent use of the method of types \cite[Chapter~1.2]{CK}.  In
particular, $\hat{P}_{x^n}$ denotes the empirical distribution (or
type) of a sequence $x^n\in\cX^n$, i.e.,\footnote{When the sequence
  that induces the empirical type is clear from context, we omit the
  subscript and write simply $\hat{P}$.}
\begin{equation*}
\hat{P}_{x^n}(x)\defeq\frac{1}{n}\sum_{i=1}^n \openone(x_i=x).
\end{equation*}
The joint empirical distribution $\hat{P}_{(x^n,y^n)}$ for a sequence
pair $(x^n,y^n)$ is defined analogously, i.e.,
\begin{equation*}
\hat{P}_{x^n,y^n}(x,y)\defeq\frac{1}{n}\sum_{i=1}^n \openone(x_i=x, y_i=y),
\end{equation*}
and, in turn, a sequence $y^n$ is said to have a conditional empirical
distribution $\hat{P}_{y^n|x^n}\in\cP^{\cY|\cX}$ given $x^n$ if for all
$(x,y)\in\cX\times\cY$,
\begin{equation*}
\hat{P}_{x^n,y^n}(x,y)\defeq\hat{P}_{x^n}(x)\,\hat{P}_{y^n|x^n}(y|x).
\end{equation*}

As additional notation, $P\in\cP^\cX$ is said to be an $n$-type if
$nP(x)$ is an integer for all $x\in\cX$.  The set of all $n$-types
over an alphabet $\cX$ is denoted using $\cP^\cX_n$.  The $n$-type
class of $P$, denoted using $\cT_P^n$, is the set of all sequences
$x^n$ that have type $P$, i.e., such that $\hat{P}_{x^n}=P$.  A set of
sequences is said to have constant composition if they belong to the
same type class.  When clear from the context, we sometimes omit the
superscript $n$ and simply write $\cT_P$.  For distributions on the
alphabet $\cX\times \cY$ the set of joint $n$-types
$\cP_n^{\cX,\cY}$ is
defined analogously. The
set of sequences $y^n$ that have a conditional type $W$ given $x^n$ is denoted
by $\cT_W(x^n)$, and $\cP_n^{\cY|\cX}$ denotes the set of empirical conditional
distributions, i.e., the set of $W\in \cP_n^{\cY|\cX} $ such that $W=\hat{P}_{y^n|x^n}(y|x)$
for some $(x^n,y^n)\in \cX^n \times \cY^n$.

Finally, the following three standard type results are often used in our
analysis.

\begin{fact}[\hspace{-.01cm}{\cite[Lemma~1.2.2]{CK}}]
\label{fact:1}
\begin{align*}
|\cP_n^{\cX}| &\leq (n+1)^{|\cX|}\\
|\cP_n^{\cX,\cY}| &\leq (n+1)^{|\cX|\cdot|\cY|}\\
|\cP_n^{\cY|\cX}| &\leq (n+1)^{|\cX|\cdot |\cY|}.
\end{align*}
\end{fact}

\begin{fact}[\hspace{-.01cm}{\protect\cite[Lemma 1.2.6]{CK}}]
\label{fact:2}
If $X^n$ is independent and identically distributed (i.i.d.) according
to $P_1\in\cP^\cX$, then 
\begin{equation*}
\frac{1}{(n+1)^{|\cX|}}e^{-nD(P_2\|P_1)}\leq \pr(X^n\in \cT_{P_2}) \leq  e^{-nD(P_2\|P_1)}.
\end{equation*}
for any $P_2\in\cP^\cX_n$.
\end{fact}

\begin{fact}[\hspace{-.01cm}{\protect\cite[Lemma 1.2.6]{CK}}]
\label{fact:3}
If the input $x^n\in \cX^n$ to a memoryless channel $Q\in
\cP^{\cY|\cX}$ has type $P\in\cP^\cX$, then the probability of observing
a channel output sequence  $Y^n$ which lies in  $\cT_W(x^n)$  satisfies
\begin{align*}
\frac{1}{(n+1)^{|\cX||\cY|}} e^{-nD(W\|Q|P)}&\leq
\pr(Y^n\in \cT_W(x^n)|x^n)\\
&\leq e^{-nD(W\|Q|P)}
\end{align*}
for any $W\in \cP^{\cY|\cX}$ such that $\cT_W(x^n)$ is non-empty.

\end{fact}

%Define $\supp(\cdot)$....

\section{Model and Performance Criterion}\label{moper}

The asynchronous communication model of interest captures the
setting where infrequent delay-sensitive data must be reliably
communicated.  For a discussion of this model and its connections
with related communication and statistical models we refer to
\cite[Section II]{TCW}.

We consider discrete-time communication without feedback over a discrete memoryless
channel characterized by its finite input and output alphabets $\cX$
and $\cY$, respectively, and transition probability matrix $Q(y|x)$, for
all $y\in \cY$ and $x\in \cX$.  Without loss of generality, we assume
that for all $y\in \cY$ there is some $x\in \cX$ for which $Q(y|x)>0$.

There are $M\ge 2$ messages $m\in \{1,2,\dots,M\}$.  For each message
$m$, there is an associated codeword
\begin{equation*}
c^n(m) \defeq  c_1(m) \, c_2(m) \, \cdots \, c_n(m),
\end{equation*}
which is a string of $n$ symbols drawn from $\cX$.  The $M$ codewords
form a codebook $\cC_n$ (whence $|\cC_n|=M$).  Communication takes
place as follows.  The transmitter selects a message $m$ randomly and
uniformly over the message set and starts sending the corresponding
codeword $c^n(m)$ at a random time $\nu$, unknown to the receiver,
independent of $c^n(m)$, and uniformly distributed over
$\{1,2,\dots,\asynclev\}$, where $\asynclev\defeq  e^{n\asyncexp}$
is referred to as the \emph{asynchronism level} of the channel, with
$\asyncexp$ termed the associated \emph{asynchronism exponent}.  The
transmitter and the receiver know the integer parameter $\asynclev\ge
1$.  The special case $\asynclev=1$ (i.e., $\asyncexp=0$) corresponds
to the classical synchronous communication scenario.

\begin{figure}
\begin{center}
\input{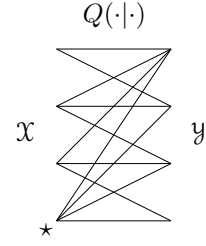}
\caption{\label{grapheess} Graphical depiction of the transmission
  matrix for an asynchronous discrete memoryless channel.  The ``no
  input'' symbol $\star$ is used to characterize the channel output
  when the transmitter is silent.}
\end{center}
\end{figure}

\begin{figure}
\begin{center}
\input{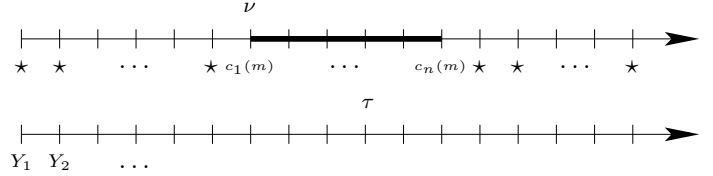}
\caption{\label{grapheesss} Temporal representation of the channel
  input sequence (upper axis) and channel output sequence (lower
  axis).  At time $\nu$ message $m$ starts being sent and decoding
  occurs at time $\tau$. Since $\nu$ is unknown at the
  receiver, the decoding time may be before the entire
  codeword has been received, potentially (but not necessarily) resulting in a
  decoding error.}
\end{center}
\end{figure}

When a codeword is transmitted, a noise-corrupted version of the
codeword is obtained at the receiver.  When the transmitter is silent,
the receiver observes only noise.  To characterize the output
distribution when no input is provided to the channel, we make use of a specially designated
``no-input'' symbol $\star$ in the input alphabet $\cX$, as depicted
in Figs.~\ref{grapheess} and~\ref{grapheesss}.  Specifically,
\begin{equation}
Q_\star \defeq  Q(\cdot|\star)
\end{equation}
characterizes the noise distribution of the channel.  Hence,
conditioned on the value of $\nu$ and on the message $m$ to be
conveyed, the receiver observes independent symbols
$Y_1,Y_2,\dots,Y_{\asynclev+n-1}$ distributed as follows.  If $$t\in
\{1,2,\dots,\nu-1\}$$ or $$t\in [\nu+n,\nu+n+1,\dots, \asynclev+n-1]\,,$$
the distribution of $Y_t$ is $Q_\star$.  If
$$t\in\{\nu, \nu+1, \dots, \nu+n-1\}\,,$$ the distribution of $Y_t$ is
$Q(\cdot|{c_{t-\nu+1}(m)})$.  Note that since the transmitter can
choose to be silent for arbitrary portions of its length-$n$
transmission as part of its message-encoding strategy, the symbol
$\star$ is eligible for use in the codebook design.

The decoder takes the form of a sequential test $(\tau,\phi)$, where
$\tau$ is a stopping time, bounded by $\asynclev+n-1$, with respect to
the output sequence $Y_1,Y_2,\dots\,$, indicating when decoding
happens, and where $\phi$ denotes a decision rule that declares the
decoded message; see Fig.~\ref{grapheesss}.  Recall that a stopping
time $\tau$ (deterministic or randomized) is an integer-valued random
variable with respect to a sequence of random variables
$\{Y_i\}_{i=1}^\infty$ so that the event $\{\tau=t\}$, conditioned on
$\{Y_i\}_{i=1}^{t}$, is independent of $\{Y_{i}\}_{i=t+1}^{\infty}$,
for all $t\ge 1$.  The function $\phi$ is then defined as any
$\cF_\tau$-measurable map taking values in $\{1,2,\dots,M\}$,
where $\cF_1,\cF_2,\dots$ is the natural filtration
induced by the process $Y_1,Y_2,\dots\,$. 

A code is an encoder/decoder pair
$(\cC,(\tau,\phi))$.\footnote{Note that the proposed asynchronous discrete-time communication
model still assumes some degree of synchronization since transmitter and receiver
are supposed to have access to clocks ticking at unison. This is sometimes referred
to as frame asynchronous symbol synchronous communication.}

The performance of a code operating over an asynchronous
channel is
quantified as follows.  First, we define the maximum (over messages),
time-averaged decoding error probability\footnote{Note that there
  is a small abuse of notation as $\pr(\cE)$ need not be a probability.}
\begin{equation} 
\label{maxerror}
\pr(\cE)=\max_{m}\frac{1}{\asynclev}\sum_{t=1}^\asynclev \pr_{m,t} (\cE),
\end{equation}
where $\cE$ indicates the event that the decoded message does not
correspond to the sent message, and where the subscripts ${m,t}$
indicate the conditioning on the event that message $m$ starts
being sent at time $\nu=t$. Note that by definition we have
$$\pr_{m,t} (\cE)=\pr_{m,t}(\phi(Y^\tau)\ne m)\,.$$

Second, we define communication rate with respect to the average
elapsed time between the time the codeword starts being sent and
the time the decoder makes a decision, i.e.,
\begin{equation} 
\label{vilag}
R = \frac{\ln M}{\delay},
\end{equation}
where
\begin{equation} 
\delay=\max_m \frac{1}{\asynclev}\sum_{t=1}^\asynclev
\ex_{m,t}( \tau-t)^+,
\label{eq:delay-def}
\end{equation}
where $x^+$ denotes $\max\{0,x\}$, and where $\ex_{m,t}$ denotes the
expectation with respect to $\pr_{m,t}$.\footnote{Note that
  $\ex_{m,t}(\tau_n-t)^+$ should be interpreted as
  $\ex_{m,t}((\tau_n-t)^+)$.}

With these definitions, the class of communication strategies of
interest is as follows.
\begin{defn}[$(R,\asyncexp)$ Coding Scheme]
\label{defcs}
A pair $(R,\asyncexp)$ with $R\geq 0$ and $\asyncexp\geq 0$ is
achievable if there exists a sequence
$\{(\cC_n,(\tau_n,\phi_n)\}_{n\geq 1}$ of codes,
indexed by the codebook length $n$, that asymptotically achieves a
rate $R$ at an asynchronism exponent $\asyncexp$.  This means that for
any $\eps>0$ and every $n$ large enough, the code
$(\cC_n,(\tau_n,\phi_n))$
\begin{enumerate}
\item operates under asynchronism level $\asynclev_n=e^{(\asyncexp-\eps)n}$;
\item yields a rate at least equal to $R-\eps$;
\item achieves a maximum error probability of at most $\eps$.
\end{enumerate}
An $(R,\asyncexp)$ coding scheme is a sequence
$\{(\cC_n,(\tau_n,\phi_n))\}_{n\geq 1}$ that achieves the rate-exponent pair $(R,\asyncexp)$.
\end{defn}

In turn, capacity for our model is defined as follows.
\begin{defn}[Asynchronous Capacity]
\label{caps}
For given $\asyncexp\geq 0$, the asynchronous capacity $R(\asyncexp)$ is
the supremum of the set of rates that are achievable at asynchronism
exponent $\asyncexp$.   Equivalently, the asynchronous capacity is
characterized by $\asyncexp(R)$, defined as the supremum of the set of
asynchronism exponents that are achievable at rate $R\geq 0$.
\end{defn}
Accordingly, we use the term ``asynchronous capacity'' to
designate either $R(\alpha)$ or $\alpha(R)$.
While $R(\asyncexp)$ may have the more natural immediate
interpretation, most of our results are more conveniently expressed in
terms of $\asyncexp(R)$.

In agreement with our notational convention, the capacity of the
synchronous channel, which corresponds to the case where
$\alpha=0$, is simply denoted by $C$ instead of $R(0)$.
Throughout the paper we only consider channels with
$C>0$.

\begin{remark} One could alternatively consider the rate with
respect to the duration the transmitter occupies the
channel and define it with respect to the block length $n$. In this
case capacity is a special case of the general asynchronous capacity per
unit cost  result \cite[Theorem $1$]{CTT}. 
\end{remark}

In \cite{TCW,CTW} it is shown that reliable communication is possible if
and only if the asynchronism exponent $\asyncexp$ does not exceed a
limit referred to as the ``synchronization threshold.''
\begin{thm}[{\cite[Theorem~2]{TCW}},{\cite{CTW}}]
\label{ulimit}
If the asynchronism exponent is strictly smaller than the
\emph{synchronization threshold}
\begin{equation*}
\thres\defeq 
\max_x D(Q(\cdot|x)\|Q_\star)=\asyncexp(R=0),
\end{equation*}
then there exists a coding scheme $\{(\cC_n,(\tau_n,\phi_n))\}_{n\geq
1}$ that achieves a maximum error probability tending to zero as
$n\rightarrow\infty$. 

 Conversely, any coding scheme
$\{(\cC_n,(\tau_n,\phi_n))\}_{n\geq 1}$ that operates at an
asynchronism exponent strictly greater than the synchronization
threshold, achieves (as $n\rightarrow\infty$) a maximum probability
of error equal to one.

 Moreover,\footnote{This claim appeared in \cite[p.~4515]{TCW}.}
$$\thres>0\quad \text{if and
only if}\quad C>0\,.$$
\end{thm}

A few comments are in order. The cause of unreliable communication above the
synchronization threshold is the following. When asynchronism is so large, with
probability approaching one pure noise mimics a codeword for {\it any} codebook
(regardless of the rate) before the actual codeword even starts being
sent.\footnote{This follows from the converse of \cite[Theorem]{CTW}, which says
that above $\thres$, even the codeword of a single codeword codebook is
mislocated with probability tending to one.} This results in an error
probability of at least $1/2$ since, by our model assumption, the message set contains at least
two messages. On the other hand, below the synchronization threshold reliable communication is possible. If
the codebook is properly chosen, the noise won't mimic any codeword with
probability tending to one, which allows the decoder to reliably detect the
sent message. 
%Moreover, if the message set is not too large, the decoder will
%also be able to reliably isolate the sent message.

Note that $$\thres=\infty$$ if and only if pure
noise can't generate all channel outputs, i.e., if and only if 
$Q_\star(y)=0$ for some $y\in \cY$. Indeed, in this case
it is possible to avoid the previously mentioned decoding
confusion by designing codewords (partly) composed of
symbols that generate channel outputs which are
impossible to generate with pure noise. 

The last claim in Theorem~\ref{ulimit} says that reliable
asynchronous communication is possible if and only if reliable
synchronous communication is possible. That the former implies
the latter is obvious since asynchronism can only hurt
communication. That the latter implies the former is perhaps
less obvious, and a high-level justification is as follows.
When $C>0$, at least two channel inputs yield different
conditional output distributions, for otherwise the
input-output mutual information is zero regardless of the
input distribution. Hence, $Q(\cdot|\star)\ne Q(\cdot|x)$ for
some $x\ne \star$. Now, by designing codewords mainly composed
of $x$ it is possible to reliably signal the codeword's
location to the decoder even under an exponential asynchronism, since the channel outputs look statistically different
than noise during the message transmission. Moreover, if the
message set is small enough, it is possible to 
guarantee reliable message location and successfully identify which
message from the message set was sent. Therefore, exponential
asynchronism can be accommodated, hence $\thres>0$. 

Finally, it should be pointed out that in \cite{TCW} all the
results are stated with respect to average (over messages) delay and
error probability in place of maximum (over messages) delay and error
probability as in this paper.  Nevertheless, the same results hold in
the latter case as discussed briefly later at the end of
Section~\ref{pfofresults}.

\iffalse In the sequel, to simplify the notation we omit the explicit dependency
on the channel and use $R(\asyncexp)$, $\asyncexp(R)$, and
$\thres$ in place of $R(\asyncexp,Q)$, $\asyncexp(R,Q)$, and
$\thres(Q)$, respectively.
\fi
\section{Main Results}
\label{results}

This section is divided into two parts. In Section~\ref{seca}, we provide
general upper and lower bounds on capacity, and derive several of its
properties.  In Section~\ref{strain}, we investigate the performance limits of
training-based schemes and establish their suboptimality in a certain
communication regime. Since both sections can be read independently, the
practically inclined reader may read Section~\ref{strain} first.

All of our results assume a uniform
distribution on $\nu$. Nevertheless, this assumption is not
critical in our proofs. The results can be extended to
non-uniform distributions by following the same arguments as
those used to establish asynchronous capacity per unit cost for
non-uniform $\nu$ \cite[Theorem $5$]{CTT}.

\subsection{General Bounds on Asynchronous Capacity}
\label{seca} 
To communicate reliably, whether synchronously or asynchronously,
the input-output mutual information induced by the
codebook should at least be equal to the desired communication
rate. 

When communication is asynchronous,   a decoder  should, in
addition, be able to  discriminate between hypothesis
``noise'' and hypothesis ``message.'' These hypothesis correspond to the
situations when the transmitter is idle and when it
transmits a codeword, respectively. Intuitively, the more these hypotheses  are
statistically far apart---by means of an appropriate codebook
design---the larger the level of asynchronism which can be
accommodated for a given communication rate. 

More specifically, a code should serve the dual purpose of
minimizing the ``false-alarm'' and ``miss'' error
probabilities. 

Since the decoder doesn't know $\nu$,  the decoder may output a
message before even a message is sent. This is the false-alarm
event and it contributes to increase the error
probability---conditioned on a false-alarm the error probability
is essentially one. However, false-alarms also contribute to
increase the rate since it is defined
with respect to the receiver's decoding delay
$\ex(\tau-\nu)^+$. As
an extreme case, by immediately decoding, {\it{i.e.}}, by setting
$\tau=1$, we get an infinite rate and and error probability
(asymptotically) equal to one. As it turns out, the false-alarm
probability should be exponentially small to allow reliable
communication under exponential asynchronism.

The miss event refers to the scenario where the decoder fails
to recognize the sent message during transmission, i.e., the
message output looks like it was generated by noise. This event
impacts the rate and, to a smaller extent, also the error
probability. In fact, when the sent message is missed, the
reaction delay is usually huge, of the order of $A$.
Therefore, to guarantee a positive rate under
exponential asynchronism the miss error probability should
also be exponentially small.  

Theorem~\ref{ach} below provides a lower bound on the
asynchronous capacity. The proof of this theorem is
obtained by analyzing a coding
scheme which performs synchronization and
information transmission jointly. The codebook is a standard i.i.d.
random code across time and messages and its performance
is governed by the  Chernoff error exponents for discriminating
hypothesis ``noise'' from hypothesis ``message.''

\begin{thm}[Lower Bound on Asynchronous Capacity]
\label{ach}
Let $\asyncexp\geq 0$ and let $P\in \cP^\cX$ be some input distribution
such that at least one of the following inequalities
\begin{align*}
D(V\|(PQ)_\cY) &\geq \asyncexp \\
D(V\|Q_\star)&\geq \asyncexp
\end{align*}
holds for all distributions
$V\in \cP^\cY$, i.e.,
\begin{equation*}
\min_{V\in \cP^\cY}\max\{D(V\|(PQ)_\cY),D(V\|Q_\star)\}\geq \asyncexp.
\end{equation*}
Then, the rate-exponent pair $(R=I(PQ),\asyncexp)$ is achievable.
Thus, maximizing over all possible input distributions, we have the
following lower bound on $\asyncexp(R)$ in Definition~\ref{caps}:

\begin{equation}
\asyncexp(R) \ge
\asyncexp_-(R)\qquad R\in (0, C]
\end{equation}
where
\begin{equation}
\asyncexp_-(R) \defeq
\max_{\substack{\{P\in \cP^\cX\,:\\I(PQ)\geq R\}}} \min_{V\in
  \cP^\cY}\max\{D(V\|(PQ)_\cY),D(V\|Q_\star)\}.
\label{caphyp}
\end{equation}
\end{thm}

\begin{figure}
\begin{center}
\input{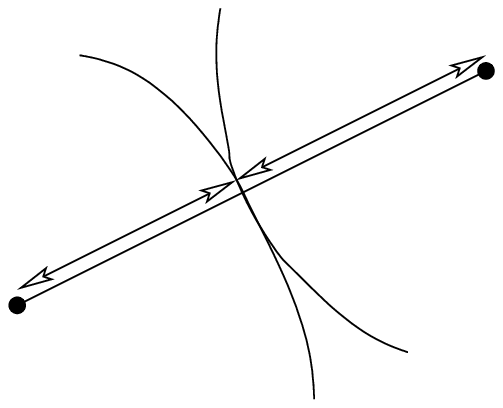}
\caption{\label{hypot} If $\alpha$ is at most the ``half-distance'' between
distributions $(PQ)_\cY$ and $Q_\star$, then $(\alpha,R)$ with $R=I(PQ)$ is
achievable.}
\end{center}
\end{figure}

Theorem~\ref{ach} provides a simple explicit lower bound on
capacity. The distribution $(PQ)_\cY$ corresponds to  the channel output
when the input to the channel is distributed
according to $P$.  The asynchronism exponent that can be accommodated
for given  $P$ and $Q_\star$ can be interpreted as being the
``equidistant point'' between
distributions $(PQ)_\cY$ and $Q_\star$, as depicted in
Fig.~\ref{hypot}. Maximizing over $P$ such that
$I(PQ)\geq R$ gives the largest such exponent that can be
achieved for rate $R$ communication.

Note that \eqref{caphyp} is much simpler to evaluate than the lower bound given
by \cite[Theorem~2]{TCW}. Moreover, the former is usually
a better bound than the latter and it exhibits  an interesting feature of
$\alpha(R)$ in the high rate regime. This feature is
illustrated in Example~\ref{exa:bsc}
to come.

Theorem~\ref{ach} extends to the following  continuous
alphabet Gaussian setting:
\begin{cor}[Asynchronous Gaussian channel]\label{tgaussian}
Suppose that for a real
input $x$ the decoder receives $Y=x+Z$, where $Z\sim\cN(0,1)$.
When there is no input to the channel, $Y=Z$, so
$Q_\star=\cN(0,1)$.  The input is power constrained so that
all codewords $c^n(m)$ must satisfy $\frac{1}{n}\sum_{i=1}^n
c_i(m)^2\leq p$ for a given constant $p>0$. For this channel we have
\begin{equation}\label{gbound}
\alpha(R)\geq \max_{\substack{P:I(PQ)\geq R\\ \ex_P X^2\leq p}}
\min_{V}\max\{D(V\|(PQ)_\cY),D(V\|Q_\star)\},
\end{equation}
for $R\in (0,C]$ where $P$ and $V$ in the optimization are distributions over the reals.
\end{cor}
If we restrict the outer maximization in \eqref{gbound} to be over Gaussian
distributions only, it can be shown that the best input has a mean
$\mu$ that is as large as possible, given the rate and power
constraints.  More precisely, $\mu$ and $R$ satisfy
\begin{equation*}
R=\frac{1}{2}\ln \left(1+p-\mu^2\right),
\end{equation*}
and the variance of the optimal Gaussian input is $p-\mu^2$.  The
intuition for choosing such parameters is that a large mean helps the
decoder to distinguish the codeword from noise---since the latter has a
mean equal to zero.  What limits the mean
is both the power constraint and the variance needed to ensure
sufficient mutual information to support communication
at rate~$R$.

\begin{IEEEproof}[Proof of Corollary~\ref{tgaussian}]
The proof uses a standard quantization argument similar to that in \cite{CN2},
and therefore we provide only a sketch of the proof. 
From the given the continuous time Gaussian
channel, we can form a discrete alphabet channel for which we can apply
Theorem~\ref{ach}. 

More specifically,  for a given constant $L>0$, the input and the output of the
channel are discretized within $[-L/2,L/2]$ into constant
size $\Delta$ contiguous intervals $\Delta_i=[l_i,l_i+\Delta)$. $L$ and $\Delta$ are
chosen so that $L\to \infty$ as $\Delta\to 0$.  To a given
input $x$ of the Gaussian channel is associated the quantized
value $\tilde{x}=l_i+\Delta/2$ where $i$ denotes the index of the
interval $\Delta_i$ which contains $x$.   If $x<-L/2 $ or $x\geq
L/2$, then $\tilde{x}$ is defined as $-L/2$ or $L/2$,
respectively. The same quantization is applied to the output of
the Gaussian channel.

For each quantized channel
we apply Theorem~\ref{ach}, then let $\Delta\to 0$ (hence $L\to
\infty$). One can
then verify that the achieved
bound corresponds to \eqref{gbound}, which  shows that Theorem~\ref{ach} also holds for the
continuous alphabet  Gaussian setting of Theorem~\ref{tgaussian}.
\end{IEEEproof}

The next result provides an upper bound to the
asynchronous capacity for channels
with finite synchronization threshold---see Theorem~\ref{ulimit}:
\begin{thm}[Upper Bound on Asynchronous Capacity]\label{convs}
For any channel $Q$ such that $\thres<\infty$, and any $R>0$,
we have that
\begin{equation}
\label{eq:convs}
\asyncexp(R) \le \max_\cS\min \{\asyncexp_1,\asyncexp_2\} \defeq \asyncexpub(R),
\end{equation}
where
\begin{align}
\asyncexp_1 &\defeq
\delta(I(P_1Q)-R+D((P_1Q)_\cY\|Q_\star))
\label{alpha1}\\
\asyncexp_2 &\defeq \min_{W\in \cP^{\cY|\cX}}
\max\{D(W\|Q|P_2),D(W\|Q_\star|P_2)\}\label{alpha2}
\end{align}
with
\begin{align} \label{ess}
\cS \defeq \Bigl\{ (P_1,P_2,&P_1',\delta)\in 
\bigl(\cP^\cX\bigr)^3\times[0,1] \,:\, \notag\\
&I(P_1Q)\ge R,\ P_2=\delta P_1 + (1-\delta)P_1' \Bigr\}.
\end{align}
If $\thres=\infty$, then
\begin{align}\label{ifin}
\asyncexp(R) \le \max_{P_2} \asyncexp_2 
\end{align}
for $R\in (0,C]$.
\end{thm}

The terms $\alpha_1$ and $\alpha_2$ in \eqref{eq:convs}
reflect the false-alarm and miss constraints alluded to above
(see discussion before Theorem~\ref{ach}). If
$\alpha>\alpha_1$, then with high probability the noise will
mimic a message before transmission starts. Instead, if
$\alpha>\alpha_2$ then reliable communication at a positive
rate is impossible since no code can guarantee a
sufficiently low probability of missing the
sent codeword. 

The parameter $\delta$ in \eqref{alpha1} and \eqref{ess} essentially
represents the ratio between the reaction delay $\ex(\tau-\nu)^+$ and
the blocklength---which need not coincide. Loosely speaking, for a
given asynchronism level a smaller $\delta$, or, equivalently, a
smaller $\ex(\tau-\nu)^+$, increases the communication rate at the
expense of a higher false-alarm error probability. The intuition for
this is that a decoder that achieves a smaller reaction delay sees, on
average, ``fewer'' channel outputs before stopping. As a consequence,
the noise is more likely to lead such a decoder into confusion. A
similar tension arises between communication rate and the miss error
probability. The optimization over the set $\cS$ attempts to strike
the optimal tradeoff between the communication rate, the false-alarm
and miss error probabilities, as well as the reaction delay as a
fraction of the codeword length.

For channels with infinite synchronization threshold,
Theorem~\ref{alphinfin} to come establishes that the bound given by
\eqref{ifin} is actually tight.

The following examples provide some useful insights. 

\begin{exa}
\label{exa:bsc}

Consider the binary symmetric channel depicted in Fig.~\ref{bssc},
which has the property that when no input is supplied to the channel,
the output distribution is asymmetric.  For this channel, in
Fig.~\ref{fig} we plot the lower bound on $\alpha(R)$ given by
\eqref{caphyp} (curve $\alpha_-(R)$) and the lower bound given by
\cite[Theorem~1]{TCW} (the dashed line LB[3]).\footnote{Due to the
  complexity of evaluating the lower bound given by
  \cite[Theorem~1]{TCW}, the curves labeled LB[3] are actually upper
  bounds on this lower bound. We believe these bounds are fairly
  tight, but in any case we see that the resulting upper bounds are
  below the lower bounds given by \eqref{caphyp}.}  The $\alpha_+(R)$ curve
correspond to the upper bound on $\alpha(R)$ given by
Theorem~\ref{convs}.  For these plots, the channel parameter is
$\veps=0.1$.

\begin{figure}
\begin{center}
\input{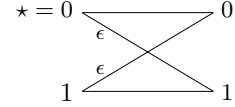}
\caption{\label{bssc} A channel for which $\asyncexp(R)$ is
  discontinuous at $R=C$.}
\end{center}
\end{figure}
\begin{figure}
\begin{center}
\input{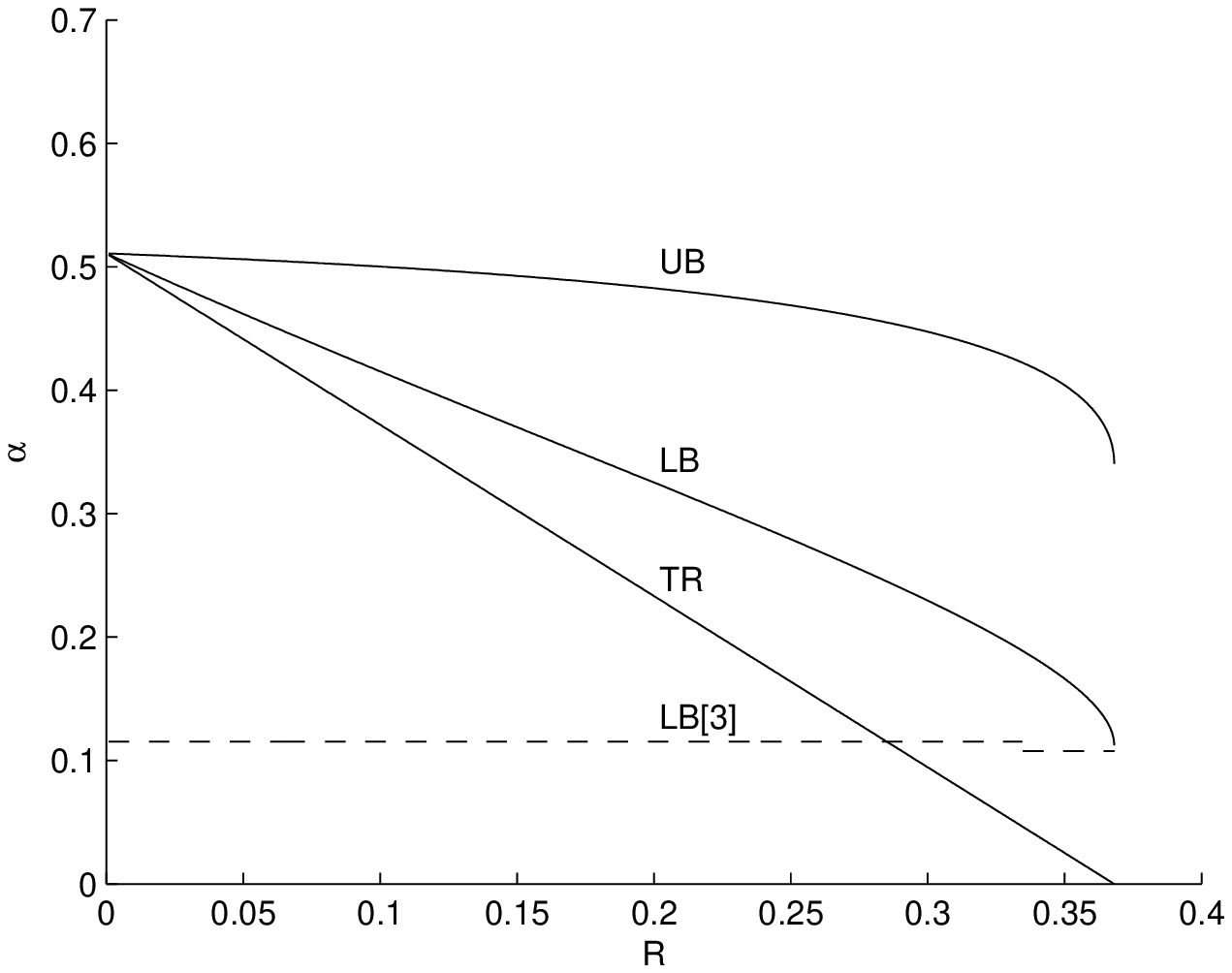}
\caption{\label{fig} Capacity upper and lower bounds on the
  asynchronous capacity of the channel of Fig.~\ref{bssc} with
  $\veps=0.1$ and $\star=0$.  $\alpha_-(R)$ represents the lower bound given by
  Theorem~\ref{ach}, LB[3] represents the lower bound obtained in
  \cite[Theorem 1]{TCW}, and $\alpha_+(R)$ represents the upper bound given by
  Theorem~\ref{convs}. }
\end{center}
\end{figure}
The discontinuity of $\alpha(R)$ at $R=C$ (since $\asyncexp(R)$ is
clearly equal to zero for $R>C$) implies that we do not need to back
off from the synchronous capacity in order to operate under
exponential asynchronism.\footnote{To have a better sense of what it
  means to be able to decode under exponential asynchronism and, more
  specifically, at $R=C$, consider the following numerical
  example. Consider a codeword length $n$ equal to $150$. Then
  $\alpha=.12$ yields asynchronism level $A=e^{n\alpha}\approx 6.5
  \times 10^{7}\,.$ If the codeword is, say, $30$ centimeters long,
  then this means that the decoder can reliably sequentially decode
  the sent message, with minimal delay (were the decoder cognizant of
  $\nu$, it couldn't achieve a smaller decoding delay since we operate
  at the synchronous capacity), within $130$ kilometers of mostly
  noisy data! }

Note next that the $\alpha_-(R)$ is better than LB[3] for all rates. In fact,
empirical evidence suggests that $\alpha_-(R)$ is better than LB[3] in general.
%FIXME: i THINK WE SHOULD WE MENTION THAT THIS
%IS ACTUALLY AN UPPER BOUND TO THE LOWER BOUND [3]. A
%FOOTNOTE SHOULD HOPEFULLY SUFFICE
Additionally, note that $\alpha_-(R)$ and $\alpha_+(R)$ are not tight.
\end{exa}

Next, we show how another binary symmetric channel has some rather
different properties.

\begin{exa}
\label{exa:bsc-sn}

Consider the binary symmetric channel depicted in
Fig.~\ref{bscextnoise}, which has the property that when no input is
provided to the channel the output distribution is symmetric.  When
used synchronously, this channel and that of Example~\ref{exa:bsc} are
completely equivalent, regardless of the crossover probability
$\veps$.  Indeed, since the $\star$ input symbol in
Fig.~\ref{bscextnoise} produces $0$ and $1$ equiprobably, this input
can be ignored for coding purposes and any code for this channel
achieves the same performance on the channel in Fig.~\ref{bssc}.
\begin{figure}
\begin{center}
\input{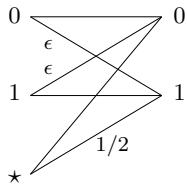}
\caption{\label{bscextnoise} Channel for which $\alpha(R)$ is
  continuous at $R=C$. }
\end{center}
\end{figure}

However, this equivalence no longer holds when the channels are used
asynchronously.  To see this, we plot the corresponding upper and
lower bounds on performance for this channel in Fig.~\ref{fig_sym}.
Comparing curve $\alpha_-(R)$ in Fig.~\ref{fig} with curve $\alpha_+(R)$ in
Fig.~\ref{fig_sym}, we see that asynchronous capacity for the channel
of Fig.~\ref{bssc} is always larger than that of the current example.
Moreover, since there is no discontinuity in exponent at $R=C$ in our
current example, the difference is pronounced at $R=C=0.368\ldots$;
for the channel of Fig.~\ref{bssc} we have $\alpha(C)\approx 0.12>0$.

\begin{figure}
\begin{center}
\input{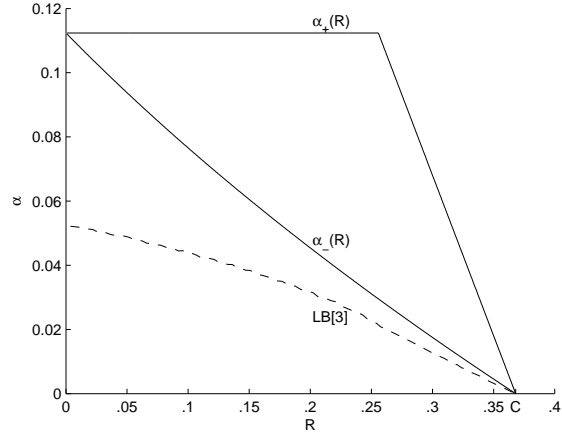}
\caption{\label{fig_sym} Capacity upper and lower bounds on the
  asynchronous capacity of the channel of Fig.~\ref{bscextnoise} with
  $\veps=0.1$.  $\alpha_-(R)$ represents the lower bound given by
  Theorem~\ref{ach}, LB[3] represents the lower bound obtained in
  \cite[Theorem 1]{TCW}, and $\alpha_+(R)$ represents the upper bound given by
  Theorem~\ref{convs}.}
\end{center}
\end{figure}

\end{exa}

The discontinuity of $\asyncexp(R)$ at $R=C$ observed in
Example~\ref{exa:bsc} is in fact typical, holding in all but one
special case.
\begin{cor}[Discontinuity of $\asyncexp(R)$ at $R=C$]
\label{limittocap}
We have $\asyncexp(C)=0$ if and only if $Q_\star$ corresponds to the
(unique) capacity-achieving output distribution of the synchronous
channel.
\end{cor}

By Corollary~\ref{limittocap}, for the binary symmetric channel of
Example~\ref{exa:bsc}, $\alpha(R)$ is discontinuous at $R=C$ whenever
$\eps \ne 1/2$. To see this, note that the capacity achieving output
distribution of the synchronous channel assigns equal weights to
$\star$ and~$1$, differently than $Q_\star$.

The justification for the discontinuity in Example~\ref{exa:bsc} is as
follows.  Since the capacity-achieving output distribution of the
synchronous channel (Bernoulli($1/2$)) is biased with respect to
the noise distribution $Q_\star$, hypothesis ``message'' and ``noise''
can be discriminated with exponentially small error
probabilities. This, in turn, enables reliable detection of the sent
message under exponential asynchronism. By contrast, for the channel
of Example~\ref{exa:bsc-sn}, this bias no longer exists and
$\alpha(R=C)=0$. For this channel, to accomodate a positive asynchronism exponent we need
to backoff from the synchronous capacity $C$ so that the codebook
output distribution can be differentiated from the noise.

\iffalse
\begin{exa}\label{exsymnoise}
Consider the channel depicted in Fig.~\ref{bscextnoise}. In this case the capacity achieving
output distribution of the synchronous channel corresponds to $Q_\star$. Therefore, $\alpha(R)$
is continuous at $R=C$ by Corollary~\ref{limittocap}.

For $\veps=.1$, Fig.~\ref{fig_sym} represents the two capacity lower bounds, given
by Theorem~\ref{ach} and \cite[Theorem 1]{TCW}, and the capacity upper bound given by
Theorem~\ref{convs}. Note that the gap between the upper and the lower
bounds vanishes as $R\to \infty$, by contrast with Example~\ref{exasym}.

A comparison of Figs.~\ref{fig} and \ref{bscextnoise} shows that, although the two channels have
the same synchronous capacity, their asynchronous capacities differ
significantly.
\end{exa}
\fi
\begin{IEEEproof}[Proof of Corollary~\ref{limittocap}]
From Theorem \ref{ach}, a strictly positive asynchronism exponent can
be achieved at $R=C$ if $Q_\star$ differs from the synchronous
capacity-achieving output distribution---\eqref{caphyp} is strictly
positive for $R=C$ whenever $Q_\star$ differs from the synchronous
capacity-achieving output distribution since the divergence between
two distributions is zero only if they are equal.

Conversely, suppose $Q_\star$ is equal to the capacity-achieving
output distribution of the synchronous channel. 
We show that for any $(R,\asyncexp)$ coding scheme where
$R=C$, $\asyncexp$ is necessarily equal to zero.  

From Theorem~\ref{convs}, 
\begin{align*}
\alpha(R)\leq\max_{\cS}\alpha_1 
\end{align*} 
where $\cS$ and $\alpha_1$ are given by
\eqref{ess} and \eqref{alpha1}, respectively. Since
$R=C$, $I(P_1Q)=C$, and since
$Q_\star=(P_1Q)_\cY$, we have
$D((P_1Q)_\cY||Q_\star)=0$. Therefore,
$\alpha_1=0$  for any $\delta$, and we conclude
that $\alpha(C)=0$. 
\end{IEEEproof}

In addition to the discontinuity at $R=C$, $\asyncexp(R)$ may also be
discontinuous at rate zero:
\begin{cor}[Discontinuity of $\asyncexp(R)$ at $R=0$]
\label{disc}
If
\begin{equation} 
\label{critdisc}
\thres > \max_{x\in \cX} D(Q_\star\|Q(\cdot|x)),
\end{equation}
then $\asyncexp(R)$ is discontinuous at rate $R=0$.
\end{cor}
\begin{exa}
Channels that satisfy \eqref{critdisc} include those for which the
following two conditions hold: $\star$ can't produce all channel
outputs, and if a channel output can be produced by $\star$, then it
can also be produced by any other input symbol. For these channels
\eqref{critdisc} holds trivially; the right-hand side term is finite
and the left-hand side term is infinite. The simplest such channel is
the Z-channel depicted in Fig.~\ref{graphdisc} with $\veps\in (0,1)$.

\begin{figure}
\begin{center}
\input{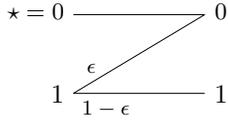}
\caption{\label{graphdisc} Channel for which $\asyncexp(R)$ is
discontinuous at $R=0$, assuming $\veps\in (0,1)$.}
\end{center}
\end{figure}

Note that if $\veps=0$, \eqref{critdisc} doesn't hold since both the
left-hand side term and the right-hand side term are infinite. In
fact, if $\veps=0$ then asynchronism doesn't impact communication;
rates up to the synchronous capacity can be achieved regardless of the
level of asynchronism, i.e., 
\begin{equation*} 
\alpha(R)=\thres=\infty\qquad R\in [0,C].
\end{equation*}
To see this, note that by prepending a 1 to each codeword suffices
to guarantee perfect synchronization without impacting rate
(asymptotically).

More generally, asynchronous capacity for channels with infinite
synchronization threshold is established in
Theorem~\ref{alphinfin} to come.
\end{exa}

An intuitive justification for the possible discontinuity of
$\asyncexp(R)$ at $R=0$ is as follows.  Consider a channel where
$\star$ cannot produce all channel outputs (such as that depicted in
Fig.~\ref{graphdisc}).  A natural encoding strategy is to start
codewords with a common preamble whose possible channel outputs differ
from the set of symbols that can be generated by $\star$.  The
remaining parts of the codewords are chosen to form, for instance, a
good code for the synchronous channel.  Whenever the decoder observes
symbols that cannot be produced by noise (a clear sign of the
preamble's presence), it stops and decodes the upcoming symbols.  For
this strategy, the probability of decoding before the message is
actually sent is clearly zero.  Also, the probability of wrong
message isolation conditioned on correct preamble location can be made
negligible by taking codewords long enough.  Similarly, the
probability of missing the preamble can be made negligible by using a
long enough preamble.  Thus, the error probability of this
training-based scheme can be made negligible, regardless of the
asynchronism level.

The problem arises when we add a positive rate constraint, which
translates into a delay constraint.   Conditioned on missing the
preamble, it can be shown that the delay $(\tau-\nu)^+$ is large, 
in fact of order $\asynclev$.   It can be shown that if \eqref{critdisc} holds, the
probability of missing the preamble is larger than $1/\asynclev$.
Therefore, a positive rate puts a limit on the maximum asynchronism
level for which reliable communication can be
guaranteed, and this limit can be smaller than
$\thres$.

We note that it is an open question whether or not $\asyncexp(R)$ may
be discontinuous at $R=0$ for channels that do not
satisfy~\eqref{critdisc}.

Theorem~\ref{alphinfin} provides an exact characterization of capacity
for the class of channels with infinite synchronization threshold,
i.e., whose noise distribution $Q_\star$ cannot produce all possible
channel outputs.
\begin{thm}[Capacity when $\thres=\infty$]\label{alphinfin}
If $\thres = \infty$, then
\begin{equation}
\asyncexp(R)=\bar{\alpha}\label{cinth}
\end{equation}
for $R\in (0,C]$,
where 
$$\bar{\alpha}\triangleq \max_{P\in \cP^\cX} \min_{W\in \cP^{\cY|\cX}}
\max\{D(W\|Q|P),D(W\|Q_\star|P)\}\,.$$
\end{thm}

\begin{figure}
\begin{center}
\input{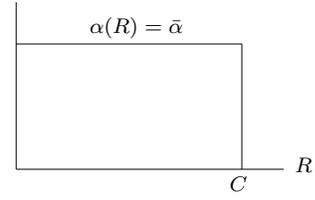}
\caption{\label{alip} Typical shape of the capacity of an asynchronous
channel $Q$ for which $\thres=\infty$.}
\end{center}
\end{figure}

Therefore, when $\thres=\infty$, $\asyncexp(R)$ is actually a constant
that does not depend on the rate, as Fig.~\ref{alip} depicts.  
Phrased
differently, $R(\alpha)=C$  up to $\alpha=\bar{\alpha}$.
For $\alpha>\bar{\alpha}$ we have $R(\alpha)=0$.

Note that when $\thres=\infty$, $\alpha(R)$ can be
discontinuous at $R=0$ since the
right-hand side of \eqref{cinth} is upper bounded by 
$$\max_{x\in \cX}D(Q_\star||Q(\cdot|x)),$$
which can be finite.\footnote{To see this choose
$W=Q_\star$ in the minimization \eqref{cinth}.}

We conclude this section with a result of independent interest related
to synchronous communication, and which is obtained as a byproduct of
the analysis used to prove Theorem~\ref{convs}. This result
essentially says that any nontrivial fixed length codebook, i.e., that
achieves a nontrivial error probability, contains a very good large
(constant composition) sub-codebook, in the sense that its rate is
almost the same as the original code, but its error probability decays
exponentially with a suitable decoder.  In the following theorem
$(\cC_n,\phi_n)$ denotes a standard code for a synchronous channel
$Q$, with fixed length $n$ codewords and decoding happening at time
$n$.
\begin{thm}
\label{prop:expurgation}
Fix a channel $Q\in \cP^{\cY|\cX}$, let $q>0$, and let $\eps,\gamma>0$ be such that $\eps+\gamma\in (0,l)$ with
$l\in (0,1)$. If $(\cC_n,\phi_n)$
is a code that achieves
an error probability $\eps$, then there exists an $n_\circ(l,\gamma,q,|\cX|,|\cY|)$ such that for all $n\geq n_\circ$
there exists $(\cC_n',\phi_n')$ such that\footnote{We
use  $ n_\circ(q)$ to denote some threshold index which could be explicitly
given as a function of $q$.}

\begin{enumerate}
\item $\ds\cC_n'\subset \cC_n$, $\cC_n'$ is constant composition;
\item the maximum error probability is less than $\eps_n$ where
$${\eps}_n=2(n+1)^{|\cX|\cdot|\cY|}\exp(-nq^2/(2\ln
2));$$
\item $\ds\frac{\ln |\cC_n'|}{n}\geq \frac{\ln |\cC_n|}{n}-\gamma$.
\end{enumerate}
\end{thm}
Theorem~\ref{prop:expurgation} is a stronger version of \cite[Corollary~1.9,
  p.~107]{CK} and its proof amounts to a tightening of some of the
arguments in the proof of the latter, but otherwise follows it closely.

\subsection{Training-Based Schemes}
\label{strain}
Practical solutions to asynchronous communication usually separate
synchronization from information transmission. We investigate a very
general class of such ``training-based schemes'' in which codewords
are composed of two parts: a preamble that is common to all codewords,
followed by information symbols. The decoder first attempts to detect
the preamble, then decodes the information symbols.  The results in
this section show that such schemes are suboptimal at least in certain
communication regimes. This leads to the conclusion that the
separation of synchronization and information transmission is in
general not optimal.

We start by defining a general class of training-based
schemes:

\begin{defn}[Training-Based Scheme]
\label{tbs}
A coding scheme $\{(\cC_n,(\tau_n,\phi_n))\}_{n\geq 1}$ is said to be
{\it{training-based}} if for some $\eta \in [0,1]$ and all $n$ large enough
\begin{enumerate}
\item \label{tba-i} there is a common preamble
across codewords of size $\eta n$;
\item \label{tba-iii} the decoding time $\tau_n$ is
such that the event $$\{\tau_n=t\},$$ conditioned on the
$\eta n$
observations $Y_{t- n+1}^{t-n+\eta n}$, is independent of all other
observations (i.e., $Y_1^{t- n}$ and $Y_{t-n+\eta
n+1}^{\asynclev+n-1}$). 
%\item[iii)]
%given that the decoder stops time $\tau=n$ upon the observation of $y^n$, the
%declares message depends only on $y_{n-\eta N+1}^n$.
\end{enumerate}
\end{defn}
Note that Definition~\ref{tbs}
is in fact very general. The only restrictions are that the codewords
all start with the same training sequence, and that the
decoder's 
decision to stop at any particular time should be based
on the processing of (at most) $\eta n$ past output symbols
corresponding to the length of the preamble.

In the sequel we use $\asyncexpt(R)$ to denote the asynchronous capacity
restricted to training based schemes.
\begin{thm}[Training-based scheme capacity bounds]\label{thm2}
Capacity
restricted to training based schemes satisfies
\begin{align}\label{bcaptbs}
\asyncexptl(R)\leq \asyncexpt(R)\leq
\asyncexptu(R) \qquad R\in (0,C]
\end{align}
where
\begin{align*}
\asyncexptl(R)&\defeq m_1\left(1-\frac{R}{C}\right)\\
\asyncexptl(R)&\defeq \min
\left\{m_2\left(1-\frac{R}{C}\right),\asyncexpub(R)\right\}\,,
\end{align*}
where the constants $m_1$ and $m_2$ are defined as
\begin{align*}
m_1&\defeq  \max_{P\in \cP^\cX}\min_{W\in
\cP^{\cY|\cX}}\max\{D(W||Q|P),D(W||Q_\star|P)\}\\
m_2&\defeq
 -\ln(\min_{y\in \cY} Q_\star(y))\,,
\end{align*}
and where $\asyncexpub(R)$ is defined in 
Theorem~\ref{convs}.

Moreover, a rate
$R\in [0,C]$ training-based scheme allocates at most a fraction
$$\eta=\left(1-\frac{R}{C}\right)$$
to the preamble.
\end{thm}
Since $m_2<\infty$ if and only $\thres<\infty$, the upper-bound in
\eqref{bcaptbs} implies:
\begin{cor}[Asynchronism in the high rate
regime]\label{cotocap}
For  training-based schemes $$\asyncexpt(R)\overset{R\to C}{\longrightarrow}
0$$ whenever $\thres<\infty$. 
\end{cor}

In general, $\alpha(C)>0$ as we saw in
Corollary~\ref{limittocap}. Hence a direct consequence of
Corollaries~\ref{limittocap} and \ref{cotocap} is that
training-based schemes are suboptimal in the high rate
regime. Specifically, we have the following result.
\begin{cor}[Suboptimality of training-based schemes]\label{subtr}
There exists a channel-dependent threshold $R_*$ such that for all
$R>R_*$, 
\begin{equation*} 
\asyncexpt(R)<\asyncexp(R)
\end{equation*}
except possibly when $Q_\star$ corresponds to the capacity-achieving
output distribution of the synchronous channel, or when the channel is
degenerate, i.e., when $\thres=\infty$.
\end{cor}
The last claim of Theorem~\ref{thm2} says that the size of the preamble
decreases (linearly) as the rate increases. This, in turn, implies that
$\asyncexpt(R)$ tends to zero as $R$ approaches $C$. Hence, in the
high rate regime most of the symbols should carry information, and the decoder should try to
detect these symbols as part of the decoding process. In other words,
synchronization and information transmission should be jointly performed;
transmitted bits should carry information while also
helping the decoder to locate the sent codeword.

If we are willing to
reduce the rate, are training-based schemes still suboptimal?
We do not have a definite answer to this question, but the
following examples provide some insights.

\begin{exa}\label{exastbs}
Consider the channel depicted in Fig.~\ref{bssc} with $\veps=0.1$.  In
Fig.~\ref{figtr}, we plot the upper and lower bounds to capacity
restricted to training-based schemes given by Theorem~\ref{thm2}.   $\alpha_-(R)$ and $\alpha_+(R)$
represent the general lower and upper bounds to capacity given by
Theorems~\ref{ach} and \ref{convs}; see Fig.~\ref{fig}.
\begin{figure}
\begin{center}
\input{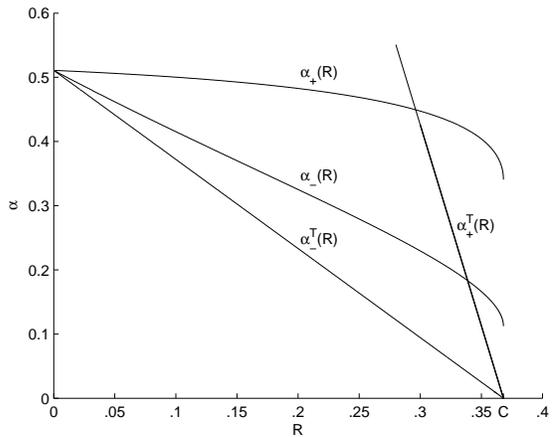}
\caption{\label{figtr} Upper and lower bounds to
capacity restricted to training-based schemes ($\asyncexptu(R)$ and
$\asyncexptl(R)$, respectively) for the binary symmetric channel
depicted in Fig.~\ref{bssc} with $\veps=0.1$. $\alpha_+(R)$ and $\alpha_-(R)$ represent the
capacity general upper and lower bounds given by
Theorems~\ref{ach} and \ref{convs}.  }
\end{center}
\end{figure}

By comparing $\asyncexp_-(R)$ with $\asyncexptu(R)$ in Fig.~\ref{figtr} we observe
that for rates above roughly $ 92\%$ of the synchronous
capacity $C$, training-based
schemes are suboptimal. 

For this channel, we observe that $\alpha_-(R)$ is always above
$\asyncexptl(R)$. This feature does not generalize to arbitrary
crossover probabilities $\veps$. Indeed, consider
the channel in Fig.~\ref{bssc}, but with an
arbitrary crossover probability $\veps$, and let
$r$ be an arbitrary constant such that $0<r<1$. From
Theorem~\ref{thm2}, training-based schemes can achieve rate asynchronism pairs $(R,\alpha)$
that satisfy 
\begin{equation*}
\asyncexp \geq  m_1(1-R/C(\veps)) \quad R\in (0,C(\veps)]\,.
\end{equation*}
For the channel at hand 
$$m_1=D_B(1/2||\veps)\,,$$
hence $\asyncexp$ tends to
infinity as $\eps\rightarrow 0$, for any fixed $R\in
(0,r)$---note that $C(\veps)\to 1$ as
$\veps\to 0$.

Now, consider the random coding scheme that yields
Theorem~\ref{ach}. This scheme, which performs synchronization and
information transmission jointly,
achieves for any given rate $R\in [0,C]$ 
asynchronism exponent\footnote{The analysis of the coding scheme that yields Theorem~\ref{ach} is actually
tight in the sense that the coding scheme achieves \eqref{caphyp} with equality (see proof of
Theorem~\ref{ach} and remark p.~\pageref{remark}.)
}
$$\asyncexp =
\max_{\substack{\{P\in \cP^\cX\,:\\I(PQ)\geq R\}}} \min_{V\in
  \cP^\cY}\max\{D(V\|(PQ)_\cY),D(V\|Q_\star)\}.$$
This expression is upper-bounded by\footnote{To
see this, choose $V=Q_\star$ in the minimization.}
\begin{equation} 
\max_{P\in \cP^\cX:I(PQ)\geq R}D(Q_\star\|(PQ)_\cY), 
\label{rcodes} 
\end{equation}
which is bounded in the limit $\eps \rightarrow 0$ as long as
$R>0$.\footnote{Let $P^*=P^*(Q)$ be an input distribution $P$ that
  maximizes \eqref{rcodes} for a given channel.  Since $R\leq
  I(P^*Q)\leq H(P^*)$, $P^*$ is uniformly bounded away from $0$ and
  $1$ for all $\eps \geq 0$.  This implies that \eqref{rcodes} is
  bounded in the limit $\eps \rightarrow 0$.} Therefore the joint
synchronization-information transmission code yielding
Theorem~\ref{ach} can be outperformed by training-based schemes at
moderate to low rate, even when the output distribution when no input
is supplied is asymmetric. This shows that the
general lower bound given by Theorem~\ref{ach} is loose in general.
\end{exa}

\begin{exa}
For the channel depicted in Fig.~\ref{bscextnoise} with $\veps=0.1$,
in Fig.~\ref{figtrsym} we plot the upper and lower bounds on capacity
restricted to training-based schemes, as given by Theorem~\ref{thm2}.
For this channel it turns out that the training-based scheme upper
bound $m_2(1-R/C)$ (see Theorem~\ref{thm2}) is loose and hence
$\asyncexptu(R)=\alpha_+(R)$ for all rates.  By contrast with the
example of Fig.~\ref{figtr}, here the general lower bound $\alpha_-(R)$ is below
the lower bound for the best training best schemes ($\asyncexptl(R)$ line).

Finally, observe that, at all rates,  $\asyncexpub(R)$ in Fig.~\ref{figtrsym} is
below $\asyncexplb(R) $ (and even $\asyncexptl(R)$) in
Fig.~\ref{figtr}. In other words, under
asymmetric noise, it is possible to accommodate a much larger
level of asynchronism than under symmetric noise, at all rates.
\begin{figure}
\begin{center}
\input{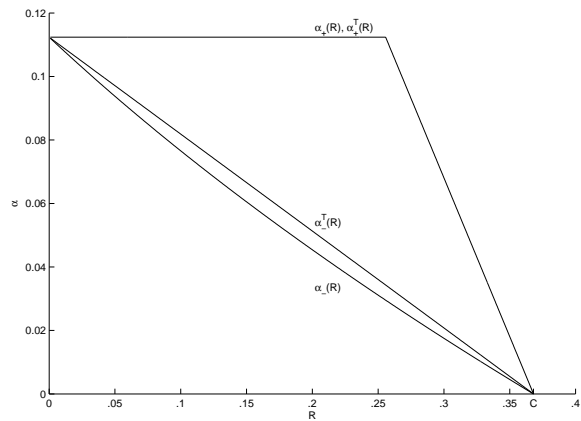}
\caption{\label{figtrsym} Lower bound ($\asyncexptl(R)$) to capacity restricted to
  training-based schemes for the channel of Fig.~\ref{bscextnoise}. $\alpha_+(R)$
  and $\alpha_-(R)$ represent the capacity general upper and lower bounds given
  by Theorems~\ref{ach} and \ref{convs}. For this channel the training
  upper bound ($\asyncexptu(R)$) coincides with $\alpha_+(R)$, and hence is not plotted separately. }
\end{center}
\end{figure}

\end{exa}

\section{Analysis}
\label{pfofresults}

In this section, we establish the theorems of Section~\ref{results}.

\subsection{Proof of Theorem~\ref{ach}}

Let $\asyncexp\geq 0$ and $P\in \cP^\cX$ satisfy the assumption of the
theorem, i.e., be such that at least one of the following inequalities holds
\begin{align}D(V||(PQ)_\cY)&\geq \alpha\notag\\
D(V||Q_\star)&\geq \alpha \label{twodiff}
\end{align}
for all distributions $V\in \cP^{\cY}$,  and let $\asynclev_n=e^{n(\alpha-\eps)}$.

The proof is based on a random coding argument associated with the
following communication strategy.  The codebook
$\cC=\{c^n(m)\}_{m=1}^M$ is randomly generated so that all $c_i(m)$,
$i\in \{1,2,\ldots,n\}$, $m\in \{1,2,\ldots,M\}$, are
i.i.d.\ according to $P$.  The sequential decoder operates according
to a two-step procedure.  The first step consists in making an coarse
estimate of the location of the sent codeword.  Specifically, at time
$t$ the decoder tries to determine whether the last $n$ output symbols
are generated by noise or by some codeword on the basis of their
empirical distribution $\hat{P}=\hat{P}_{y_{t-n+1}^t}$.  If
$D(\hat{P}\|Q_\star)< \alpha$, $\hat{P}$ is declared a ``noise type,''
the decoder moves to time $t+1$, and repeats the procedure, i.e.,
tests whether $\hat{P}_{y_{t-n+2}^{t+1}}$ is a noise type.  If,
instead, $D(\hat{P}\|Q_\star)\geq \alpha$, the decoder marks the
current time as the beginning of the ``decoding window,'' and proceeds
to the second step of the decoding procedure.

The second step consists in exactly locating and identifying the sent
codeword.   Once the beginning of the decoding window has been marked,
the decoder makes a decision the first time that the previous $n$ symbols
are jointly typical with one of the codewords.   If no
such time is found within $n$ successive time steps, the decoder stops
and declares a random message.   The typicality decoder operates as
follows.\footnote{In the literature this decoder is often referred to
as the ``strong typicality'' decoder.} Let $P_m$ be the probability
measure induced by codeword $c^n(m)$ and the channel, i.e.,
\begin{equation}\label{pmemp}
{P}_m(a,b)\defeq  \hat{P}_{c^n(m)}(a)Q(b|a) \quad(a,b)\in \cX\times\cY.
\end{equation}
At time $t$, the decoder computes the empirical
distributions $\hat{P}_m$ induced by $c^n(m)$ and the $n$ output
symbols $y_{t-n+1}^t$ for all $m\in\{1,2,\ldots,M\}$.   If
$$|\hat{P}_{c^n(m),y^{t}_{t-n+1}}(a,b)-P_m(a,b)|\leq
\mu$$ for all
$(a,b)\in \cX\times \cY$ and a unique index $m$, the decoder declares
message $m$ as the sent message.   Otherwise, it moves one step ahead
and repeats the second step of the decoding procedure on the basis of
$y^{t+1}_{t-n+2}$, i.e., it tests whether $y^{t+1}_{t-n+2}$ is typical with a
codeword. 

At the end of the asynchronism time window, i.e., at time
$\asynclev_n+n-1$, if $\hat{P}_{A_n}^{A_n+n-1}$ is either a noisy type or if it is typical with 
none of the codewords, the decoder declares a message
at random.

Throughout the argument we assume that the
typicality parameter $\mu$ is a negligible, strictly
positive quantity.

\iffalse
The divergence condition in the theorem ---which is
independent of the number of messages $M$---guarantees that the above decoding
strategy satisfies $\ex (\tau_n -\nu)^+=O(n)$ and that it correctly locates the
sent codeword within a window of size $O(n)$ with high probability.   Then, by
imposing that $M$ satisfies $\ln M/n\leq I(PQ)-\eps$, we guarantee
correct message location and identification with high probability
given that the
codeword has been approximatively located with high probability at the end of
the first phase.
\fi

We first show that, on average, a randomly chosen
codebook combined with the sequential decoding
procedure described above achieves the rate-exponent
pairs $(R,\asyncexp)$ claimed by the theorem. This, as
we show at the end of the proof, implies the existence
of a nonrandom codebook that, together with the above
decoding procedure, achieves any pair $(R,\asyncexp)$
claimed by the theorem.

Let $\ln M/n =
I(PQ)-\eps$, $\eps>0$. We first compute
the average, over messages and codes, expected reaction delay and
probability of error.   These quantities, by symmetry of the encoding
and decoding procedures, are the same as the average over codes
expected reaction delay and probability of error conditioned on the
sending of a particular message.   Below, expected reaction delay and
error probability are computed conditioned on the sending of
message~$m=1$.

Define the following events:
\begin{align*}
\cE_1&=\{D(\hat{P}_{Y_{\nu}^{\nu+n-1}}\|Q_\star)<\alpha
\text{ , i.e., }
\hat{P}_{Y_{\nu}^{\nu+n-1}} 
\text{ is a ``noise type''}\},\\ 
\cE_2&=\{\text{${Y_{\nu}^{\nu+n-1}}$
is not typical with $C^n(1)$}\},\\ 
\cE_3&=\{D(\hat{P}_{Y_{t-n+1}^{t}}\|Q_\star)\geq \alpha
\text{ for some } t<\nu\}.
\end{align*}
For the reaction delay we have
\begin{align}
\label{expe}
& \ex_1(\tau_n-\nu)^+ \notag\\
&\quad= \ex_1[(\tau_n-\nu)^+\openone(\tau_n\geq \nu+ 2n)]\notag\\
&\qquad{}+\ex_1[(\tau_n-\nu)^+\openone(\nu+n \leq \tau_n<\nu+ 2n)] \notag \\
&\qquad{}+\ex_1[(\tau_n-\nu)^+\openone(\tau_n< \nu+ n)]\notag \\
&\quad\leq (A_n+n-1) \pr_1(\tau_n\geq \nu+ 2n)\notag \\
&\qquad{} + 2n \pr_1(\nu+n \leq \tau_n< \nu+2n) +n,
\end{align}
where the subscript $1$ in $\ex_1$ and $\pr_1$ indicates
conditioning on the event that message $m=1$ is sent.  The two
probability terms on the right-hand side of the second inequality of
\eqref{expe} are bounded as follows.

The term $\pr_1(\tau_n\geq \nu+ 2n)$ is upper bounded by the probability
that the decoding window starts after time $\nu+n-1$.   This, in turn,
is upper bounded by the probability of the event that, at time
$\nu+n-1$, the last $n$ output symbols induce a noise type.
Therefore, we have
\begin{align}
\label{ntype}
\pr_1(\tau_n\geq \nu+ 2n)&\leq \pr_1(\cE_1) \notag \\
&\leq \sum_{\{V\in \cP^\cY_n:\; D(V\|Q_\star)\leq \alpha\}}e^{-nD(V\|(PQ)_\cY)}\notag
\\
&\leq \sum_{\{V\in \cP^\cY_n:\; D(V\|Q_\star)\leq \alpha\}}e^{-n\alpha}\notag \\
&\leq \poly(n)e^{-n\alpha},
\end{align}
where the second inequality follows from the definition of the event
$\cE_1$ and Fact~\ref{fact:2}; where the third inequality follows from
\eqref{twodiff}  (which implies that if $D(V\|Q_\star)\leq \alpha$ then necessarily
$D(V\|(PQ)_\cY)\geq \alpha$ ); and
where the fourth inequality follows from Fact~\ref{fact:1}.

The probability $\pr_1(\nu+n \leq \tau_n< \nu+2n)$ is at most the
probability that the decoder has not stopped by time $\nu+n-1$.   This
probability, in turn, is at most the probability that, at time
$\nu+n-1$, the last $n$ output symbols either induce a noisy type, or
are not typical with the sent codeword $C^n(1)$ (recall that message
$m=1$ is sent). By union bound we get
\begin{align}
\pr_1 (\nu+n \leq \tau_n< \nu+2n)&\leq \pr_1 (\tau_n\geq  \nu+n)\notag \\
&\leq
\pr_1(\cE_1)+\pr_1(\cE_2)\notag\\
&\leq \poly(n)e^{-n\alpha}+o(1)\notag \\
&=o(1)\quad (n\rightarrow\infty),
\label{ntype2}
\end{align}
where we used the last three computation steps of \eqref{ntype} to bound $\pr_1(\cE_1)$, and where we used
\cite[Lemma~2.12, p.~34]{CK} to show that $\pr_1(\cE_2)$ tends to
zero as $n$ tends to infinity.   From \eqref{expe}, \eqref{ntype}, and
\eqref{ntype2}, we deduce that
\begin{equation*}
\ex_1 (\tau_n-\nu)^+ \leq n(1+o(1))\quad (n\rightarrow\infty)
\end{equation*}
since $A_n=e^{n(\alpha-\eps)}$, by assumption.

We now compute $\pr_1(\cE)$, the average error probability conditioned on sending
message $m=1$. We have
\begin{align}
&\pr_1(\cE)\notag\\ 
&\ =\pr_1(\cE\cap \{\tau_n<\nu\})\notag\\
    &\qquad{} +\pr_1(\cE\cap \{\nu\leq \tau_n \leq \nu +n-1\})\notag\\
&\qquad{} +\pr_1(\cE\cap \{\tau_n\geq \nu+n\})\notag\\
&\leq\pr_1(\tau_n<\nu)+\pr_1(\cE\cap \{\nu\leq \tau_n\leq\nu+n-1\}) \notag\\
&\qquad{} +\pr_1(\tau_n\geq \nu+n)\notag\\ 
&\leq \pr_1(\cE_3)+\pr_1(\cE\cap \{\nu\leq
\tau_n \leq \nu +n-1\})\notag\\
&\qquad{}+o(1)\quad (n\rightarrow\infty),
\label{errore}
\end{align}
where for the last inequality we used the definition of $\cE_3$ and
upper bounded $\pr_1(\tau\geq \nu+n)$ using the last three
computation steps of \eqref{ntype2}.

For $\pr_1(\cE_3)$, we have
\begin{align}
\label{errore2}
\pr_1(\cE_3) &= \pr(\cup_{t<\nu}\{D(\hat{P}_{Y_{t-n+1}^{t}}\|Q_\star)\geq
\alpha\})\notag\\
&\leq A_n\sum_{\{V\in \cP^\cX_n:\; D(V\|Q_\star)\geq
\alpha\}}e^{-nD(V||Q_\star)}\notag \\
&\leq A_n \sum_{\{V\in \cP^\cX_n:\; D(V\|Q_\star)\geq \alpha\}}e^{-n\alpha}\notag \\
&\leq A_n e^{-n\alpha}\poly(n)\notag\\
&=o(1)\quad (n\rightarrow\infty)
\end{align}
where the first inequality in \eqref{errore2} follows from the union bound over
time and
Fact~\ref{fact:2}; where the third inequality follows from
Fact~\ref{fact:1}; and where the last equality holds since 
$A_n=e^{n(\alpha-\eps)}$, by assumption.

We now show that
\begin{equation} 
 \label{ultimo}
 \pr_1(\cE\cap \{\nu\leq \tau_n \leq \nu +n-1\}) = o(1) 
\quad (n\rightarrow\infty),
\end{equation}
which, together with \eqref{errore} and \eqref{errore2}, shows that
$\pr_1(\cE)$ goes to zero as $n\rightarrow\infty$.

We have
\begin{align}
\label{penultimo}
\pr_1(\cE&\cap \{\nu\leq \tau_n \leq \nu +n-1\})\notag\\
&= \pr_1(\cup_{t=\nu}^{\nu+n-1}\{\cE \cap\{\tau_n=t\}\cap \cE_3\})\nonumber \\
&\quad{}+\pr_1(\cup_{t=\nu}^{\nu+n-1}\{\cE \cap
\{\tau_n=t\}\cap \cE_3^\comp\})\notag \\
&\leq
\pr_1(\cE_3)+\pr_1(\cup_{t=\nu}^{\nu+n-1}\{\cE
\cap \{\tau_n=t\}\cap \cE_3^\comp\}) \notag \\
&\leq o(1) + \pr_1(\{\cE \cap
\{\tau_n=\nu+n-1\})\nonumber\\
&\quad{}+\pr_1(\cup_{t=\nu}^{\nu+n-2}\{\cE \cap
\{\tau_n=t\}\cap \cE_3^\comp\})\notag \\
&\leq o(1) + o(1)\nonumber \\
&\quad{}+
 \pr_1(\cup_{t=\nu}^{\nu+n-2}\{\cE \cap
\{\tau_n=t\}\cap \cE_3^\comp\})\quad (n\rightarrow\infty)
\end{align}
where the second inequality follows from
\eqref{errore2}; where the fourth inequality
follows from the definition of event $\cE_2$; and
where the
third inequality follows from the fact that, given the
correct codeword location, i.e., $\tau_n=\nu+n-1$, the typicality decoder guarantees vanishing
error probability since we assumed that $\ln M/n=I(PQ)-\eps$
(see \cite[Chapter 2.1]{CK}).   

The event $\{\cE \cap \{\tau_n=t\}\cap
\cE_3^\comp\}$, with $\nu\leq t \leq \nu+n-2$, happens when a block of $n$
 consecutive symbols, received between $\nu-n+1$ and $\nu+n-2$, is jointly typical
with a codeword other than the sent codeword $C^n(1)$.   Consider a
block $Y^n$ in this range, and let $J\in \cP^{\cX, \cY}_n$ be a
typical joint type, i.e. $$|J(x,y)-P(x)Q(y|x)|\leq
\mu$$ for all
$(x,y)\in \cX\times \cY$---recall that $\mu>0$ is the typicality
parameter, which we assume to be a negligible quantity throughout the
proof.

For some $1 \leq k \leq n-1$, the first $k$
symbols of block $Y^n$ are
generated by noise, and the remaining $n-k$ symbols are generated by
the sent codeword, i.e., corresponding to $m=1$.
Thus, $Y^n$ is
independent of any unsent codeword $C^n(m )$. 
The probability that $C^n(m )$,
$m\ne 1$, together with $Y^n$ yields a particular type $J$
is upper bounded as
follows:
\begin{align}
&\pr(\hat{P}_{C^n(m),Y^n}=J)\notag\\
&\quad=\sum_{y^n\in \cY^n}\pr(Y^n=y^n)\sum_{x^n :
\hat{P}_{x^n,y^n}=J} \pr(X^n=x^n)\notag \\
&\quad= \sum_{y^n\in \cY^n}\pr(Y^n=y^n)\sum_{x^n:
\hat{P}_{x^n,y^n}=J} e^{-n(H(J_\cX)+D(J_\cX\|P))}\notag \\
&\quad\leq  \sum_{y^n\in \cY^n}\pr(Y^n=y^n) e^{-n H(J_X)}|\{x^n:
\hat{P}_{x^n,y^n}=J\}|\notag \\
&\quad\leq \sum_{y^n\in \cY^n}\pr_1(Y^n=y^n) e^{-n H(J_\cX)}e^{n H
(J_{\cX|\cY})}\notag \\
&\quad\leq e^{-n I(J)}, \label{mezzomezzo}
\end{align}
where $H(J_\cX)$ denotes the entropy of the left marginal of
$J$,
\begin{equation*}
H(J_{\cX|\cY})\defeq  -\sum_{y\in \cY} J_\cY(y)\sum_{x\in \cX}
J_{\cX|\cY}(x|y)\ln J_{\cX|\cY}(x|y),
\end{equation*}
and where $I(J)$ denotes the mutual information induced by $J$.  

The first
equality in \eqref{mezzomezzo} follows from the independence of
$C^n(m)$ and $Y^n$, the second
equality follows from \cite[Theorem~11.1.2, p.~349]{CT}, and the
second inequality follows from \cite[Lemma~2.5, p.~31]{CK}.  

It follows that the probability that an unsent
codeword $C^n(m)$ together with $Y^n$ yields a
type $J$ that is typical, i.e., close to $PQ$, is
upper bounded as 
\begin{equation*}
\pr_1(\hat{P}_{C^n(m),Y^n}=J)\leq e^{-n (I(PQ)-\eps/2)}
\end{equation*}
for all $n$ large enough, by continuity of the
mutual information.\footnote{The typicality
parameter $\mu=\mu(\eps)>0$ is chosen small enough so that
this inequality holds.}

Note that the set of inequalities \eqref{mezzomezzo} holds for any
block of $n$ consecutive output symbols $Y^n$ that is independent of
codeword $C^n(m)$.\footnote{Note that the fact
that $Y^n$ is partly
generated by noise and partly by the sent codeword
$C^n(1)$ is not used to establish
\eqref{mezzomezzo}.}  Hence, from the union
bound, it follows that
\begin{align}\label{wei}
&\pr_1(\cup_{t=\nu}^{\nu+n-2}\{\cE \cap
\{\tau_n=t\}\cap \cE_3^\comp\}) \nonumber\\
&\quad\leq n\sum_{m\ne 1}\sum_{\substack{\{J\in
\cP_{\cX,\cY}^n:\;\; \forall (x,y)\in \cX\times\cY,\\
|J(x,y)-P(x)Q(y|x)|\leq \mu\}}}
\pr(\hat{P}_{C^n(m),Y^n}=J) \notag \nonumber\\
&\quad\leq n M e^{-n (I(PQ)-\eps/2)}\poly(n)\notag\\
&\quad\leq e^{-n\eps/2}\poly(n),
\end{align}
where the second inequality follows from
Fact~\ref{fact:1}, and where the
third inequality follows from the assumption that $\ln M/n=
I(PQ)-\eps$.  Combining \eqref{wei} with \eqref{penultimo}
yields \eqref{ultimo}.

So far, we have proved that a random codebook has a decoding delay
averaged over messages that is at most $n(1+o(1))$ ($n \rightarrow
\infty$), and an error probability averaged over messages that
vanishes as $n\rightarrow\infty$, whenever $A_n=e^{n(\alpha-\eps)}$,
$\eps>0$. This, as we now show, implies the
existence of nonrandom codebooks achieving the same performance,
yielding the desired result. The expurgation arguments we use are
standard and in the same spirit as those given in \cite[p.~203-204]{CT}
or \cite[p.~151]{G}.

For a particular codebook $\cC_n$, let $\pr(\E|\cC_n)$ and
$\ex((\tau_n-\nu)^+|\cC_n)$ be the average, over messages, error
probability and reaction delay, respectively. We have proved that
for any $\eps>0$,
\begin{equation*}
\ex(\ex(\tau_n-\nu)^+|\cC_n))\leq n(1+\eps)
\end{equation*}
and
\begin{equation*}
\ex(\pr(\E|\cC_n))\leq \eps
\end{equation*}
for all $n$ large enough.

Define events
\begin{equation*}
\cA_1=\{\ex(\tau_n-\nu)^+|\cC_n)\leq
n(1+\eps)^2\},
\end{equation*}
and
\begin{equation*}
\cA_2=\{\pr(\E|\cC_n)\leq \eps k\}
\end{equation*}
where $k$ is arbitrary.

From Markov's inequality it follows
that\footnote{Probability here is averaged over
randomly generated codewords.}
\begin{equation*}
\pr(\cA_1\cap \cA_2)\geq
1-\frac{1}{1+\eps}-\frac{1}{k}\,.
\end{equation*}
Letting $k$ be large enough so that the right-hand side of the above
inequality is positive, we deduce that there exists a particular code
$\cC_n$ such that
\begin{equation*}
\ex(\tau_n-\nu)^+|\cC_n)\leq n(1+\eps)^2
\end{equation*}
and
\begin{equation*}
\pr(\E|\cC_n)\leq \eps k.
\end{equation*}
We now remove from $\cC_n$ codewords with poor reaction delay and error
probability.  Repeating the argument above with the fixed code $\cC_n$,
we see that a positive fraction of the codewords of $\cC_n$ have
expected decoding delay at most $n(1+\eps)^3$ and error probability
at most $\eps k^2$. By only keeping this set of codewords, we conclude that for any
$\eps>0$ and all $n$ large enough, there exists a rate $R=I(PQ)-\eps$
code operating at asynchronism level $\asynclev=e^{(\asyncexp-\eps)
n}$ with maximum error probability less than~$\eps$.
\hfill\IEEEQEDclosed

\begin{remark} \label{remark} It is possible to somewhat strengthen 
the conclusion of Theorem~\ref{ach} in two ways.  First, it can be
strenthened by observing that what we actually proved is that the
error probability not only vanishes but does so exponentially in
$n$.\footnote{Note that the error probability of the typicality
  decoder given the correct message location, i.e., $\pr(\E\cap
  \{\tau_n=\nu+n-1\}\})$, is exponentially small in $n$ \cite[Chapter
    2]{CK}.}  Second, it can be strengthened by showing that the
proposed random coding scheme achieves \eqref{caphyp} with equality. A
proof is deferred to Appendix~\ref{remarksection}.
\end{remark}

\subsection{Proof of Theorem~\ref{convs}}

We show that any rate $R>0$ coding scheme operates at an
asynchronism $\alpha$ bounded from above by
$\max_{\cS}\min\{\alpha_1,\alpha_2\}$, where $\cS$, $\alpha_1$, and
$\alpha_2$ are defined in the theorem's statement.

We prove Theorem~\ref{convs} by establishing the following four claims. 

The first claim says that, without loss of generality, we may restrict ourselves
to constant composition codes.  Specifically, it is possible to expurgate an
arbitrary code to make it of constant composition while impacting
(asymptotically) neither the rate
nor the asynchronism exponent the original code is operating at.  In more
detail, the expurgated codebook is such that all codewords have the same type,
and also so that all codewords have the same type over the first $\delay_n$
symbols (recall that $\delay_n\defeq  \max_m \ex(\tau_n-\nu)^+$). The
parameter $\delta$ in Theorem~\ref{convs} corresponds to the ratio $\delay_n
/n$, and $P_1$ and $P_2$ correspond to the empirical types over the first
$\delay_n$ symbols and the whole codeword (all $n$ symbols), respectively.

{\it Fix an arbitrarily small constant $\veps>0$.}
\begin{claim}
\label{claim:i}
 Given any coding scheme $\{(\cC_n,(\tau_n,\phi_n))\}_{n\geq 1}$
achieving $(R,\asyncexp)$ with $R>0$ and $\asyncexp>0$, there exists a second coding scheme
$\{(\cC_n',(\tau_n,\phi_n))\}_{n\geq 1}$ achieving $(R,\asyncexp)$
that is obtained by expurgation, i.e., $\cC_n'\subset\cC_n$,
$n=1,2,\ldots$, and that has constant
composition with respect to some distribution $P_n^1$ over the first
\begin{equation}\label{dn}
d(n)\defeq  \min \{\lfloor(1+\eps)\delay_n\rfloor,n\}
\end{equation}
symbols, and constant composition with respect to some distribution $P_n^2$ over $n$
symbols. (Hence, if $\lfloor(1+\eps)\delay_n\rfloor\geq
n$, then $P_n^1=P_n^2$.) Distributions $P_n^1$
and $P_n^2$ satisfy Claims $2-4$ below.
\end{claim}
Distribution $P^1_n$ plays the same role as the codeword distribution for
synchronous communication. As such it should induce a large enough input-output
channel mutual information to support rate $R$ communication.
\begin{claim}
\label{claim:ii}
For all $n$ large enough $$R\leq I(P_n^1Q)(1+\eps)\,.$$
\end{claim}

Distribution $P^2_n$ is specific to asynchronous communication. Intuitively,
$P^2_n$ should induce an output distribution that is sufficiently different
from pure noise so that to allow a decoder to distinguish between noise and any
particular transmitted message when the asynchronism level corresponds to
$\alpha$. Proper message detection means that the decoder should not overreact
to a sent codeword (i.e., declare a message before even it is sent), but
also not miss the sent codeword. As an extreme case, it is possible to achieve
a reaction delay $\ex(\tau-\nu)^+$ equal to zero by setting $\tau=1$, at the
expense of a large probability of error. In contrast, one clearly minimizes the
error probability by waiting until the end of the asynchronism window,  {\it
i.e.}, by setting $\tau=A_n+n-1$, at the expense of the rate, which will be
negligible in this case.

The ability to properly detect only a single codeword with type $P_n^2$ is captured by condition
$\alpha\leq\alpha_2$ where $\alpha_2$ is defined in the theorem's statement. 
This condition is equivalently stated as:
\begin{claim}
\label{claim:iii}
For any $W\in \cP^{\cY|\cX}$ and for all $n$ large enough, at least
one of the following two inequalities holds
\begin{align*}
&\asyncexp < D(W\|Q_\star|P_n^2)+\eps,\\
&\asyncexp < D(W\|Q|P_n^2)+\eps.
\end{align*}
\end{claim}

As it turns out, if the synchronization threshold is finite, $P_n^1$ plays also
a role in the decoder's ability to properly detect the transmitted message.
This is captured by condition $\alpha \leq \alpha_1$ where $\alpha_1$ is
defined in the theorem's statement. Intuitively, $\alpha_1$ relates to the
probability that the noise produces a string of length $n$ that looks typical
with the output of a {\emph{randomly selected codeword}}. If $\alpha>\alpha_1$, the
noise produces many such strings with high probability, which implies a large
probability of  error. 
\begin{claim}
\label{claim:iv}
For all $n$ large enough, 
\begin{equation*} 
\asyncexp \leq \frac{d(n)}{n} \left(
I(P_n^1Q)-R+D((P_n^1Q)_\cY\|Q_\star) \right) + \eps
\end{equation*}
provided that $\thres<\infty$.
\end{claim}

Note that, by contrast with the condition in Claim~\ref{claim:iii}, the
condition in Claim~\ref{claim:iv} depends also on the communication rate since
the error yielding to the latter condition depends on the number of codewords.

Before proving the above claims, we show how they imply Theorem~\ref{convs}. 
The first part of the Theorem, i.e., when $\thres<\infty$,
follows from Claims \ref{claim:i}-\ref{claim:iv}. To see this, note that the
bounds $\alpha_1$ and $\alpha_2$ in the Theorem correspond to the bounds of
Claims~\ref{claim:iii} and \ref{claim:iv}, respectively, maximized over $P_n^1$ and $P_n^2$. The
maximization is subjected to the two constraints given by Claims \ref{claim:i}
and \ref{claim:ii}:
$P_n^1$ and $P_2^n$ are  the empirical distributions of the codewords of
$\cC_n'$ over the first $\delta n$ symbols ($\delta\in [0,1]$), and over the
entire codeword length, respectively, and condition $R\leq I(P_n^1Q)(1+\eps)$
must be satisfied. 
Since $\eps>0$ is arbitrary, the result then follows by taking the limit $\eps\downarrow 0$ on the above derived
bound on $\alpha$.

Similarly, the second part of Theorem~\ref{convs}, i.e., when
$\thres=\infty$, is a consequence of Claim~\ref{claim:iii} only.

We now prove the claims. As above, $\veps>0$ is supposed to be an arbitrarily
small constant.
\begin{IEEEproof}[Proofs of Claims~\ref{claim:i} and ~\ref{claim:ii}]
We show that for all $n$
large enough,
we have
\begin{align}\label{aux1}
\frac{R-\eps}{1+\eps}\leq \frac{\ln |{\cal{C}}_n'|}{d(n)}\leq I(P_n^1Q)+\eps\,, 
\end{align}
where ${\cal{C}}_n'$ is a subset of codewords from ${\cal{C}}_n$ that have
constant composition $P_n^1$ over the first $d(n)$ symbols, where $d(n)$ is
defined in \eqref{dn}, and
constant composition $P_n^2$ over $n$ symbols.  This is done via an expurgation
argument in the spirit of \cite[p.~151]{G} and \cite[p.~203-204]{CT}.

We first show the left-hand side inequality of \eqref{aux1}. Since  $\{({\cal{C}}_n,(\tau_n,\phi_n))\}_{n\geq 1}$ achieves a rate $R$, by
definition (see Definition~\ref{defcs}) we have  
$$\frac{\ln |{\cal{C}}_n|}{\delay_n}\geq R-\eps/2$$ for
all $n$ large enough. Therefore, 
$$\frac{\ln |{\cal{C}}_n|}{d(n)}\geq
\frac{R-\eps/2}{1+\eps}$$
for all $n$ large enough.

\iffalse We show that given $Q\in \cP^{\cY|\cX}$, with $\thres<\infty$,
and an $(R,\asyncexp)$ coding scheme
$\{(\cC_n,(\tau_n,\phi_n))\}_{n\geq 1}$ with $R>0$ and $\asyncexp>0$,
one can extract a coding scheme whose codebooks have constant
compositions over length $n$ and over length $d(n)$, and that satisfy Claim 2-4.

Let $\eps>0$ be some arbitrarily small constant.
Since
$\{(\cC_n,(\tau_n,\phi_n))\}_{n\geq 1}$ achieves a rate $R$, by
definition (see Definition~\ref{defcs}) we have
\begin{equation*}
\frac{\ln |\cC_n|}{\delay_n}\geq R-\eps/2
\end{equation*}
for $n$ large enough.   Therefore, we have \begin{equation*} \frac{\ln
|\cC_n|}{d(n)}\geq \frac{R-\eps/2}{1+\eps}
\end{equation*}
for $n$ large enough.
\fi
Now, group the codewords of $\cC_n$ into families such that elements
of the same family have the same type over the first $d(n)$ symbols.
Let $\cC_n''$ be the largest such family and let $P_n^1$ be its type.
Within $\cC_n''$, consider the largest subfamily $\cC_n'$ of
codewords that have constant composition over $n$ symbols, and let
$P_n^2$ be its type (hence, all the codewords in $\cC_n'$ have common
type $P_n^1$ over $d(n)$ symbols and common type $P_n^2$ over $n$
symbols).

By assumption, $R>0$, so $\cC_n$ has a number of codewords that is
exponential in $\delay_n$.  Due to Fact~\ref{fact:1}, to establish the left-hand
side inequality of \eqref{aux1}, i.e., to show that
$\cC_n'$ achieves essentially the same rate as $\cC_n$, it suffices to show that
the number of subfamilies in $\cC_n'$ is bounded by a polynomial in
$\delay_n$.  We do this assuming that $\thres<\infty$ and that
Claim~\ref{claim:iv} (to be proved) holds.  

By assumption, $\thres<\infty$, and thus from Theorem~\ref{ulimit} we have
that $D((PQ))_\cY\|Q_\star)<\infty$ for any input distribution $P$.
Using Claim~\ref{claim:iv} and the assumption that $\asyncexp>0$, we
deduce that $\liminf_{n\rightarrow\infty}d(n)/n>0$, which implies
that $n$ cannot grow faster than linearly in $\delay_n$.  Therefore,
Fact~\ref{fact:1} implies that the number of subfamilies of $\cC_n'$
is bounded by a polynomial in~$\delay_n$.

We now prove the right-hand side inequality of \eqref{aux1}. 
Letting $\cE^\comp$ denote the event of a correct decoding,
Markov's inequality implies that for every message index $m$,
\begin{align} &\pr_m(\{(\tau_n-\nu)^+\leq
(1+\eps)\delay_n\}\cap \cE^\comp) \notag\\ &\ \geq
1-\frac{\ex_m(\tau_n-\nu)^+}{\delay_n}\frac{1}{1+\eps}-\pr_m(\cE)\notag\\
&\ \geq 1-\frac{1}{1+\eps}-\pr_m(\cE), \label{ineqdetcor}
\end{align} since $\delay_n\defeq 
\max_m\ex_m(\tau_n-\nu)^+$.   The right-hand side of
\eqref{ineqdetcor} is strictly greater than zero for $n$ large
enough because an $(R,\asyncexp)$ coding scheme achieves a
vanishing maximum error probability as $n\rightarrow\infty$.  
This means that $\cC_n'$ is a good code for the synchronous
channel, i.e., for $\asynclev=1$.   More precisely, the
codebook formed by truncating each codeword in $\cC_n'$ to
include only the first $d(n)$ symbols achieves a probability
of error (asymptotically) bounded away from one with a
suitable decoding function.   This implies that the right-hand
side of \eqref{aux1} holds for $n$ large enough by
\cite[Corollary~1.4, p.~104]{CK}. \end{IEEEproof} In
establishing the remaining claims of the proof, unless
otherwise stated, whenever we refer to a codeword it is
assumed to belong to codebook $\cC_n'$.  Moreover, for
convenience, and with only minor abuse of notation, we let $M$
denote the number of codewords in $\cC_n'$.
\begin{IEEEproof}[Proof of Claim~\ref{claim:iii}] We fix $W\in
\cP^{\cY|\cX}$ and show that for all $n$ large enough, at
least one of the two inequalities \begin{equation*}
D(W\|Q|P_n^2)>\asyncexp -\eps, \end{equation*}
\begin{equation*} D(W\|Q_\star|P_n^2)>\asyncexp -\eps,
\end{equation*} must hold. To establish this, it may be
helpful to interpret $W$ as the true channel behavior during
the information transmission period, i.e., as the conditional
distribution induced by the transmitted codeword and the
corresponding channel output. With this interpretation,
$D(W\|Q|P_n^2)$ represents the large deviation exponent of the
probability that the underlying channel $Q$ behaves as $W$
when codeword distribution is $P_n^2$, and
$D(W\|Q_\star|P_n^2)$ represents the large deviation exponent
of the probability that the noise behaves as $W$ when codeword
distribution is $P_n^2$. As it turns out, if both the above
inequalities are reversed for a certain $W$, the asynchronism
exponent is too large. In fact, in this case both the
transmitted message and pure noise are very likely to produce
such a $W$. This, in turn will confuse the decoder. It will
either miss the transmitted codeword or stop before even
the actual codeword is sent.

 In the sequel, we often use the shorthand notation
$\cT_W(m)$ for $\cT_W^n(c^n(m))$.

Observe first that if $n$ is such that
\begin{equation} 
\label{msg0}
\pr_{m}(Y_\nu^{\nu+n-1}\in \cT_W(m))=0,
\end{equation}
then
\begin{equation*}
D(W\|Q|P_n^2)=\infty,
\end{equation*}
by Fact~\ref{fact:3}.   Similarly, observe that if $n$ is such that
\begin{equation} 
\label{noise0}
\pr_\star(Y_\nu^{\nu+n-1}\in \cT_W(m))=0,
\end{equation}
where $\pr_\star$ denotes the probability under pure noise (i.e., the
$Y_i$'s are i.i.d.\ according to $Q_\star$), then
\begin{equation*}
D(W\|Q_\star|P_n^2)=\infty.
\end{equation*}
Since the above two observations hold regardless of $m$ (because all
codewords in $\cC_n'$ have the same type), Claim~\ref{claim:iii}
holds trivially for any value of $n$ for which \eqref{msg0} or
\eqref{noise0} is satisfied.

In the sequel, we thus restrict our attention to values of
$n$ for which
\begin{equation} 
\label{msg00}
\pr_{m}(Y_\nu^{\nu+n-1}\in \cT_W(m))\ne 0
\end{equation}
and
\begin{equation} 
\label{noise00}
\pr_\star(Y_\nu^{\nu+n-1}\in\cT_W(m))\ne 0.
\end{equation}

Our approach is to use a change of measure to show that
if Claim~\ref{claim:iii}  does not hold, then the expected reaction
delay grows exponentially with $n$, implying that the rate is
asymptotically equal to zero.  To see this, note that any coding
scheme that achieves vanishing error probability cannot have $\ln M$
grow faster than linearly with $n$, simply because of the limitations
imposed by the capacity of the synchronous channel.  Therefore, if
$\ex (\tau_n-\nu)^+$ grows exponentially with $n$, the rate goes to
zero exponentially with $n$.  And note that for $\ex (\tau_n-\nu)^+$
to grow exponentially, it suffices that $\ex_m(\tau_n-\nu)^+$ grows
exponentially for at least one message index $m$, since
$\delay_n=\max_m\ex_m(\tau_n-\nu)^+$ by definition.

To simplify the exposition and avoid heavy notation, in the following
arguments we disregard discrepancies due to the rounding of
noninteger quantities.   We may, for instance, treat $\asynclev/n$ as an
integer even if $\asynclev$ is not a multiple of $n$.   This has no
consequences on the final results, as these discrepancies vanish when
we consider code with blocklength $n$ tending to
infinity.

We start by lower bounding the reaction delay as\footnote{Recall that
the subscripts $m,t$ indicate conditioning on the event that
message $m$ starts being sent at time $t$.}
\begin{align}
\delay_n&\defeq \max_m\frac{1}{\asynclev}
\sum_{t=1}^\asynclev
\ex_{m,t}(\tau_n-t)^+\notag \\ &\geq
\frac{1}{3}\sum_{t=1}^{\asynclev_n/3}\pr_{m,t}((\tau_n-t)^+\geq
A_n/3)\notag \\ &\geq
\frac{1}{3}\sum_{t=1}^{\asynclev_n/3}\pr_{m,t}(\tau_n\geq t+
A_n/3)\notag\\
&\geq \frac{1}{3}\sum_{t=1}^{\asynclev_n/3}\pr_{m,t}(\tau_n\geq 2A_n/3), \label{markov}
\end{align}
where for the first inequality we used Markov's inequality.
The message index $m$ on the right-hand side of \eqref{markov} will be
specified later; for now it may correspond to any message.

  We lower
bound each term $\pr_{m,t}(\tau_n \geq 2A_n/3)$ in the above sum as
\begin{align}
\pr_{m,t}(\tau_n&\geq 2A_n/3)&\notag\\
&\ \geq \pr_{m,t}(\tau_n\geq 2A_n/3 \mid Y_t^{t+n-1}\in
\cT_W(m))\notag\\
& \hspace{1.5cm}\times\pr_{m,t}(Y_t^{t+n-1}\in \cT_W(m))\notag \\
&\ \geq \pr_{m,t}(\tau_n\geq 2A_n/3 \mid Y_t^{t+n-1}\in \cT_W(m)) \notag\\
&\hspace{1.5cm}\times e^{-n D_1}\poly(n),
\label{chgemeas}
\end{align}
where $D_1\defeq  D(W\|Q|P_n^2)$, and where the second inequality
follows from Fact~\ref{fact:3}.\footnote{Note that the right-hand side of
the first inequality in \eqref{chgemeas} is well-defined because of
\eqref{msg00}.   }

The key step is to apply the change of measure
\begin{align} 
\pr_{m,t}&(\tau_n\geq 2A_n/3|Y_t^{t+n-1}\in
\cT_W(m))\nonumber \\
&=\pr_\star(\tau_n\geq 2A_n/3|Y_t^{t+n-1}\in\cT_W(m))\label{key}\,.
\end{align}
To see that \eqref{key} holds, first note that
for any $y^n$ 
\begin{align*} 
\pr_{m,t}&(\tau_n\geq 2A_n/3|Y_t^{t+n-1}=y^n)\\
&=\pr_\star(\tau_n\geq
2A_n/3|Y_t^{t+n-1}=y^n)
\end{align*}
since distribution $\pr_{m,t}$ and $\pr_\star$ differ only over channel outputs
$Y_t^{t+n-1}$.

Next, since sequences inside $\cT_W(m)$ are permutations of each other
\begin{align*}
\pr_{m,t}(Y_t^{t+n-1}=y^n|Y_t^{t+n-1}\in
\cT_W(m))&=\frac{1}{|\cT_W(m)|}\\
=\pr_\star(Y_t^{t+n-1}=y^n|Y_t^{t+n-1}\in
\cT_W(m)),
\end{align*}
we get
\begin{align*}
&\pr_{m,t}(\tau_n\geq 2A_n/3|Y_t^{t+n-1}\in \cT_W(m))\\
&\ =\sum_{y^n\in \cT_W(m)}\pr_{m,t}(\tau_n\geq
2A_n/3|Y_t^{t+n-1}=y^n)\\
&\  \qquad \qquad{}\times \pr_{m,t}(Y_t^{t+n-1}=y^n|Y_t^{t+n-1}\in
\cT_W(m))\notag \\
&\ =\sum_{y^n\in \cT_W(m)}\pr_\star(\tau_n \geq
2A_n/3|Y_t^{t+n-1}=y^n)\\
&\ \qquad  \qquad{} \times \pr_\star(Y_t^{t+n-1}=y^n|Y_t^{t+n-1}\in
\cT_W(m))\notag \\
&\ =\pr_\star(\tau_n\geq 2A_n/3|Y_t^{t+n-1}\in \cT_W(m)).
\end{align*}
This proves \eqref{key}.   Substituting \eqref{key} into the right-hand side
of \eqref{chgemeas} and using \eqref{markov}, we get
\begin{align*}
\delay_n& \geq
e^{-nD_1}\poly(n)\\
&\hspace{.5cm}\times \sum_{t=1}^{\asynclev/3} \pr_\star(\tau_n\geq
2A_n/3|Y_t^{t+n-1}\in \cT_W(m))\notag\\
&\geq e^{-
n(D_1-D_2)}\poly(n)\\
&\hspace{.5cm}\times \sum_{t=1}^{\asynclev/3} \pr_\star(\tau_n\geq
2A_n/3,Y_t^{t+n-1}\in \cT_W(m)),
\end{align*}
where $D_2\defeq  D(W\|Q_\star|P_n^2)$, and where the last
inequality follows from Fact~\ref{fact:3}.   By summing only over the
indices that are multiples of $n$, we obtain the weaker inequality
\begin{align}
\label{und}
\delay_n &\geq e^{- n(D_1-D_2)}\poly(n)\notag\\
&\times \sum_{j=1}^{\asynclev/3n} \pr_\star(\tau_n\geq
2A_n/3,Y_{jn}^{jn+n-1}\in \cT_W(m)).
\end{align}
Using \eqref{und}, we show that $\ex (\tau_n-\nu)^+$ grows
exponentially with $n$ whenever $D_1$ and $D_2$ are both upper bounded
by $\asyncexp-\eps$.  This, as we saw above, implies that the rate is
asymptotically equal to zero, yielding Claim~\ref{claim:iii}.

Let $\asynclev= e^{\asyncexp n}$, and let $\mu\defeq \eps/2$ .   We
rewrite the above summation over $\asynclev/3n$ indices as a sum of
$\asynclev_1=
e^{n(\asyncexp-D_2-\mu)}/3n$ superblocks of $\asynclev_2= e^{n(D_2+\mu)}$
indices.   We have
\begin{multline*} 
\sum_{j=1}^{\asynclev/3n} \pr_\star(\tau_n\geq 2A_n/3,Y_{jn}^{jn+n-1}\in \cT_W(m)) \\
=\sum_{s=1}^{\asynclev_1}\sum_{j\in I_s}\pr_\star(\tau_n\geq 2A_n/3, 
Y_{jn}^{jn+n-1}\in \cT_W(m)),
\end{multline*}
where $I_s$ denotes the $s$\/th superblock of $\asynclev_2$ indices.   Applying
the union bound (in reverse), we see that
\begin{multline*}
\sum_{s=1}^{\asynclev_1}\sum_{j\in I_s}\pr_\star(\tau_n\geq 2A_n/3, Y_{jn}^{jn+n-1}\in
\cT_W(m)) \\
\geq  \sum_{s=1}^{\asynclev_1}\pr_\star\bigg(\tau_n\geq
2A_n/3,\cup_{j\in I_s} \{Y_{jn}^{jn+n-1}\in \cT_W(m)\}\big).
\end{multline*}
We now show that each term
\begin{equation}
\label{alesia}
\pr_\star\big(\tau_n\geq 2A_n/3,\cup_{j\in I_s} \{Y_{jn}^{jn+n-1}\in
\cT_W(m)\}\big)
\end{equation}
in the above summation is large, say greater than $1/2$, by showing
that each of them involves the intersection of two large probability
events.   This, together with \eqref{und}, implies that
\begin{align}\label{abor}
\delay_n&=\poly(n)\Omega(e^{n(\asyncexp -D_1-\mu)})\notag \\
&\geq \Omega(\exp(n\eps/2))
\end{align}
since $D_1\leq \asyncexp -\eps$, yielding the desired
result.\footnote{Our proof shows that for all indices $n$ for which
$D_1\leq \asyncexp-\eps$ and $D_2\leq \asyncexp-\eps$,
\eqref{abor} holds.   Therefore, if $D_1\leq \asyncexp-\eps$ and
$D_2\leq \asyncexp-\eps$ for every $n$ large enough, the reaction
delay grows exponentially with $n$, and thus the rate vanishes.   In
the case where $D_1\leq \asyncexp-\eps$ and $D_2\leq
\asyncexp-\eps$ does not hold for all $n$ large enough, but still
holds for infinitely many values of $n$, the corresponding asymptotic
rate is still zero by Definition~\ref{defcs}.}

Letting $\cE$ denote the decoding error event, we have for all $n$ large enough
\begin{align}\label{errbor}
\eps&\geq \pr_m(\cE)\notag \\
&\geq \pr_m(\cE| \nu> 2A_n/3,\tau_n\leq 2A_n/3 )\notag\\
&\hspace{1cm}\times\pr_m(\nu> 2A_n/3,\tau_n\leq 2A_n/3
)\notag \\
&\geq \frac{1}{2}\pr_m(\nu>  2A_n/3)\pr_m(\tau_n\leq 2A_n/3|\nu> 2A_n/3)\notag\\
&\geq \frac{1}{6}\pr_m(\tau_n\leq 2A_n/3|\nu> 2A_n/3).
\end{align}
The third inequality follows by noting that the event $\{\nu>
2A_n/3,\tau_n\leq 2A_n/3\}$ corresponds to the situation where the decoder
stops after observing only pure noise.   Since a codebook consists of
at least two codewords,\footnote{By assumption, see
Section~\ref{moper}.} such an event causes an error with probability
at least $1/2$ for at least one message $m$.   Thus,
inequality~\eqref{errbor} holds under the assumption that $m$
corresponds to such a message.\footnote{Regarding the fourth
inequality in \eqref{errbor}, note that $\pr_m(\nu> 2A_n/3)$ should be
lower bounded by $1/4$ instead of $1/3$ had we taken into account
discrepancies due to rounding of noninteger quantities.   As mentioned
earlier, we disregard these discrepancies as they play no role
asymptotically.}

Since the event $\{\tau_n\leq 2A_n/3 \}$ depends on the channel outputs only up to
time $2A_n/3$, we have
\begin{equation}\label{ze} 
\pr_m(\tau_n\leq 2A_n/3|\nu> 2A_n/3)=\pr_\star (\tau_n\leq 2A_n/3).
\end{equation}
Combining \eqref{ze} with \eqref{errbor} we get
\begin{equation} 
\label{abor1}
\pr_\star (\tau_n>2A_n/3)\geq 1- 6\eps.
\end{equation}
Now, because the $Y_{jn}^{jn+n-1}$, $j\in I_s$, are i.i.d.\ under $\pr_\star$,
\begin{align*} 
\pr_\star&\bigg(\cup_{j\in I_s} \{Y_{jn}^{jn+n-1}\in \cT_W(m)\}\bigg)\\
&= 1-(1-\pr_\star(
Y^n\in \cT_W(m)))^{|I_s|}.
\end{align*}
From Fact~\ref{fact:3} it follows that
\begin{equation*} 
\pr_\star(Y^{n}\in \cT_W(m))\geq \poly(n) \exp (-n D_2),
\end{equation*}
and by definition $|I_s|=e^{n(D_2+\mu)}$, so
\begin{equation} 
\pr_\star\bigg(\cup_{j\in I_s} \{Y_{jn}^{jn+n-1}\in
\cT_W(m)\}\bigg)=1-o(1) 
\quad (n\rightarrow\infty).
\label{abor2}
\end{equation}
Combining \eqref{abor1} and \eqref{abor2}, we see that each term
\eqref{alesia} involves the intersection of large probability events
for at least one message index $m$.   For such a message index, by
choosing $\eps$ sufficiently small, we see that for all
sufficiently large $n$, every single term \eqref{alesia}, $s\in
\{1,2,\ldots,A_1\}$ is bigger than $1/2$.
% Hence using \eqref{und}, if
%$D_1<\asyncexp$ and $D_2<\asyncexp$, then
%$\delay_n$ grows exponentially with $n$, implying that the rate is
%asymptotically equal to zero.   We conclude that, if the coding scheme achieves a
%strictly positive rate, then either $D_1\geq \asyncexp$, or $D_2 \geq \asyncexp$, or
%both inequalities hold.   Condition ii) follows.
\end{IEEEproof}

Finally, to establish the remaining Claim~\ref{claim:iv}, we make use of
Theorem~\ref{prop:expurgation}, whose proof is provided in
Appendix~\ref{app:expurgation}. This theorem implies that any
nontrivial codebook contains a (large) set of codewords whose rate is almost
the same as the original codebook and whose error
probability decays faster than polynomially, say as $ e^{-\sqrt{n}}$, with a
suitable decoder. Note that we don't use
the full implication of Theorem~\ref{prop:expurgation}.

\begin{IEEEproof}[Proof of Claim~\ref{claim:iv}]
The main idea behind the proof is that if Claim~\ref{claim:iv} does
not hold, the noise is likely to produce an output that is ``typical''
with a codeword before the message is even sent, which means that any
decoder must have large error probability. Although the idea is fairly simple,
it turns out that a suitable definition for ``typical'' set and its related
error probability analysis make the proof somewhat lengthy.

\iffalse
To formalize this argument, we need a definition of typical outputs that is
powerful enough
to translate into a lower bound on the probability of error for any decoder.
Since we are trying
to prove a lower bound, we first optimistically assume that the decoder is
revealed
the entire output sequence (of length $\asynclev+n-1$).   If the decoder
achieves a low
error probability, and the reaction delay $\ex(\tau-\nu)^+\leq d$ on the
original
output sequence, then it follows that the decoder can identify the sent codeword
within a window of
size essentially equal to $d$.   Roughly speaking,  if $d=ln$, then the decoder
can output a list of
$l$ blocks, one of these corresponding to the sent message.   Typicality is then
defined with respect to the maximum likelihood $l$-list decoder.   The formal
argument follows.
\fi

Proceeding formally, consider  inequality \eqref{ineqdetcor}.   This
inequality says that, with nonzero probability, the decoder makes a correct
decision and stops soon after the beginning of the information transmission
period.   This motivates the definition of a new random process, which we call
the modified output process. With a slight abuse of notation, in the remainder
of the proof we use $Y_1,Y_2,\ldots,Y_{\asynclev+n-1}$ to denote the modified
output process. The modified output process is generated as if the sent
codeword were truncated at the position $\nu+d(n)$, where $d(n)$ is defined in
\eqref{dn}. Hence, this process can be thought of as the random process
``viewed'' by the sequential
decoder.

Specifically,  the distribution of the modified output process is as follows. If
\begin{equation*}
n\geq \lfloor\delay_n(1+\eps)\rfloor,
\end{equation*}
then the $Y_i$'s for
\begin{equation*}
i\in \left\{1,\ldots,\nu-1\right\}
\cup\left\{\nu+\lfloor\delay_n(1+\eps)\rfloor,\ldots,A_n+n-1\right\}
\end{equation*}
are i.i.d.\ according to $Q_\star$, whereas the block
\begin{equation*}
Y_\nu,Y_{\nu+1},\ldots,Y_{\nu+\lfloor\delay_n(1+\eps)\rfloor-1}
\end{equation*}
is distributed according to $Q(\cdot|c^{d(n)})$, the output
distribution given that a \emph{randomly selected} codeword has been
transmitted.   Note that, in the conditioning, we use $c^{d(n)}$
instead of $c^{d(n)}(m)$ to emphasize that the output distribution is
averaged over all possible messages, i.e., by definition
\begin{equation*}
Q(y^{d(n)} |c^{d(n)})=\frac{1}{M}\sum_{m=1}^M Q(y^{d(n)}
|c^{d(n)}(m)).
\end{equation*}
Instead, if
\begin{equation*}
n<\lfloor\delay_n(1+\eps)\rfloor,
\end{equation*}
then the modified output process has the same distribution as the
original one, i.e., the $Y_i$'s for
\begin{equation*}
i\in \left\{1,\ldots,\nu-1\right\}\cup\left\{\nu+n,\ldots,A_n+n-1\right\}
\end{equation*}
are i.i.d.\ according to $Q_\star$, whereas the block
\begin{equation*}
Y_\nu,Y_{\nu+1},\ldots,Y_{\nu+n-1}
\end{equation*}
is distributed according to $Q(\cdot|c^n)$.

Consider the following augmented decoder that, in addition to
declaring a message, also outputs the time interval
\begin{equation*}
[\tau_n-\lfloor\delay_n(1+\eps)\rfloor+1,\tau_n-\lfloor\delay_n(1+\eps)\rfloor+2,\ldots,\tau_n],
\end{equation*}
of size $\lfloor\delay_n(1+\eps)\rfloor$.   A simple consequence
of the right-hand side of \eqref{ineqdetcor} being (asymptotically) bounded
away from zero is that, for $n$ large enough, if the augmented decoder
is given a modified output process instead of the original one, with
a strictly positive probability it declares the correct message, and
the time interval it outputs contains $\nu$.

Now, suppose the decoder is given the modified output process and
that it is revealed that the (possibly truncated) sent codeword was
sent in one of the
\begin{equation} 
\label{rn}
r_n=\left\lfloor \frac{(A_n+n-1)-(\nu \bmod d(n))}{d(n)}\right\rfloor
\end{equation}
consecutive blocks of duration $d(n)$, as shown in
Fig.~\ref{graphees2}.   Using this additional knowledge, the decoder
can now both declare the sent message and output a list of
\begin{equation}
\label{list}
\ell_n=\lceil\lfloor\delay_n(1+\eps)\rfloor/d(n)\rceil
\end{equation}
block positions, one of which corresponding to the sent message, with
a probability strictly away from zero for all $n$ large enough.   To do
this the decoder, at time $\tau_n$, declares the decoded message and
declares the $\ell_n$ blocks that overlap with the time indices in
\begin{equation*}
\{\tau_n-\lfloor\delay_n(1+\eps)\rfloor+1,\tau_n-\lfloor\delay_n(1+\eps)\rfloor+2,\dots,\tau_n\}.
\end{equation*}

\begin{figure}
\begin{center}
\input{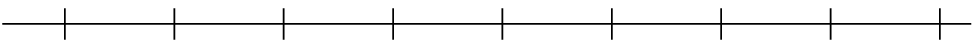}
\caption{\label{graphees2} Parsing of the entire received sequence of
size $\asynclev+n-1$ into $r_n$ blocks of length $d(n)$, one of which being
generated by the sent message, and the others being generated by
noise.}
\end{center}
\end{figure}

We now show that the above task that consists of declaring the sent
message and producing a list of $\ell_n$ blocks of size $d(n)$, one of
which being the output of the transmitted message, can be performed
only if $\asyncexp$ satisfies Claim~\ref{claim:iv}.  To that aim we
consider the performance of the (optimal) maximum likelihood decoder
that observes output sequences of maximal length
\begin{equation*}
d(n)\cdot r_n.
\end{equation*}

Given a sample $y_1,y_2,\ldots,y_{\asynclev+n-1}$ of the modified output
process, and its parsing into consecutive blocks of duration $d(n)$,
the optimal decoder outputs a list of $\ell_n$ blocks that are most
likely to occur.   More precisely, 
the maximum likelihood $\ell_n$-list decoder operates as follows. For
each message $m$, it finds a list of $\ell_n$ blocks $y^{d(n)}$ (among
all $r_n$ blocks) that maximize the ratio
\begin{equation}
\label{e0}
\frac{Q(y^{d(n)}|c^{d(n)}(m))}{Q(y^{d(n)}|\star)},
\end{equation}
and computes the sum of these ratios. The maximum likelihood
$\ell_n$-list decoder then outputs the list whose sum is maximal, and
declares the corresponding message.\footnote{To see this, consider a channel output $y^{d_n\cdot r_n}$ that is composed of
$r_n$ consecutive blocks of size $d_n$, where the $j$\/th block is generated by
codeword $c^{d(n)}$ and where all the other blocks are generated by noise. The
probability of this channel output is 
$$\pr(y^{d_n\cdot r_n}|m,j)=Q(y^{d(n)}(j)|c^{d(n)})\prod_{i\ne
j}Q_\star(y^{d(n)}(i))$$
where $y^{d(n)}(j)$, $j\in \{1,2,\ldots,r_n\}$, denotes the $j$\/th bloc of $y^{d_n\cdot r_n}$
}

The rest of the proof consists in deriving an upper bound on the
probability of correct maximum likelihood $\ell_n$-list decoding, and
show that this bound tends to zero if Claim~\ref{claim:iv} is not
satisfied.  To that aim, we first quantify the probability that the
noise distribution $Q_\star$ outputs a sequence that is typical with a
codeword, since the performance of the maximum likelihood $\ell_n$-list
decoder depends on this probability, as we show below.

By assumption, $(\cC_n',(\tau_n,\phi_n))$ achieves a probability of
error ${\eps}_n'\rightarrow0$ as $n\rightarrow\infty$ at the
asynchronism exponent $\asyncexp$.  This implies that $\cC_n'$ can
also achieve a nontrivial error probability on the synchronous channel
(i.e., with $\asynclev=1$).  Specifically, by using the same argument
as for \eqref{ineqdetcor}, we deduce that we can use $\cC_n'$ on the
synchronous channel, force decoding to happen at the fixed time
$$d(n)=\min\{n,\lfloor(1+\eps)\delay_n\rfloor\}\,,$$ where $\delay_n$
corresponds to the reaction delay obtained by
$(\cC_n',(\tau_n,\phi_n))$ in the asynchronous setting, and guarantee
a (maximum) probability of error $\eps_n''$ such that
\begin{equation*}
\eps_n'' \leq \frac{1}{1+\eps}+\eps_n'
\end{equation*}
with a suitable decoder. Since the right-hand side of the above inequality is
strictly below one for $n$ large enough, 
Theorem~\ref{prop:expurgation} with $q=1/4$ implies that the code
$\cC_n'$ has a large subcode $\tilde{\cC}_n$, i.e., of almost the same
rate with respect to $d(n)$, that, together with an appropriate
decoding function $\tilde{\phi}_n$, achieves a maximum error
probability at most equal to
\begin{equation}\label{epsn}
{\eps}_n=2(n+1)^{|\cX|\cdot|\cY|}\exp(-\sqrt{n}/(2\ln
2))
\end{equation}
for all $n$ large enough.

We now start a digression on the code
$(\tilde{\cC}_n,\tilde{\phi}_n)$ when used on channel $Q$
synchronously.  The point is to exhibit a set of ``typical output
sequences'' that cause the decoder $\tilde{\phi}_n$ to make an error
with ``large probability.'' We then move back to the asynchronous
channel $Q$ and show that when Claim~\ref{claim:iv} does not hold, the
noise distribution $Q_\star$ is likely to produce typical output
sequences, thereby inducing the maximum likelihood $\ell_n$-list decoder
into error.

Unless stated otherwise, we now consider
$(\tilde{\cC}_n,\tilde{\phi}_n)$ when used on the synchronous
channel. In particular error events are defined with respect to this
setting.

The set of typical output sequences is obtained through a few steps.
We first define the set $\cA_m$ with respect to codeword
$c^{d(n)}(m)\in \tilde{\cC}_n$ as
\begin{align} 
\cA_m = &\bigl\{ y^{d(n)}\in \cT_W(c^{d(n)}(m))\;\text{with } W\in
\cP^{\cY|\cX}\::\notag\\
&\qquad \pr(\cT_W(c^{d(n)}(m))|c^{d(n)}(m))\geq \sqrt{{\eps}_{d(n)}}
\bigr\}
\label{am}
\end{align}
where $\eps_n$ is defined in \eqref{epsn}.
%and defining the ``atypical event'' $\cE_1$ as
%$$\cE_1=\{Y_{\nu}^{\nu+d(n)-1}\notin \cA\},$$
%we have
%$$\pr_1(\cE_1)\leq  \eps_n\ln N\overset{n\to \infty}{\longrightarrow} 0.$$

Note that, by using Fact~\ref{fact:3}, it can easily be checked that
$\cA_m$ is nonempty for $n$ large enough (depending on $|\cX|$ and
$|\cY|$), which we assume throughout the argument.   For a fixed $m$,
consider the set of sequences in $\cA_m$ that maximize \eqref{e0}.
These sequences form a set $\cT_{\bar{Q}}(c^{d(n)}(m))$, for some
$\bar{Q}\in \cP^{\cY|\cX}_n$.   It follows that for every message index
$m$ for which $c^{d(n)}(m)\in \tilde{\cC}_n$, we have
\begin{align}
&{\eps}_{d(n)}\geq \pr_m(\E) \notag\\
 &\geq \pr_m(\cE|\{Y_\nu^{\nu+d(n)-1}\in
 \cT_{\bar{Q}}(c^{d(n)}(m))\})\nonumber\\
 &\times\pr_m(\{Y_\nu^{\nu+d(n)-1}\in
 \cT_{\bar{Q}}(c^{d(n)}(m))\})\notag \\
&\geq \pr_m(\cE|\{Y_\nu^{\nu+d(n)-1}\in
 \cT_{\bar{Q}}(c^{d(n)}(m))\})\sqrt{{\eps}_{d(n)}} \notag \\
&\geq \pr_m(\cE| \{Y_\nu^{\nu+d(n)-1}\in \cB_m\})\times\nonumber \\
&\pr_m(\{Y_\nu^{\nu+d(n)-1}\in
\cB_m\}|\{Y_\nu^{\nu+d(n)-1}\in
\cT_{\bar{Q}}(c^{d(n)}(m))\})\notag\\
&\times\sqrt{{\eps}_{d(n)}}\notag  \\
&\geq \frac{1}{2}\times \notag\\
&\pr_m(\{Y_\nu^{\nu+d(n)-1}\in \cB_m\}|\{Y_\nu^{\nu+d(n)-1}\in
 \cT_{\bar{Q}}(c^{d(n)}(m))\})\notag \\
 &\times\sqrt{{\eps}_{d(n)}}
\label{seteq}
\end{align}
where for the third inequality we used the definition of $\bar{Q}$;
where on the right-hand side of the fourth inequality we defined the set
\begin{align*}
 &\cB_m\defeq \notag\\
 &\bigl\{y^{d(n)}\in
 \cT_{\bar{Q}}(c^{d(n)}(m))\cap\left(\cup_{m'\ne m}\cT_{\bar{Q}}(c^{d(n)}(m'))\right)
\bigr\};
\end{align*}
and where the fifth inequality follows from this
definition.\footnote{Note that, given that message $m$ is sent, if
  the channel produces a sequence in $\cB_m$ at its output, the
  (standard) optimal maximum likelihood decoder makes an error with
  probability at least half. Hence the decoding rule
  $\tilde{\phi}_n$ also makes an error with probability at least
  half.}

From \eqref{seteq} we get
\begin{multline}
\label{mago}
\pr_m(\{Y_\nu^{\nu+d(n)-1}\in \cB_m\}|\{Y_\nu^{\nu+d(n)-1}\in
 \cT_{\bar{Q}}(c^{d(n)}(m))\})\\
\leq 2\sqrt{{\eps}_{d(n)}}.
\end{multline}
Therefore, by defining $\tilde{\cB}_m$ as 
\begin{align*}
 \tilde{\cB}_m \defeq  \cT_{\bar{Q}}(c^{d(n)}(m))\backslash \cB_m
\end{align*}
the complement of $\cB_m$ in
$\cT_{\bar{Q}}(c^{d(n)}(m))$, it follows from \eqref{mago} that
\begin{equation*}
| \tilde{\cB}_m|>
(1-2\sqrt{{\eps}_{d(n)}})|\cT_{\bar{Q}}(c^{d(n)}(m))|,
\end{equation*}
since under $\pr_m$ all the sequences in $\cT_{\bar{Q}}(c^{d(n)})(m)$
are equiprobable.

The set $\cup_{m'\ne m}^M\tilde{\cB}_{m'}$ is the sought set of
``typical output sequences'' that causes the decoder make an error with
``high probability'' conditioned on the sending of message $m$ and
conditioned on the channel outputting a sequence in
$\cT_{\bar{Q}}(c^{d(n)}(m))$.   This ends our digression on
$(\tilde{\cC}_n,\tilde{\phi}_n)$.

We now compute a lower bound on the probability under $Q_\star$ of
producing a sequence in $\cup_{m=2}^M\tilde{\cB}_m$.   Because the sets
$\{\tilde{\cB}_m\}$ are disjoint, we deduce that
\begin{align}
 |\cup_{m=2}^M\tilde{\cB}_m|&\geq
 (1-2\sqrt{{\eps}_n})\sum_{m=2}^M |\cT_{\bar{Q}}(c^{d(n)}(m))|\notag\\
&\geq \frac{(1-2\sqrt{{\eps}_n})}{(d(n)+1)^{|\cX|\cdot|\cY|}}(M-1)
e^{d(n)H(\bar{Q}|P^1_n)}\notag \\
&\geq \frac{1}{(4n)^{|\cX||\cY|}}e^{d(n)(H(\bar{Q}|P_1^n)+\ln M /d(n))}
\label{di}
\end{align}
for all $n$ large enough. For the second inequality we used \cite[Lemma~2.5,
p.~31]{CK}. For the third inequality we used the fact that $d(n)\leq n$,
$M\geq 2$, $(1-2\sqrt{{\eps}_{d(n)}})\geq 1/2$ for $n$ large
enough,\footnote{Note that $d(n)\overset{n\to \infty}{\longrightarrow } \infty$
since the coding scheme under consideration achieves a
strictly positive rate. } and that, without loss of
generality, we may assume that $|\cX|\cdot|\cY|\geq 2$
since the synchronous capacity $C$ is
non-zero---as we assume throughout the paper. Hence we get
\begin{align*}
Q_\star(\cup_{m=2}^M\tilde{\cB}_m)&=\sum_{y^{d(n)}\in
\cup_{m=2}^M\tilde{\cB}_m}Q_\star(y^{d(n)})\notag\\
&\geq |\cup_{m=2}^M\tilde{\cB}_m|\min_{y^{d(n)}\in
\cup_{m=2}^M\tilde{\cB}_m}Q_\star(y^{d(n)})\notag\\
&\geq \frac{1}{(4n)^{|\cX||\cY|}}
e^{d(n)(H(\bar{Q}|P_n^1)+(\ln M) /d(n))}\notag\\
&\qquad{}\times
e^{-d(n)(D((P_n^1\bar{Q})_\cY\|Q_\star)+H((P_n^1\bar{Q})_\cY))}
\end{align*}
for all $n$ large enough, where for the second inequality we used
\eqref{di} and \cite[Theorem~11.1.2, p.~349]{CT}.   Letting
\begin{equation*}
e_n\defeq  \ln
I(P_n^1\bar{Q})-(\ln M)/d(n)+D((P_n^1\bar{Q})_\cY\|Q_\star),
\end{equation*}
we thus have
\begin{equation} 
\label{probbmtilde}
Q_\star(\cup_{m=2}^M\tilde{\cB}_m)
\geq \frac{1}{(4n)^{|\cX||\cY|}}e^{-e_n \cdot d(n)}
\end{equation}
for $n$ large enough.

Using \eqref{probbmtilde}, we now prove Claim~\ref{claim:iv} by
contradiction.  Specifically, assuming that
\begin{equation}
\label{hypalpha}
\asyncexp > \frac{d(n)}{n}e_n+\eps/2\quad \text{for infinitely many indices
$n$},
\end{equation}
we prove that, given message $m=1$ is sent, the probability of error
of the maximum likelihood $\ell_n$-list decoder does not converge to
zero.  As final step, we prove that the opposite of \eqref{hypalpha}
implies Claim~\ref{claim:iv}.

Define the events
\begin{align*}
\cE_1&=\{Y_{\nu}^{\nu+n-1}\notin \cA_1\},\\
\cE_2&=\{Z\leq  \frac{1}{2}\frac{1}{(4n)^{2|\cX||\cY|}}e^{\asyncexp n-e_n\cdot
d(n)}\},
\end{align*}
where $\cA_1$ is defined in \eqref{am}, and where $Z$ denotes the
random variable that counts the number of blocks generated by
$Q_\star$ that are in $\cup_{m=2}^M\tilde{\cB}_m$.   Define also the
complement set
\begin{equation*}
\cE_3\defeq  (\cE_1\cup \cE_2)^\comp.
\end{equation*}
The probability that the maximum likelihood $\ell_n$-list decoder makes a
\emph{correct} decision given that message $m=1$ is sent is upper
bounded as
\begin{align}
\pr_1(\cE^\comp)&=\sum_{i=1}^3 \pr_1(\cE^\comp|\cE_i)\pr_1(\cE_i)\notag \\
&\leq \pr_1(\cE_1)+\pr_1(\cE_2)+\pr_1(\cE^\comp|\cE_3).
\label{tsum}
\end{align}
From the definition of $\cA_1$, we have
\begin{equation} 
\label{e1}
\pr_1(\cE_1)=o(1) \quad (n\rightarrow\infty).
\end{equation}
Now for $\pr_1(\cE_2)$.   There are $r_n-1$ blocks independently
generated by $Q_\star$ ($r_n$ is defined in \eqref{rn}).   Each of
these blocks has a probability at least equal to the right-hand side of
\eqref{probbmtilde} to fall within $\ds \cup_{m=2}^M\tilde{\cB}_m$.
Hence, using \eqref{probbmtilde} we get
\begin{align}
\ex_1{Z} &\geq (r_n-1) \frac{1}{(4n)^{|\cX||\cY|}}e^{-e_n d(n) }\notag \\
&\geq \frac{1}{(4n)^{2|\cX||\cY|}}e^{\asyncexp n -e_n d(n) }
\label{lboundz} 
\end{align}
since $r_n\geq e^{\asyncexp n}/n$.   Therefore,
\begin{align}
\label{e2}
\pr_1\big(\cE_2\big)&\leq \pr_1(Z\leq (\ex_1 Z)/2)\notag \\
&\leq \frac{4}{\ex_1 Z}\notag \\
& \leq \poly (n) e^{-\asyncexp n +e_n d(n)}
\end{align}
where the first inequality follows from \eqref{lboundz} and the
definition of $\cE_2$; where for the second inequality we used
Chebyshev's inequality and the fact that the variance of a binomial is
upper bounded by its mean; and where for the third inequality we used
\eqref{lboundz}.

%Using \eqref{hypalpha} and the fact that $d(n)\leq n$ we conclude from
%\eqref{lboundz} that
%\begin{align}\pr_1\big(\cE_2\big)=o(1) \quad (n\rightarrow\infty).
%\end{align}

Finally for $\pr_1\big(\cE^\comp|\cE_3\big)$.   Given $\cE_3$, the decoder
sees at least
\begin{equation*}
\frac{1}{2}\frac{1}{(4n)^{2|\cX||\cY|}}e^{\asyncexp n-e_n\cdot d(n)}
\end{equation*}
time slots whose corresponding ratios \eqref{e0} are at least as large
as the one induced by the correct block $Y_{\nu}^{\nu+d(n)-1}$.
Hence, given $\cE_3$, the decoder produces a list of $\ell_n$ block
positions, one of which corresponds to the sent message, with
probability at most
\begin{align}\label{lzbound}
\pr_1(\cE^\comp|\cE_3)
&\leq \ell_n\left(\frac{1}{2}\frac{1}{(4n)^{2|\cX||\cY|}}
  e^{\asyncexp n-e_n\cdot d(n)}\right)^{-1}\notag\\
&=\poly(n)e^{-\asyncexp n+e_n\cdot d(n)},
\end{align}
where the first inequality follows from union bound, and where for the
equality we used the fact that finite rate implies
$\ell_n=\poly(n)$.\footnote{This follows from the definition of rate
$R=\ln M/\ex(\tau-\nu)^+$, the fact that $\ln M/n\leq C$ for reliable
communication, and the definition of $\ell_n$ \eqref{list}.   }

From \eqref{tsum}, \eqref{e1}, \eqref{e2}, and \eqref{lzbound}, the
probability that the maximum likelihood $\ell_n$-list decoder makes a
correct decision, $\pr_1\left(\cE^\comp\right)$, is arbitrarily small for
infinitely many indices $n$ whenever \eqref{hypalpha} holds.
Therefore to achieve vanishing error probability we must have, for
all $n$ large enough,
\begin{align}
\label{alcond}
\asyncexp \leq  \frac{d(n)}{n} &\left(
I(P_n^1\bar{Q})-(\ln M)/d(n)+D((P_n^1\bar{Q})_\cY\|Q_\star)\right)\notag\\
&\hspace{1cm}+\eps/2.
\end{align}
We now show, via a continuity argument, that the above condition
implies Claim~\ref{claim:iv}.  Recall that $\bar{Q}\in \cP^{\cY|\cX}$,
defined just after \eqref{am}, depends on $n$ and has the property
\begin{equation}
\label{b1}
\pr(\cT_{\bar{Q}}(c^{d(n)}(m)|c^{d(n)}(m)))\geq \sqrt{{\eps}_{d(n)}}.
\end{equation}
Now, from Fact~\ref{fact:3} we also have the upper bound
\begin{equation}
\label{b2}
\pr(\cT_{\bar{Q}}(c^{d(n)}(m)|c^{d(n)}(m)))\leq e^{-d(n)D(\bar{Q}\|Q|P_n^1)}.
\end{equation}
Since $\sqrt{{\eps}_{d(n)}}=\Omega(e^{-\sqrt{_{d(n)}}})$, from
\eqref{b1} and \eqref{b2} we get
\begin{equation*}
D(\bar{Q}\|Q|P_1^n)\rightarrow0\quad\text{as}\quad n\rightarrow\infty,
\end{equation*}
and therefore
\begin{equation*}
\|P_n^1\bar{Q}-P_n^1Q\|\rightarrow0\quad\text{as}\quad n\rightarrow\infty,
\end{equation*}
where $\|\cdot\|$ denotes the $L_1$ norm.  Hence, by continuity of the
divergence, condition \eqref{alcond} gives, for all $n$ large enough,
\begin{align} 
\label{vcond}
\asyncexp \leq  \frac{d(n)}{n} &\left( I(P_n^1Q)-(\ln M)/d(n)+D((P_n^1{Q})_\cY\|Q_\star) \right)\nonumber \\
&\hspace{1cm}   + \eps 
\end{align}
which yields Claim~\ref{claim:iv}.
\end{IEEEproof}

%Now, if $\sqrt{\eps_n}\ne\Omega(e^{-\sqrt{n}})$, one just repeats the
%entire argument above starting with the definition of the set
%\eqref{am}, by replacing
%$\sqrt{\eps_n}$ by $\max\{\eps_n,e^{-\sqrt{n}}\}$.   One then concludes
%that \eqref{vcond} also holds if $\sqrt{\eps_n}\ne\Omega(e^{-\sqrt{n}})$.

\subsection{Proof of Corollary~\ref{disc}}
By assumption $\thres$ is nonzero since divergence is always non-negative.  This implies that the synchronous
capacity is nonzero by the last claim of Theorem~\ref{ulimit}. This, in
turn, implies that $(R,\asyncexp)$ is achievable for some sufficiently small $R>0$ and
$\asyncexp>0$  by \cite[Corollary~1]{TCW}.

Using Theorem~\ref{convs}, 
\begin{align}\label{kiko}
\alpha\leq \alpha(R)\leq \max_\cS \alpha_2
\end{align}
where $\alpha_2$ is given by expression
\eqref{alpha2}. In this expression, by
letting $W=Q_\star$ in the minimization, we deduce that
$\alpha_2\leq D(Q_\star||Q|P_2)$, and therefore
\begin{align*}\max_\cS \alpha_2&\leq \max_\cS
D(Q_\star||Q|P_2)\\
&= \max_{P_2}D(Q_\star||Q|P_2)\\
&= \max_{x\in \cX} \sum_{y\in \cY}
Q_\star(y)\ln\frac{Q_\star(y)}{Q(y|x)}\\
&= \max_{x\in\cX}D(Q_\star\|Q(\cdot|x)),
\end{align*}
and from \eqref{kiko} we get
$$\alpha\leq
\max_{x\in\cX}D(Q_\star\|Q(\cdot|x))\,.$$
Since, by assumption,
\begin{equation*} 
\thres>\max_{x\in \cX} D(Q_\star\|Q(\cdot|x)),
\end{equation*}
and since $\thres=\asyncexp(R=0)$ by Theorem~\ref{ulimit},
it follows that $\asyncexp(R)$ is discontinuous at $R=0$.\hfill\IEEEQEDclosed

\subsection{Proof of Theorem~\ref{alphinfin}}

We first exhibit a coding scheme that achieves any $(R,\asyncexp)$ with $R\leq
C$ and 
\begin{equation*}
\asyncexp \leq \max_{P\in \cP^\cX} \min_{W\in \cP^{\cY|\cX}}
\max\{D(W\|Q|P),D(W\|Q_\star|P)\}.
\end{equation*}

All codewords start with a common preamble that is composed of
$(\ln(n))^2$ repetitions of a symbol $x$ such that $D(Q(\cdot |x) \|
Q_\star) = \infty$ (such a symbol exists since
$\thres=\infty$).   The next $(\ln(n))^3$ symbols of each
codeword are drawn from a code that achieves a rate equal to
$R-\eps$ on the synchronous channel.   Finally, all the codewords end with
a common large suffix $s^l$ of size $l=n-(\ln(n))^2-(\ln(n))^3$ that has an
empirical type $P$ such that, for all $W\in \cP^{\cY|\cX}$, at least one of the
following two inequalities holds:
\begin{align*}
D(W\|Q|P)&\geq \asyncexp\\
D(W\|Q_\star|P)&\geq \asyncexp.
\end{align*}
The receiver runs two sequential decoders in parallel, and makes a
decision whenever one of the two decoder declares a message.   If the
two decoders declare different messages at the same time, the receiver
declares one of the messages at random.

The first decoder tries to identify the sent message by first locating
the preamble.   At time $t$ it checks if the channel output $y_t$ can
be generated by $x$ but cannot be generated by noise, i.e., if
\begin{equation} 
\label{c1}
Q(y_t|x)>0\quad \text{and}\quad Q(y_t|\star)=0.
\end{equation}
If condition \eqref{c1} does not hold, the decoder moves one-step ahead
and checks condition \eqref{c1} at time $t+1$.   If condition
\eqref{c1} does hold, the decoder marks the current time as the
beginning of the ``decoding window'' and proceeds to the second step.
The second step consists in exactly locating and identifying the sent
codeword.   Once the beginning of the decoding window has been marked,
the decoder makes a decision the first time it observes $(\ln n)^3$
symbols that are typical with one of the codewords.   If no such time
is found within $(\ln(n))^2+(\ln(n))^3$ time steps from the time the
decoding window has been marked, the decoder declares a random
message.

The purpose of the second decoder is to control the average reaction
delay by stopping the decoding process in the rare event when the
first decoder misses the codeword.   Specifically, the second ``decoder''
is only a stopping rule based on the suffix $s^l$.   At each time $t$
the second decoder checks whether $D(\hat{P}_{Y^{t}_{t-l+1}}\|Q|P) <
\asyncexp$.   If so, the decoder stops and declares a random message.   If
not, the decoder moves one step ahead.

The arguments for proving that the coding scheme described above
achieves $(R,\alpha)$ provided 
\begin{equation}
\label{condalpinfi}
\asyncexp \leq \max_P \min_W \max\{D(W\|Q|P),D(W\|Q_\star|P)\}
\end{equation}
closely parallel those used to prove
Theorem~\ref{ach}, and are therefore omitted.\footnote{In particular,
note that the first decoder never stops before time $\nu$.}

The converse is the second part of Theorem~\ref{convs}. \hfill\IEEEQEDclosed

\subsection{Proof of Theorem~\ref{thm2}}
\subsubsection{Lower bound}
To establish the lower bound in Theorem~\ref{thm2}, we exhibit a training based
scheme with preamble size $\eta n$ with
\begin{equation}\label{eta}
\eta =(1-R/C),
\end{equation}
and that achieves any rate asynchronism pair $(R,\alpha)$ such that 
\begin{align}
\label{alr}
\asyncexp &\leq m_1\left(1-\frac{R}{C}\right)\qquad R\in (0,C]
\end{align}
where
$$m_1\defeq \max_{P\in \cP^\cX}\min_{W\in
\cP^{\cY|\cX}}\max\{D(W\|Q|P), D(W\|Q_\star|P)\}.$$

Fix $R\in (0,C]$ and let $\alpha$
satisfy \eqref{alr}.  Each codeword starts with a
common preamble of size $\eta n$ where $\eta$ is given
by \eqref{eta} and whose empirical distribution is equal to\footnote{$P_p$ need not be a valid
type for finite values of $n$, but this small discrepancy plays no role
asymptotically since $P_p$ can be approximated arbitrarily well with types of
order sufficiently large.}
\begin{align*}
&P_p\defeq \notag \\
&\arg \max_{P\in \cP^\cX}\big(\min_{W\in
\cP^{\cY|\cX}}\max\{D(W\|Q|P), D(W\|Q_\star|P)\}\big).
\end{align*}
The remaining $(1-\eta) n$ symbols of each codeword are i.i.d.\
generated according to a distribution $P$ that almost achieves
capacity of the synchronous channel, i.e., such that $I(PQ)=C-\eps$ for some small
$\eps>0$.

Note that by \eqref{alr}  and \eqref{eta}, $\alpha$ is such that for any $W\in \cP^{\cY|\cX}$ at least one of the following two
inequalities holds:
\begin{align} D(W||Q|P_p)&\geq \alpha/\eta\notag\\
 D(W||Q_\star|P_p)&\geq \alpha/\eta\,. \label{condeux}
 \end{align}

The preamble detection rule is to stop the first time when last
$\eta n$ output symbols $Y_{t-\eta n+1}^t$ induce an empirical conditional
probability $\hat{P}_{Y_{t-\eta n+1}^t|x^{\eta n}}$ such that
\begin{align}\label{divru}
D(\hat{P}_{Y_{t-\eta n+1}^t|x^{\eta n}}||Q|P_p)\leq D(\hat{P}_{Y_{t-\eta n+1}^t|x^{\eta n}}||Q_\star|P_p)
\end{align}
where $x^{\eta n}$ is the preamble. 

   When the preamble is
located, the decoder makes a decision on the basis of the upcoming
$(1-\eta)n$ output symbols using maximum likelihood decoding. If no preamble has
been located by time $A_n+n-1$, the decoder declares a message at random.

We compute the reaction delay and the error probability. 
For notational convenience, instead of the decoding time, we consider
the time $\tau_n$ that the decoder detects the preamble, i.e., the first time $t$
such that \eqref{divru} holds. The actual
decoding time occurs $(1-\eta)n$ time instants after the preamble has
been detected, i.e., at time $\tau_n+(1-\eta)n$.  

For the reaction delay we have
\begin{align}\label{bnz}
\ex(\tau_n-\nu)^+&=\ex_1(\tau_n-\nu)^+\notag\\
&= \ex_1[(\tau_n-\nu)^+\openone(\tau_n\geq \nu+\eta n)]\notag\\
&\quad{}+\ex_1[(\tau_n-\nu)^+\openone(\tau_n\leq \nu+\eta n-1)] \notag \\
&\leq (A_n+n-1) \pr_1(\tau_n\geq \nu+ \eta n)+\eta n
\end{align}
where, as usual, the subscript $1$ in $\ex_1$ and $\pr_1$ indicates
conditioning on the event that message $m=1$ is sent. 
A similar computation as in \eqref{ntype} yields
\begin{align}
\label{ntype22}
&\pr_1(\tau_n\geq \nu+ \eta n)\nonumber \\
&\leq \pr_1(D(\hat{P}_{Y^{\nu+\eta n-1}_\nu|x^{\eta
n}}||Q|P_p) \geq
\alpha/\eta) \notag \\
&\leq \sum_{W\in \cP^{\cY|\cX}_n:\; D(W||Q|P_p) 
\geq \alpha/\eta }e^{-\eta nD(W\|Q|P_p)}\notag
\\
&\leq \poly(n)e^{- n\alpha}\,.
\end{align}
The first inequality follows from the fact that event $\{\tau_n\geq \nu +n\}$ is
included into event 
$$\{D(\hat{P}_{Y^{\nu+\eta n-1}_\nu|x^{\eta
n}}||Q|P_p) > D(\hat{P}_{Y^{\nu+\eta n-1}_\nu|x^{\eta
n}}||Q_\star|P_p))\}$$
which, in turn, is included into event
$$\{D(\hat{P}_{Y^{\nu+\eta n-1}_\nu|x^{\eta
n}}||Q|P_p) \geq
\alpha/\eta\}$$
because of \eqref{condeux}. The second inequality follows from
Fact~\ref{fact:2}.
Hence, from \eqref{bnz} and \eqref{ntype22} 
\begin{align}
\ex(\tau_n-\nu)^+\leq \eta n+o(1)
\end{align}
whenever $A_n=e^{n(\alpha-\eps)}$, $\eps>0$. Since the actual decoding
time occurs $(1-\eta)n$ time instants after $\tau_n$, where $\eta=(1-R/C)$, and that
the code used to transmit information achieves the capacity of the synchronous
channel,  the above strategy operates at
rate $R$.

To show that the above strategy achieves vanishing error probability, one uses arguments similar to those used
to prove Theorem~\ref{ach} (see from paragraph after \eqref{ntype2} onwards), so the proof is omitted. 
There is one little caveat in the analysis that concerns the event
when the preamble is located somewhat earlier than its actual timing,
i.e., when the decoder locates the preamble over a time period
$[t-\eta n+1,\ldots,t]$ with $\nu\leq t\leq \nu+\eta n-2$.   One way to
make the probability of this event vanish as $n\rightarrow\infty$, is
to have the preamble have a ``sufficiently large'' Hamming distance with
any of its shifts.   To guarantee this, one just needs to modify the
original preamble in a few (say, $\log n$) positions.   This modifies
the preamble type negligibly.   For a detailed
discussion on how to make this modification, we refer the reader to
\cite{CTW}, where the problem is discussed in the context of sequential
frame synchronization.

Each instance of the above random coding strategy satisfies the conditions of
Definition~\ref{tbs}; there is a common preamble of size $\eta n$ and the
decoder decides to stop at any particular time $t$ based on
$Y_{t-n+1}^{t-n+\eta n}$. We now show that there exists a particular
instance yielding the desired  rate and error probability.

First note that the above rate analysis only depends on the preamble, and not on the codebook that follows the preamble. Hence, because the error probability,
averaged over codebooks and messages, vanishes, we deduce that there exists at least one
codebook that achieves rate $R$ and whose average over messages error
probability tends to zero.

From this code, we remove codewords with poor
error probability, say whose error probabilities are at least twice the
average error probability. The resulting expurgated code has a
rate that tends to $R$ and a vanishing maximum error
probability.

\subsubsection{Upper bound}
To establish the upper bound it suffices to show that for training based schemes $(R,\alpha)$ with $R>0$ must satisfy
\begin{align}
\alpha\leq m_2\left(1-\frac{R}{C}\right)\,.
\label{ltbw}
\end{align}
The upper bound in Theorem~\ref{thm2} then follows from
\eqref{ltbw} and the general upper bound derived in
Theorem~\ref{convs}.

The upper bound \eqref{ltbw} follows from the following lemma:
\begin{lem}\label{thm1}
A rate $R>0$ coding scheme whose decoder operates according to a sliding window stopping rule with window
size $\eta n$ cannot achieve an asynchronism exponent larger than $\eta m_2$.\end{lem}
Lemma~\ref{thm1} says that any coding scheme with a limited
memory stopping rule capable of processing only $\eta n$ symbols at a time
achieves an asynchronism exponent at most $O(\eta)$, unless $R=0$ or if the channel is
degenerate, i.e., $\thres=m_2=\infty$, in which case Lemma~\ref{thm1} is trivial
and we have the asynchronous capacity
expression given by Theorem~\ref{alphinfin}. 

To deduce \eqref{ltbw} from Lemma~\ref{thm1}, consider a training-based scheme which achieves a delay
$\Delta$ with a non-trivial error probability (i.e., bounded away from $0$). Because the preamble
conveys no information, the rate is at most
$$C\frac{\min\{\delay,n\}-\eta n}{\delay} \leq C(1-\eta)$$
by the channel coding
theorem for a synchronous channel. Hence, for a rate $R>0$ training-based scheme
the training fraction $\eta$ is upper bounded as
$$\eta\leq 1-\frac{R}{C}\,.$$
This implies \eqref{ltbw} by Lemma~\ref{thm1}.\hfill\IEEEQEDclosed

\begin{IEEEproof}[Proof of Lemma~\ref{thm1}]The lemma holds trivially if $m_2=\infty$. We thus assume that $m_2<\infty$.
Consider a training-based scheme $\{(\cC_n,(\tau_n,\phi_n))\}_{n\geq
1}$ in the sense of Definition~\ref{tbs}. For notational convenience, we consider $\tau_n$ to be the time when
the decoder detects the preamble. The actual decoding time (in the sense of
Definition~\ref{tbs} part~\ref{tba-iii}) occurs $(1-\eta)n$ times instants after
the preamble has been detected,
i.e., at time $\tau_n+(1-\eta)n$. This allows us
to write $\tau_n$ as 
$$\tau_n=\inf\{t\geq  1: \: S_t=1\},$$
where $$S_t=S_t(Y^{t}_{t-\eta n+1})\quad 1\leq t\leq A_n+n-1,$$ referred to as the ``stopping rule at
time $t$,'' is a binary random variable such that $\{S_t=1\}$ represents the set of output sequences $y_{t-\eta n+1}^t$ which make $\tau_n$ stop at
time $t$, assuming that $\tau_n$ hasn't stopped before time $t$.

%, i.e., by definition of $S_t$
%$$\pr(S_t=1)=\pr(\tau_n=t|\tau_n> t-1)\,,$$
%$$ \pr(S_t=0)=\pr(\tau_n>t|\tau_n> t-1)\,.$$

Now, every sequence
$y^{\eta n}\in \cY^{\eta n}$ satisfies
$$Q_\star(y^{\eta n}) \geq e^{-m_2\eta n }.$$
Therefore, any deterministic stopping rule stops at any particular
time either with probability zero or with probability at least $e^{-m_2 \eta n}$, i.e., for all $t$, either the
stopping rule $S_t$ satisfies $\pr(S_t=1)\geq e^{-m_2\eta n }$ or it is trivial in the sense that $\pr(S_t=1)=0$. 
For now, we assume that the stopping rule is deterministic; the randomized case follows easily as we describe
at the end of the proof.

Let $\cal{S}$ denote the subset of indices $t\in \{1,2,\ldots,A_n/4\}$ such that $S_t$ is non-trivial, and let
$\bar{\cal{S}}_k$ denote the subset of indices in $\cal{S}$ that are congruent to $k$ mod $\eta n$, i.e.,
$$\bar{\cal{S}}_k=\{t:t\in {\cal{S}}, \: t=j\cdot \eta n +k,\:j= 0,1,\ldots\}\,.$$
Note that for each $k$, the set of stopping rules $S_t$, $t\in
{\cal{\bar{S}}}_k$ are independent since $S_t$ depends
only on $Y_{t-\eta n+1}^t$.

 By repeating the same argument as in \eqref{errbor}-\eqref{ze}, for any
 $\veps>0$, for all $n$ large enough and any message index $m$ the error probability $\pr_m(\cE)$
 satisfies 
\begin{align}\label{errbor11}
\eps&\geq \pr_m(\cE)\notag \\
&\geq\frac{1}{4} \pr_\star (\tau_n\leq A_n/2).
\end{align}
Since $\veps>0$ is arbitrary, we deduce
 \begin{align}\label{stopstar}
 \pr_\star(\tau_n\geq A_n/2)\geq 1/2
 \end{align}
 i.e., a coding scheme achieves a vanishing error probability only if the
 probability  of stopping after time $A_n/2$ is at least $0.5$ when the channel input is
all $\star$'s. Thus, assuming that our coding scheme achieves
vanishing error probability, we have $$|{\cal{S}}|<\eta n e^{m_2 \eta n}\,.$$ To see
this, note that if $|{\cal{S}}|\geq \eta n e^{m \eta n}$,  then there exists a
value $k^*$
such that $|{\cal{\bar{S}}}_{k^*}|\geq  e^{m_2 \eta n}$, and hence
\begin{align*}
\pr_\star(\tau_n\geq A_n/2)&\leq \pr_\star(S_t=0, \: t\in {\cal{S}})\\
&\leq \pr_\star(S_t=0, \: t\in {\cal{\bar{S}}}_{k^*})\\
&= (1- e^{- m_2\eta n})^{|{\cal{\bar{S}}}_{k^*}|}\\
&\leq (1- e^{- m_2\eta n})^{e^{m_2 \eta n}}\,.
\end{align*}
Since the above last term tends to $1/e<1/2$ for $n$ large enough,
$\pr_\star(\tau_n\geq A_n/2)<1/2$ for $n$ large enough, which is in conflict with the assumption that the coding scheme achieves
vanishing error probability. 

The fact that $|{\cal{S}}|<\eta n e^{m_2 \eta n}$ implies, as we shall prove
later, that 
\begin{align}\label{kekey}
\pr(\tau_n \geq A_n/2|\nu\leq A_n/4) \geq \frac{1}{2}\left(1-\frac{8\eta^2 n^2
e^{m_2\eta
n}}{A_n}\right).
\end{align}
Hence,  
\begin{align}\label{del}
\ex(\tau_n-\nu)^+&\geq \ex((\tau_n-\nu)^+ | \tau_n\geq A_n/2, \nu\leq A_n/4)\notag\\
&\hspace{1cm}\times \pr(\tau_n\geq A_n/2, \nu\leq A_n/4)\nonumber\\
&\geq \frac{A_n}{16}\pr(\tau_n\geq A_n/2|\nu\leq A_n/4)\nonumber \\
&\geq \frac{A_n}{32}\left(1-\frac{8\eta n^2 e^{m_2\eta n}}{A_n}\right)\,.
\end{align}
where for the second inequality we used the fact that $\nu$ is uniformly
distributed, and where the third inequality holds by \eqref{kekey}.
Letting $A_n=e^{\alpha n}$, from \eqref{del} we deduce that if $\alpha > m\eta$,
then $\ex(\tau_n-\nu)^+$ grows exponentially with $n$, implying that the rate is
asymptotically zero.\footnote{Any coding scheme that achieves vanishing error
probability cannot have $\ln M$ grow faster than linearly with $n$, because of
the limitation imposed by the capacity of the synchronous channel. Hence, if
$\ex(\tau_n-\nu)^+$ grows exponentially with $n$, the rate goes to zero
exponentially with $n$.} 
Hence a sliding window stopping rule which operates on a window of size $\eta n$ cannot
accommodate a positive rate while achieving an asynchronism exponent larger than
$\eta m$. This establishes the desired result.

We now show~\eqref{kekey}. Let $\cal{N}$ be the subset of indices in
$\{1,2,\ldots,A_n/4\}$ with the following property. For any
$t\in \cal{N}$, the
$2 n$ indices  $\{t,t+1,\ldots,t+2 n-1\}$ do not belong to $\cal{S}$, i.e.,
all $2n$ of the associated stopping rules are trivial. Then we have
\begin{align}
\pr(\tau_n\geq A_n/2|\nu \leq A_n/4)&\geq \pr(\tau_n\geq A_n/2|\nu \in
{\cal{N}})\notag \\
&\hspace{.3cm}\times \pr(\nu
\in {\cal{N}}|\nu\leq A_n/4)\nonumber\\
&\hspace{-2cm}= \pr(\tau_n\geq A_n/2|\nu \in
{\cal{N}})\frac{|{\cal{N}}|}{A_n/4}\label{forgot}
\end{align}
since $\nu$ is uniformly distributed.
Using that $|{\cal{S}}|<\eta n e^{m_2\eta n}$, 
$$|{\cal{N}}|\geq (A_n/4-2\eta n^2 e^{m_2\eta n}),$$
hence from \eqref{forgot}
\begin{align}\label{elis}
\pr(\tau_n&\geq A_n/2|\nu \leq A_n/4)\nonumber \\
&\geq \pr(\tau_n\geq A_n/2|\nu \in {\cal{N}})\left(1-\frac{8\eta n^2 e^{m_2\eta
n}}{A_n}\right)\,.
\end{align}
Now, when $\nu \in {\cal{N}}$, all stopping times that could
potentially depend on the transmitted codeword 
symbols are actually trivial, so the event $\{\tau_n\geq A_n/2\}$ is independent
of the symbols sent at times $\nu, \nu+1, \ldots, \nu+N-1$. Therefore,
\begin{align}
\pr(\tau_n\geq A_n/2|\nu \in {\cal{N}})&=\pr_\star(\tau_n\geq A_n/2). \label{gew}
\end{align}
Combining \eqref{gew} with \eqref{elis} gives the desired claim \eqref{kekey}.

Finally, to see that randomized stopping rules also cannot achieve asynchronism
exponents larger than $\eta m$, note that a randomized stopping rule can be 
viewed as simply a probability
distribution over deterministic stopping rules. The previous analysis
shows that for any deterministic stopping rule, and any asynchronism exponent
larger than $\eta m$, either the probability of error is large (e.g., at least $1/8$),
or the expected delay is exponential in $n$. Therefore, the same holds for randomized
stopping rules. 
\end{IEEEproof}

\subsection{Comments on Error Criteria}

We end this section by commenting on maximum versus average rate/error
probability criteria.   The results in this paper consider the rate
defined with respect to maximum (over messages) reaction delay and
consider maximum (over messages) error probability.   Hence all the
achievability results also hold when delay and error probability are
averaged over messages.

To see that the converse results in this paper also hold for the
average case, we use the following standard expurgation argument.
Assume $\{(\cC_n,(\tau_n,\phi_n))\}$ is an $(R,\asyncexp)$ coding scheme
where the error probability and the delay of $(\cC_n,(\tau_n,\phi_n))$
are defined as
\begin{equation*}
\eps_n\defeq  \frac{1}{M}\sum_{m=1}^M\pr_m(\cE),
\end{equation*}
and
\begin{equation*}
\bar{\delay}_n\defeq  \frac{1}{M}\sum_{m=1}^M \ex_m(\tau_n-\nu)^+,
\end{equation*}
respectively.  By definition of an $(R,\asyncexp)$ coding scheme, this
means that given some arbitrarily small $\eps>0$, and for all $n$
large enough,
\begin{equation*}
\eps_n\leq \eps
\end{equation*}
and
\begin{equation*}
\frac{\ln M}{\bar{\delay}_n}\geq R-\eps.
\end{equation*}
Hence, for $n$ large enough and any
$\delta>1$, one can find a (nonzero) constant fraction of codewords
${\cC_n}'\subset\cC_n$ (${\cC_n}'$ is the ``expurgated'' ensemble) that
satisfies the following property: the rate defined with respect to
maximum (over ${\cC_n}'$) delay is at least $(R-\eps)/\delta$
and the maximum error probability is less than $\eta \eps$,
where $\eta=\eta(\delta)>0$.   One then applies the converse results
to the expurgated ensemble to derive bounds on $(R/\delta,\asyncexp)$, and
thus on $(R,\asyncexp)$, since $\delta>1$ can be chosen arbitrarily.

\section{Concluding Remarks}
\label{concludingrem}

We analyzed a model for asynchronous communication which captures the situation when
information is emitted infrequently.   General upper and lower bounds on capacity were
derived, which coincide in certain cases.  The forms of these bounds are similar and
have two parts: a mutual information part and a divergence part. The mutual
information part is reminiscent of synchronous communication: to achieve a certain
rate, there must be, on average, enough mutual information between the time information
is sent and the time it is decoded.   The divergence part is novel, and comes from
asynchronism.   Asynchronism introduces two additional error events that must be
overcome by the decoder.   The first event happens when the noise produces a channel
output that looks as if it was generated by a codeword.   The larger the level of
asynchronism, the more likely this event becomes. The second event happens when the
channel behaves atypically, which results in the decoder missing the codeword. When
this event happens, the rate penalty is huge, on the order of the asynchronism level.  
As such, the second event contributes to increased average reaction delay, or
equivalently, lowers the rate.  The divergence part in our upper and lower bounds on
capacity strikes a balance between these two events.

An important conclusion of our analysis is that, in general, training-based
schemes are not optimal in the high rate, high asynchronism regime.
In this regime, training-based architectures are unreliable,
whereas it is still possible to achieve an arbitrarily low probability
of error using strategies that combine synchronization with
information transmission.

Finally, we note that further analysis is possible when we restrict
attention to a simpler slotted communication model in which the
possible transmission slots are nonoverlapping and contiguous.  In
particular, for this more constrained model \cite{WCCW} develops a
variety of results, among which is that except in somewhat
pathological cases, training-based schemes are strictly suboptimal at
all rates below the synchronous capacity.  Additionally, the
performance gap is quantified for the special cases of the binary
symmetric and additive white Gaussian noise channels, where it is seen
to be significant in the high rate regime but vanish in the limit of
low rates.  Whether the characteristics observed for the slotted model
are also shared by unslotted models remains to be determined, and is a
natural direction for future research.

\section*{Acknowledgments}
The authors are grateful to the reviewers for their
insightful and detailed comments which very much contributed
to improve the paper. The authors would also like to
thank the associate editors Suhas Diggavi and Tsachy Weissman and
the editor-in-chief Helmut B\"olcskei for their care in handling
this paper. This paper also benefited from useful
discussions with Sae-Young Chung and Da Wang.

\appendices

%\section{Appendix}
\section{Proof of Remark~\ref{remark} (p.~\pageref{remark})}
\label{remarksection}
To show that the random coding scheme proposed in the proof of
Theorem~\ref{ach} achieves \eqref{caphyp} with equality, we show
that
\begin{align}\label{tisho}
\alpha\leq \max_{P:I(PQ)\geq
R}\min_{V\in\cP^\cY}\max\{D(V\|(PQ)_\cY),D(V\|Q_\star)\}.
\end{align}

Recall that, by symmetry of the
encoding and decoding procedures, the average reaction delay is the
same for any message. Hence
$$\Delta_n=\ex_1(\tau_n-\nu)^+,$$
where $\ex_1$ denotes expectation under the proability measure
$\pr_1$, the channel output distribution when message $1$ is sent, averaged over time and codebooks.

Suppose for the moment that
\begin{align}\label{delayn}
\ex_1(\tau_n-\nu)^+\geq n(1-o(1))\qquad n\to \infty\,.
\end{align}
It then follows from Fano's inequality that the
input distribution $P$ must satisfy $I(PQ)\geq R$. Hence,
to establish \eqref{tisho}
we will show that at least one of the following
inequalities
\begin{align}
D(V\|(PQ)_\cY) &\geq \asyncexp \notag \\
D(V\|Q_\star)&\geq\asyncexp
\label{igny}
\end{align}
holds for any $V\in \cP^\cY$. The arguments are similar to
those used to establish Claim~\ref{claim:iii} of Theorem~\ref{convs}.  Below we provide the
key steps.

We proceed by contradiction and show that if both the inequalities in
\eqref{igny} are reversed, then the asymptotic rate is zero.   To that
aim we provide a lower bound on $\ex_1(\tau_n-\nu)^+$.

Let $\tau'_n$ denote the time of the beginning of the
decoding window, i.e., the first time when the previous
$n$ output symbols have empirical distribution $\hat{P}$
such that $D(\hat{P}||Q_\star)\geq \alpha$. By definition,
$\tau_n\geq \tau'_n$, so
\begin{align} 
\ex_1(\tau_n-\nu)^+&\geq \ex_1(\tau_n'-\nu)^+ \nonumber \\
&\geq
\frac{1}{3}\sum_{t=1}^{\asynclev/3}\pr_{1,t}(\tau_n'\geq 2A_n/3), 
\label{markov2}
\end{align}
where the second inequality follows from Markov's
inequality, and where $\pr_{1,t}$ denotes the probability measure at the output of the
channel conditioned on the event that message $1$ starts being sent at
time $t$, and averaged over codebooks. Note that,  because
$\tau'_n$ is not a function of
the codebook, there is no averaging on the stopping
times.\footnote{For different codebook
realizations, stopping rule $\tau'_n$ is the same, by contrast
with $\tau_n$ which depends on the codebook via
the joint typicality criterion of the second phase.}

Fix $V\in \cP_{\cY}$.   We lower bound
each term $\pr_{1,t}(\tau_n'\geq 2A_n/3)$ in the above sum as
\begin{align}
&\pr_{1,t}(\tau_n'\geq 2A_n/3)\notag\\
&\ \geq \pr_{1,t}(\tau_n'\geq 2A_n/3| Y_t^{t+n-1}\in \cT_V)\,
  \pr_{1,t}(Y_t^{t+n-1}\in \cT_V)\notag \\
&\ \geq \pr_{1,t}(\tau_n\geq 2A_n/3|Y_t^{t+n-1}\in \cT_V) e^{-n D_1}\poly(n),
\label{chgemeas2}
\end{align}
where $D_1\defeq  D(V\|(PQ)_\cY)$, and where the second inequality
follows from Fact~\ref{fact:2}.

The key change of measure step \eqref{key} results now in the equality
\begin{align} 
\label{key2}
\pr_{1,t}(\tau_n'&\geq 2A_n/3|Y_t^{t+n-1}\in \cT_V)\notag \\
&=\pr_\star(\tau_n'\geq 2A_n/3|Y_t^{t+n-1}\in \cT_V),
\end{align}
which can easily be checked by noticing that the probability of any
sequence $y_t^{t+n-1}$ in $\cT_V$ is the same under $\pr_{1,t}$.
Substituting \eqref{key2} into the right-hand side of \eqref{chgemeas2}, and
using \eqref{markov2} and Fact~\ref{fact:2}, we get
\begin{align} 
\ex_1(\tau_n-\nu)^+ &\geq e^{-
n(D_1-D_2)}\poly(n)\notag\\
&\hspace{.3cm}\times \sum_{t=1}^{\asynclev/3} \pr_\star(\tau_n\geq
2A_n/3,Y_t^{t+n-1}\in \cT_V),
\end{align}
where $D_2\defeq  D(V\|Q_\star)$.  The rest of the proof consists
in showing that if the two inequalities in \eqref{igny} are reversed,
then the right-hand side of the above inequality grows exponentially
with $n$, which results in an asymptotic rate equal to zero.  The
arguments closely parallel the ones that prove Claim~\ref{claim:iii}
of Theorem~\ref{convs} (see from \eqref{und} onwards), and hence are
omitted. 

To conclude the proof we show \eqref{delayn}.  Using the
alternate form of expectation for non-negative random variables
$\ex X=\sum_{k\geq 0}\pr(X\geq k)$, we have
\begin{align}
\ex_1(\tau_n-\nu)^+&\geq \sum_{i=1}^{g(n)}\pr_1(\tau_n\geq
\nu+k)\nonumber \\
&\geq \sum_{i=1}^{g(n)}(1-\pr_1(\tau_n<\nu+i))\nonumber \\
&\geq g(n)(1-\pr_1(\tau_n\leq \nu+g(n)))\,,\nonumber
\end{align}
where we defined $$g(n)\defeq  n-\lceil n^{3/4}
\rceil\,,$$ and where the last
inequality follows from the fact that $\pr_1(\tau_n<\nu+i)$ is
a non-decreasing function of $i$.
Since $g(n)=n(1-o(1))$, to establish \eqref{delayn} it
suffices to show that 
\begin{align}\label{eq:tso}
\pr_1(\tau_n
\leq \nu+g(n))=o(1)\qquad
(n\to \infty)\,.
\end{align}

 Since 
$$\pr_1(\tau_n<\nu)=o(1)\qquad (n\to \infty)\,,$$
as follows from computation steps in \eqref{errore} and \eqref{errore2},
to establish \eqref{eq:tso} it suffices to show that
\begin{align}
\label{gxz}\pr_1(\nu\leq \tau_n\leq \nu+g(n))=o(1)\qquad
(n\to \infty)\,.
\end{align}

For $i\in \{0,1,\ldots,g(n)\}$ we have
\begin{align}
&\pr_1(\tau_n=\nu+i)\notag\\
&\leq
\pr_1\left(||\hat{P}_{C^n(1),Y^{\nu+i}_{\nu+i-n+1}}PQ||\leq
\mu \cdot |\cX|\cdot |\cY|\right)\nonumber \\
 &= \sum_J
\pr_1\left(\hat{P}_{C^n(1),Y^{\nu+i}_{\nu+i-n+1}}=J\right)\label{suterm}
 \end{align}
where the above summation is over all typical joint types, i.e., all $J\in
\cP^{\cX,\cY}_n$ such that 
\begin{align}\label{ttype}
|\hat{P}_{C^n(1),Y^{\nu+i}_{\nu+i-n+1}}(a,b)-J(a,b)|\leq
\mu
\end{align}
for all $(a,b)\in \cX\times \cY$.

We upper bound each term in this summation. First observe that event 
$$\{ \hat{P}_{C^n(1),Y^{\nu+i}_{\nu+i-n+1}}=J\}\,,$$
for $i\in \{0,1,\ldots,g(n)\}$, involves random vector $Y^{\nu+i}_{\nu+i-n+1}$ which is partly generated by
noise and partly generated by the transmitted codeword corresponding to message
$1$. In the following computation $k$ refers to first symbols of
$Y^{\nu+i}_{\nu+i-n+1}$ which are generated by noise, i.e., by
definition $k=n-(i+1)$. Note that since $0\leq i\leq g(n)$, we have 
$$ \lceil n^{3/4}\rceil -1\leq k\leq n-1\,.$$ 
We have
\begin{align}\label{o1}
&\pr_{1}(\hat{P}_{C^n(1),Y_{\nu+i-n+1}^{\nu+i}}=J)\notag\\
&=\sum_{\substack{J_1\in \cP_k \\ J_2\in
\cP_{n-k}\\
kJ_1+(n-k)J_2=nJ}}\left(\sum_{(x^k,y^k):\hat{P}_{x^k,y^k}=J_1}P(x^k)Q_\star(y^k)\right)\nonumber\\
&\hspace{.5cm}\times\left(\sum_{(x^{n-k},y^{n-k}):\hat{P}_{x^{n-k},y^{n-k}}=J_2}\pr(x^{n-k},y^{n-k})\right)\,,
\end{align}
where we used the following shorthand notations for probabilities
\begin{align*}
P(x^k)&\defeq  \prod_{j=1}^kP(x_j)\nonumber \\
Q_\star(y^k)&\defeq  \prod_{j=1}^kQ_\star(y_j)\nonumber\\
\pr(x^{n-k},y^{n-k})&\defeq  \prod_{j=1}^kP(x_j)Q(y_j|x_j)\,.
\end{align*}
Further, using Fact~\ref{fact:2}
\begin{align}\label{o2}
\sum_{(x^k,y^k):\hat{P}_{x^k,y^k}=J_1}&P(x^k)P_\star(y^k)\nonumber \\
&\hspace{-.3cm}=\sum_{x^k:\hat{P}_{x^k}=J_{1,\cX}}P(x^k)\sum_{y^k:\hat{P}_{y^k}=J_{1,\cY}}Q_\star(y^k)\nonumber
\\
&\hspace{-.3cm}\leq e^{-k(D(J_{1,\cX}||P)+D(J_{1,\cY}||Q_\star))}\nonumber\\
&\hspace{-.3cm}\leq e^{-kD(J_{1,\cY}||Q_\star)}\end{align}
where $J_{1,\cX}$ and $J_{1,\cY}$ denote the left and
right marginals of $J$, respectively, and where the second inequality follows by
non-negativity of divergence. 

A similar calculation yields
\begin{align}\label{o3}
\sum_{(x^{n-k},y^{n-k}):\hat{P}_{x^{n-k},y^{n-k}}=J_2}&\pr(x^{n-k},y^{n-k})\notag\\
&\leq
e^{-(n-k)D(J_2||PQ)}
\end{align}
From \eqref{o1}, \eqref{o2}, \eqref{o3} and Fact~\ref{fact:1}
we get
\begin{align}\label{bh}
&\pr_{1}(\hat{P}_{C^n(1),Y_{\nu+i-n+1}^{\nu+i}}=J)\nonumber \\
&\leq \poly(n)\notag\\
&\hspace{.5cm}\times \max_{\substack{J_1\in \cP_k^{\cX,\cY} \\ J_2\in
\cP_{n-k}^{\cX,\cY}\\
kJ_1+(n-k)J_2=nJ\\
 k:\lceil n^{3/4}\rceil -1\leq k\leq n-1}}\exp\Big[-
 k(D(J_{1,\cY}||Q_\star))\notag \\
 &\hspace{4cm}-(n-k)D(J_2||PQ)\Big].
 \end{align}
 
The maximum on the right-hand side of \eqref{bh} is equal to
 \begin{align}\label{marci}
&\max_{\substack{J_1\in \cP_k^{\cY} \\
J_2\in
\cP_{n-k}^{\cY}\\
kJ_1+(n-k)J_2=nJ_\cY\\
  k:\lceil n^{3/4}\rceil -1\leq k\leq n-1}}\exp\Big[-kD(J_{1}||Q_\star)\notag\\
  &\hspace{2cm}-(n-k)D(J_{2}||(PQ)_\cY\big)\Big]\,.
  \end{align}
We upper bound the argument of the above exponential via the log-sum inequality 
to get
\begin{align}\label{marci2}
&-kD(J_{1}||Q_\star)-(n-k)D(J_{2}||(PQ)_\cY\big)\nonumber\\
&\leq - nD \big(J_\cY\big|\big|\delta Q_\star+(1-\delta) (PQ)_\cY\big),
\end{align}
where $\delta\defeq k/n$. Using \eqref{marci2}, we upper-bound expression
\eqref{marci} by
  \begin{align}
&\max_{
  \delta :n^{-1/4} -n^{-1}\leq \delta\leq  1}\nonumber\\
  &\hspace{1cm}\exp\Big[-
  n D \big(J_\cY||\delta Q_\star+(1-\delta) (PQ)_\cY\big)\Big]\nonumber\\
&\hspace{1cm}\leq \max_{
  \delta :n^{-1/4} -n^{-1}\leq \delta\leq  1}\exp\left[-
  n\Omega(\delta^2)\right]\nonumber\\
&\hspace{1cm}\leq \exp\left[-\Omega(n^{1/2})\right],
\end{align}
where for the first inequality we used Pinsker's inequality \cite[Problem $17$ p.
58]{CK}
$$D(P_1||P_2)\geq \frac{1}{2\ln 2}||P_1-P_2||^2,$$
and assume that $\mu$ is small enough and $n$ is large
enough for this inequality to be valid.
Such $\mu$ and $n$ exist whenever the distributions 
$Q_\star$ and $(PQ)_\cY$ are different.

It then follows from \eqref{bh} that
\begin{align*}
&\pr_{1}(\hat{P}_{C^n(1),Y_{\nu+i-n+1}^{\nu+i}}=J)\leq
\exp\left[-\Omega(n^{1/2})\right]\,,
\end{align*}
hence, from \eqref{suterm} and Fact~\ref{fact:1} we
get
$$\pr_1(\tau_n=\nu+i)\leq
\exp\left[-\Omega(n^{1/2})\right]$$
for $ i\in
\{0,1,\ldots,g(n)\}$. Finally a union bound over times yields the desired result
\eqref{eq:tso} since $g(n)=O(n)$.

\section{Proof of Theorem~\ref{prop:expurgation}}
\label{app:expurgation}

The desired Theorem is a stronger version of \cite[Corollary~1.9,
  p.~107]{CK}, and its proof closely follows the proof of the latter. 

Before proceeding, we recall the definitions of $\eta$-image and
$l$-neighborhood of a set of sequences.
\begin{defn}[$\eta$-image, \cite{CK}Definition 2.1.2 p. 101] A set ${\mathcal{B}} \subset
{\mathcal{Y}}^n$ is an $\eta$-image of a set ${\mathcal{A}}\subset
{\mathcal{X}}^n$ if $Q({\mathcal{B}}|x)\geq \eta$ for all $x\in
{\mathcal{A}}$. The minimum cardinality of $\eta$-images of
${\mathcal{A}}$  is denoted $g_Q({\mathcal{A}},\eta)$.
\end{defn}
\begin{defn}[$l$-neighborhood, \cite{CK} p. 86] The~$l$-neighborhood of a set ${\mathcal{B}} \subset
{\mathcal{Y}}^n$ is the set
$$\Gamma^l {\mathcal{B}}\defeq  \{y^n\in {\mathcal{Y}}^n:
d_H(\{y^n\},{\mathcal{B}} )\leq l\}$$
where $d_H(\{y^n\},{\mathcal{B}} )$ denotes the Hamming distance
between $y^n$ and ${\mathcal{B}}$, i.e.,
$$d_H(\{y^n\},{\mathcal{B}} )=\min_{\tilde{y}^n\in {\mathcal{B}}}
d_H(y^n,\tilde{y}^n)\,.$$
\end{defn}
As other notation, for a given conditional probability $Q(y|x)$,
$(x,y)\in \cX\times \cY$, and $x^n\in \cX^n$, we define the set
\begin{align*} 
&\cT_{[Q]}^n(x^n)= \Bigl\{ y^n\in \cY^n:\\
&\ |\hat{P}_{x^n,y^n}(a,b)-\hat{P}_{x^n}(a)Q(b|a)|>
q,\ \forall (a,b)\in \cX\times \cY \Bigr\}
\end{align*}
for a constant $q>0$.
To establish Theorem~\ref{prop:expurgation}, we make use of the
following three lemmas. Since we restrict attention to block coding schemes, i.e., coding
scheme whose decoding happens at the fixed time $n$, we denote them
simply by $(\cC_n,\phi_n)$ instead of
$(\cC_n,(\gamma_n,\phi_n))$.  

In the following, $\eps_n$ is always given by 
\begin{equation*}
\eps_n=(n+1)^{|\cX|\cdot|\cY|}\exp(-nq^2/(2\ln 2)).
\end{equation*}

\begin{lem}\label{ap1}
Given $\gamma\in (0,1)$, $Q\in \cP^{\cY|\cX}$, $P\in \cP^\cX_n$, and $\cA\subset
\cT^n_P$, there exist
$(\cC_n,\phi_n)$ for each $n\geq n_\circ(\gamma,q,|\cX|,|\cY|)$ such that
\begin{enumerate}
\item \label{lemi:i}
$c^n(m)\in \cA$, for all $c^n(m)\in \cC_n$
\item \label{lemi:ii}
$\phi^{-1}_n(m)\subset \cT^n_{[Q]}(c^n(m))$, $m\in \{1,2,\ldots,M\}$
\item \label{lemi:iii}
the maximum error probability is upper bounded by $2\eps_n$
\item \label{lemi:iv} the rate satisfies
\begin{equation*}
\frac{1}{n}\ln |\cC_n|\geq
\frac{1}{n}\ln g_{Q}(\cA,\eps_n)-H(Q|P)-\gamma.
\end{equation*}
\end{enumerate}
\end{lem}

\begin{IEEEproof}[Proof of Lemma~\ref{ap1}]
The proof closely follows the proof of \cite[Lemma~1.3, p.~101]{CK} since it essentially suffices to replace $\eps$ and $\gamma$ in the proof of
\cite[Lemma~1.3, p.~101]{CK} with $2\eps_n$ and
$\eps_n$, respectively. We therefore omit the details here.  

One of the steps of the proof consists in showing that
\begin{equation} 
\label{erpr}
Q(\cT_{[Q]}^n(x^n)|x^n)\geq 1-\eps_n
\end{equation}
for all $x^n\in \cX^n$. To establish this, one proceeds as follow.  Given $P\in
\cP^\cX_n$ let $\cD$ denote the set of empirical conditional
distributions $W(y|x)\in \cP^{\cY|\cX}_n$ such that
\begin{equation*}
|\hat{P}_{x^n}(a)W(b|a)-\hat{P}_{x^n}(a)Q(b|a)|>
q
\end{equation*}
for all $(a,b)\in \cX\times\cY$.   We have
\begin{align}
&1-Q(\cT_{[Q]}^n(x^n)|x^n)\notag\\
&\ = \sum_{W\in \cD\cap
\cP^{\cY|\cX}_n} Q(\cT_{W}^n(x^n)|x^n) \\
&\ \leq \sum_{W\in \cD\cap \cP^{\cY|\cX}_n}e^{-nD(W\|Q|P)}
\label{eq:first-Qbar}\\
&\ \leq (n+1)^{|\cX|\cdot|\cY|}\exp(-n\min_{W\in
\cD}D(W\|Q|P))\label{eq:second-Qbar} \\
&\ \leq (n+1)^{|\cX|\cdot|\cY|}\exp(-n\min_{W\in
  \cD}\|PW-PQ\|^2/2\ln 2) \label{eq:third-Qbar}\\
&\ \leq (n+1)^{|\cX|\cdot|\cY|}\exp(-nq^2/(2\ln 2)\label{eq:fourth-Qbar}\\
&\ = \eps_n,\nonumber
\end{align}
which shows \eqref{erpr}. Inequality \eqref{eq:first-Qbar} follows
from Fact~\ref{fact:3}, \eqref{eq:second-Qbar} follows from Fact \ref{fact:1}, \eqref{eq:third-Qbar} follows from Pinsker's
inequality (see, e.g., \cite[Problem~17, p.~58]{CK}), and
\eqref{eq:fourth-Qbar} follows from the definition of $\cD$.
\end{IEEEproof}

\begin{lem}[{\cite[Lemma~1.4, p.~104]{CK}}]
\label{ap2}
For every $\eps, \gamma\in (0,1)$, if $(\cC_n,\phi_n)$ achieves
an error probability $\eps$ and $\cC_n\subset \cT_P^n$, then
\begin{equation*}
\frac{1}{n} \ln|\cC_n| <
\frac{1}{n}\ln g_{Q}(\cC_n,\eps+\gamma) - H(Q|P) + \gamma
\end{equation*}
whenever $n\geq n_\circ(|\cX|,|\cY|,\gamma)$.
\end{lem}
Since this lemma is established in \cite[Lemma~1.4, p.~104]{CK}, we
omit its proof.

\begin{lem}\label{ap3}
For every $\gamma>0$, $\eps \in (0,1)$, $Q\in
\cP^{\cY|\cX}$, and $\cA\subset \cX^n$
\begin{equation*}
\big| \frac{1}{n}\ln g_{Q}(\cA,\eps)-\frac{1}{n}\ln
g_{Q}(\cA,\eps_n)\big|<\gamma
\end{equation*}
whenever $n\geq n_\circ(\gamma,q,|\cX|,|\cY|)$.
\end{lem}

\begin{IEEEproof}[Proof of Lemma~\ref{ap3}]
By the Blowing Up Lemma \cite[Lemma~1.5.4, p.~92]{CK} and
\cite[Lemma~1.5.1, p.~86]{CK}, given the sequence $\{\eps_n\}_{n\geq
  1}$, there exist $\{l_n\}$ and $\{\eta_n\}$ such that
  $l_n/n\overset{n\to \infty}{\longrightarrow}0$ and
  $\eta_n\overset{n\to\infty}{\longrightarrow} 1$, and such that the
  following two properties hold.

For any $\gamma>0$ and $n\geq n_\circ(\gamma, q,|\cX|,|\cY|)$
\begin{equation} 
\label{gammaapprox}
\frac{1}{n}\ln |\Gamma^{l_n}\cB|-\frac{1}{n}\ln |\cB|<\gamma 
\quad \text{for every }\cB\subset \cY^n,
\end{equation}
and for all $x^n\in \cX^n$,
\begin{equation} 
\label{bllemma}
Q(\Gamma^{l_n}\cB|x^n)\geq \eta_n\quad \text{whenever}\quad
Q(\cB|x^n)\geq \eps_n.
\end{equation}
Now, assuming that $\cB$ is an $\eps_n$-image of $\cA$ with
$|\cB|=g_{Q}(\cA,\eps_n)$, the relation \eqref{bllemma} means
that $\Gamma^{l_n}\cB$ is an $\eta_n$-image of $\cA$.   Therefore we
get
\begin{align}
\frac{1}{n}\ln g_{Q}(\cA,\eta_n)&\leq \frac{1}{n}\ln |\Gamma^{l_n}\cB|\notag
\\
&\leq \gamma+ \frac{1}{n}\ln |\cB|\notag \\
&=\gamma+ \frac{1}{n}\ln g_{Q}(\cA,\eps_n)
\label{lb}
\end{align}
where the second inequality follows from \eqref{gammaapprox}.
Finally, since $\eta_n\rightarrow1$ and $\eps_n\rightarrow0$ as
$n\rightarrow\infty$, for $n$ large enough we have
\begin{equation*}
g_{Q}(\cA,\eps)\leq g_{Q}(\cA,\eta_n)\quad
\text{and}\quad \eps_n\leq \eps,
\end{equation*}
and therefore from \eqref{lb} we get
\begin{equation*}
\frac{1}{n}\ln g_{Q}(\cA,\eps)\leq \frac{1}{n}\ln
g_{Q}(\cA,\eps_n)\leq  \gamma+\frac{1}{n}\ln
g_{Q}(\cA,\eps)
\end{equation*}
yielding the desired result.
\end{IEEEproof}

We now use these lemmas to establish
Theorem~\ref{prop:expurgation}.  Choose $\eps,\gamma>0$ such that
$\eps+\gamma<l$.  Let $(\cC_n,\phi_n)$ be a coding scheme that achieves
maximum error probability $\eps$. Without loss of generality, we assume
that $\cC_n\subset \cT_P^n$ (If not, group codewords into families of
common type. The largest family of codewords has error probability no
larger than $\eps$, and its rate is essentially the same as the rate
of the original code $\cC_n$.)
Therefore
\begin{align} \label{aab}
\frac{1}{n}\ln |\cC_n|&\leq \frac{1}{n}\ln
g_{Q}(\cC_n,\eps+\gamma)-H(Q|P)+\gamma\notag \\ &\leq
\frac{1}{n}\ln g_{Q}(\cC_n,l)-H(Q|P)+\gamma\notag \\ &\leq
\frac{1}{n}\ln g_{Q}(\cC_n,\eps_n)-H(Q|P)+2\gamma
\end{align}
for $n\geq n_\circ(\gamma, l,|\cX|,|\cY|)$, where the first and third inequalities follow
from Lemmas~\ref{ap2} and~\ref{ap3}, respectively, and where the
second inequality follows since $g_{Q}(\cC_n,\eps)$ is
nondecreasing in $\eps$.   On the other hand, by
Lemma~\ref{ap1}, there exists a coding scheme $(\cC_n',\phi_n')$, with
$\cC_n'\subset \cC_n$ that achieves a probability of error
upper bounded by $2\eps_n$ and such that its rate satisfies
\begin{equation}\label{bbai}
\frac{1}{n}\ln |\cC_n'| \geq \frac{1}{n}\ln
g_{Q}(\cC_n,\eps_n)-H(Q|P)-\gamma
\end{equation}
for $n\geq n_\circ(\gamma, q,|\cX|,|\cY|)$.  From \eqref{aab} and
\eqref{bbai} we deduce the rate of $\cC_n'$ is lower bounded as
$$\frac{1}{n}\ln |\cC_n'| \geq \frac{1}{n}\ln |\cC_n|
-3\gamma$$
whenever $n\geq n_\circ(\gamma,l, q,|\cX|,|\cY|)$. This  yields the desired
result.
\hfill\IEEEQEDclosed

\bibliographystyle{IEEEtran}
\bibliography{../../../../../common_files/bibiog}

\begin{thebibliography}{10}
\providecommand{\url}[1]{#1}
\csname url@samestyle\endcsname
\providecommand{\newblock}{\relax}
\providecommand{\bibinfo}[2]{#2}
\providecommand{\BIBentrySTDinterwordspacing}{\spaceskip=0pt\relax}
\providecommand{\BIBentryALTinterwordstretchfactor}{4}
\providecommand{\BIBentryALTinterwordspacing}{\spaceskip=\fontdimen2\font plus
\BIBentryALTinterwordstretchfactor\fontdimen3\font minus
  \fontdimen4\font\relax}
\providecommand{\BIBforeignlanguage}[2]{{%
\expandafter\ifx\csname l@#1\endcsname\relax
\typeout{** WARNING: IEEEtran.bst: No hyphenation pattern has been}%
\typeout{** loaded for the language `#1'. Using the pattern for}%
\typeout{** the default language instead.}%
\else
\language=\csname l@#1\endcsname
\fi
#2}}
\providecommand{\BIBdecl}{\relax}
\BIBdecl

\bibitem{TCW2}
A.~Tchamkerten, V.~Chandar, and G.~Wornell, ``On the capacity region of
  asynchronous channels,'' in \emph{Proc. IEEE Int. Symp. Information Theory
  (ISIT)}, 2008.

\bibitem{CTW2}
V.~Chandar, A.~Tchamkerten, and G.~Wornell, ``Training-based schemes are
  suboptimal for high rate asynchronous communication,'' in \emph{Proc. IEEE
  Information Theory Work. (ITW)}, Taormina, October 2009.

\bibitem{TCW}
A.~Tchamkerten, V.~Chandar, and G.~Wornell, ``Communication under strong
  asynchronism,'' \emph{IEEE Trans.~Inform.~Th.}, vol.~55, no.~10, pp.
  4508--4528, October 2009.

\bibitem{Sha2}
C.~E. Shannon, ``A mathematical theory of communication,'' \emph{Bell Sys.~
  Tech. J.}, vol.~27, pp. 379--423, October 1948.

\bibitem{CTT}
V.~Chandar, A.~Tchamkerten, and D.~Tse, ``Asynchronous capacity per unit
  cost,'' in \emph{Proc. IEEE Int. Symp. Information Theory (ISIT)}, june 2010,
  pp. 280 --284.

\bibitem{CTTj}
------, ``Asynchronous capacity per unit cost,'' \emph{CoRR}, vol.
  abs/1007.4872, 2010.

\bibitem{CK}
I.~Csisz\`ar and J.~K\"orner, \emph{Information Theory: Coding Theorems for
  Discrete Memoryless Channels}.\hskip 1em plus 0.5em minus 0.4em\relax New
  York: Academic Press, 1981.

\bibitem{CLRS}
T.~H. Cormen, C.~E. Leiserson, R.~L. Rivest, and C.~Stein, \emph{Introduction
  to Algorithms, 2nd edition}.\hskip 1em plus 0.5em minus 0.4em\relax {MIT}
  Press, McGraw-Hill Book Company, 2000.

\bibitem{CTW}
V.~Chandar, A.~Tchamkerten, and G.~Wornell, ``Optimal sequential frame
  synchronization,'' \emph{IEEE Trans.~Inform.~Th.}, vol.~54, no.~8, pp.
  3725--3728, 2008.

\bibitem{CN2}
I.~Csisz{\'a}r and P.~Narayan, ``Arbitrarily varying channels with constrained
  inputs and states,'' \emph{IEEE Transactions on Information Theory}, vol.~34,
  no.~1, pp. 27--34, 1988.

\bibitem{CT}
T.~Cover and J.~Thomas, \emph{Elements of information theory}.\hskip 1em plus
  0.5em minus 0.4em\relax New York: Wiley, 2006.

\bibitem{G}
R.~G. Gallager, \emph{Information Theory and Reliable Communication}.\hskip 1em
  plus 0.5em minus 0.4em\relax Budapest: Wiley, 1968.

\end{thebibliography}

\end{document}